\documentclass[12pt]{article}
\usepackage{amssymb,amscd,array}
\usepackage[active]{srcltx}
\catcode `\@=11
\@addtoreset{equation}{section}

\newtheorem{thm}{Theorem}[section]
\newtheorem{rem}[thm]{Remark}

\newtheorem{prop}[thm]{Proposition}

\def\qed{\blacksquare}
\newcommand{\be}{\begin{equation}}
\newcommand{\ee}{\end{equation}}
\newcommand{\bea}{\begin{eqnarray}}
\newcommand{\eea}{\end{eqnarray}}
\newcommand{\R}{\mathbb{R}}

\newcommand{\C}{\mathbb{C}}

\begin{document}
\begin{titlepage}

\begin{center}
{\bf \Large{Gravity in Causal Perturbative Quantum Field Theory \\}}
\end{center}
\vskip 1.0truecm
\centerline{D. R. Grigore, 
\footnote{e-mail: grigore@theory.nipne.ro}}
\vskip5mm
\centerline{Department of Theoretical Physics,}
\centerline{Institute for Physics and Nuclear Engineering ``Horia Hulubei"}
\centerline{Bucharest-M\u agurele, P. O. Box MG 6, ROM\^ANIA}

\vskip 2cm
\bigskip \nopagebreak
\vskip 1cm
\begin{abstract}
\noindent
We extend the general framework of perturbative quantum field theory developped for the pure Yang-Mills model to gravity. 
First we present a variant of the elimination procedure of the anomalies in the second order of perturbation theory. 
After that we prove that gauge invariance restricts severely the possible finite renormalizations, so at least in the second
order of the perturbation theory, the model is renormalizable.
\end{abstract}

\end{titlepage}

\section{Introduction}

The most natural way to arrive at the Bogoliubov axioms of perturbative quantum field theory (pQFT) is by analogy with non-relativistic 
quantum mechanics \cite{Gl}, \cite{H}, \cite{D}, \cite{DF}: in this way one arrives naturally at Bogoliubov axioms 
\cite{BS}, \cite{EG}, \cite{Sc1}, \cite{Sc2}. We prefer the formulation from \cite{DF} and as presented in \cite{algebra}; 
for every set of monomials 
$ 
A_{1}(x_{1}),\dots,A_{n}(x_{n}) 
$
in some jet variables (associated to some classical field theory) one associates the operator-valued distributions
$ 
T^{A_{1},\dots,A_{n}}(x_{1},\dots,x_{n})
$  
called chronological products; it will be convenient to use another notation: 
$ 
T(A_{1}(x_{1}),\dots,A_{n}(x_{n})). 
$ 

The Bogoliubov axioms, presented in Section \ref{Bogoliubov} express essentially some properties of the scattering matrix understood as a 
formal perturbation
series with the ``coefficients" the chronological products: 
(1)  
(skew)symmetry properties in the entries 
$ 
A_{1}(x_{1}),\dots,A_{n}(x_{n}) 
$;
(2)
Poincar\'e invariance; 
(3)
causality; 
(4)
unitarity; 
(5)
the ``initial condition" which says that
$
T(A(x)) 
$
is a Wick polynomial.

So we need some basic notions on free fields and Wick monomials. One can supplement these axioms by requiring 
(6) 
power counting;
(7)
Wick expansion property. 

It is a highly non-trivial problem to find solutions for the Bogoliubov axioms, 
even in the simplest case of a real scalar field. 

There are, at least to our knowledge, three rigorous ways to do that; for completeness we remind them following 
\cite{ano-free}: 
(a)
{\it Hepp axioms} \cite{H};
(b)
{\it Polchinski flow equations} \cite{P}, \cite{S};
(c)
{\it the causal approach} due to Epstein and Glaser \cite{EG}, \cite{Gl} which we prefer. 

The procedure of Epstein and Glaser is a recursive construction for the basic objects
$ 
T(A_{1}(x_{1}),\dots,A_{n}(x_{n}))
$
and reduces the induction procedure to a distribution splitting of some distributions with causal support.  
In an equivalent way, one can reduce the induction procedure to the process of extension of distributions \cite{PS}. 

An equivalent point of view uses retarded products \cite{St1} instead of chronological products. For gauge models one has to deal 
with non-physical fields (the so-called ghost fields) and impose a supplementary axiom namely  gauge invariance, which guarantees that the 
physical states are left invariant by the chronological products.

We will extend the analysis from \cite{wick+hopf} to the case of (pure and massless) gravity. We will consider only the  second order 
of the perturbation theory i.e. only chronological products of the type 
$
T(T(x_{1}),T(x_{2})).
$

All these chronological products can be split in the loop and tree contributions. The loop contributions have been analysed in detail in
\cite{sr-gr} where it was proved that they are trivial i.e. the are coboundaries.

In the next Section we give the basic facts on perturbative quantum field theory of quantum gravity: the construction of Wick monomials, 
Bogoliubov axioms and the framework of perturbative quantum gravity.  
In Section \ref{tree} we investigate the tree contributions. As it is known, they produce anomalies; we compute in detail these anomalies
for the basic chronological products of the type
$
T(T(x_{1}),T(x_{2})). 
$
Then we show an elementary way of eliminating the anomalies by redefinitions of the chronological products. In Section \ref{finite} 
we consider the question of the renormalizability of quantum gravity. It is asserted in the literature that quantum gravity is a
non-renormalizable theory, this meaning that the arbitrariness of the chronological products (given by quasi-local operators) increases
with the order of the perturbation theory. We consider this problem in the second order of perturbation theory. After fixing
gauge invariance in the form
\be
sT(T(x_{1}),T(x_{2})) \equiv d_{Q}T(T(x_{1}),T(x_{2})) 
- i~\partial_{\mu}^{1}T(T^{\mu}(x_{1}),T(x_{2})) - i~\partial_{\mu}^{2}T(T(x_{1}),T^{\mu}(x_{2})) = 0 
\ee
we are left with an arbitrariness of the form
\bea
R(T^{I}(x_{1}),T^{J}(x_{2})) = \delta(x_{1} - x_{2})~N(T^{I},T^{J})(x_{2})
+ \partial_{\mu}\delta(x_{1} - x_{2})~N(T^{I},T^{J})^{\mu}(x_{2})
\nonumber\\
+ \partial_{\mu}\partial_{\nu}(x_{1} - x_{2})~N(T^{I},T^{J})^{\mu\nu}(x_{2}).
\label{R-tt}
\eea
However this arbitrariness is constrained by the requirement that it preserve the gauge invariance just established so it must 
verify:
\be
sR(T(x_{1}),T(x_{2})) = 0.
\label{s-TT}
\ee
Also, not all solutions af this (cohomological) equation are physically semnificative: we have trivial solutions (coboundaries) 
of the form
\bea
R(T(x_{1}),T(x_{2})) = \bar{s}B(T(x_{1}),T(x_{2})) = 
\nonumber\\
d_{Q}B(T(x_{1}),T(x_{2})) 
+ i~\partial_{\mu}^{1}B(T^{\mu}(x_{1}),T(x_{2})) + i~\partial_{\mu}^{2}B(T(x_{1}),T^{\mu}(x_{2}))
\label{R-B}
\eea
because such finite renormalizations give null contributions when restricted to the physical subspace. The new result is that 
any solution of the type (\ref{R-tt}) quadri-linear in the fields (and their derivatives) of the cocyle equation (\ref{s-TT}) is a 
coboundary i.e. of the form (\ref{R-B}). So we are left only with the solution tri-linear in the variables - see (\ref{T}), (\ref{T-int}).
Such terms can be eliminated by a redefinition of the coupling constant.

It natural to expect that such a result will be true in arbitrary orders of the perturbation theory, making perturbative quantum
gravity a bona fidae theory like the standard model. 
\newpage
\section{Perturbative Quantum Field Theory\label{pQFT}}
There are two main ingrediants in the contruction of a perturbative quantum field theory (pQFT): the construction of the Wick monomials 
and the Bogoliubov axioms. For a pQFT of Yang-Mills theories one needs one more ingrediant, namely the introduction of ghost fields and
gauge charge.

\subsection{Wick Products\label{wick prod}}

We follow the formalism from \cite{algebra}. We consider a classical field theory on the Minkowski space
$
{\cal M} \simeq \R^{4}
$
(with variables
$
x^{\mu}, \mu = 0,\dots,3
$
and the metric $\eta$ with 
$
diag(\eta) = (1,-1,-1,-1)
$)
described by the Grassmann manifold 
$
\Xi_{0}
$
with variables
$
\xi_{a}, a \in {\cal A}
$
(here ${\cal A}$ is some index set) and the associated jet extension
$
J^{r}({\cal M}, \Xi_{0}),~r \geq 1
$
with variables 
$
x^{\mu},~\xi_{a;\mu_{1},\dots,\mu_{n}},~n = 0,\dots,r;
$
we denote generically by
$
\xi_{p}, p \in P
$
the variables corresponding to classical fields and their formal derivatives and by
$
\Xi_{r}
$
the linear space generated by them. The variables from
$
\Xi_{r}
$
generate the algebra
$
{\rm Alg}(\Xi_{r})
$
of polynomials.

To illustrate this, let us consider a real scalar field in Minkowski space ${\cal M}$. The first jet-bundle extension is
$$
J^{1}({\cal M}, \R) \simeq {\cal M} \times \R \times \R^{4}
$$
with coordinates 
$
(x^{\mu}, \phi, \phi_{\mu}),~\mu = 0,\dots,3.
$

If 
$
\varphi: \cal M \rightarrow \R
$
is a smooth function we can associate a new smooth function
$
j^{1}\varphi: {\cal M} \rightarrow J^{1}(\cal M, \R) 
$
according to 
$
j^{1}\varphi(x) = (x^{\mu}, \varphi(x), \partial_{\mu}\varphi(x)).
$

For higher order jet-bundle extensions we have to add new real variables
$
\phi_{\{\mu_{1},\dots,\mu_{r}\}}
$
considered completely symmetric in the indexes. For more complicated fields, one needs to add supplementary indexes to
the field i.e.
$
\phi \rightarrow \phi_{a}
$
and similarly for the derivatives. The index $a$ carries some finite dimensional representation of
$
SL(2,\C)
$
(Poincar\'e invariance) and, maybe a representation of other symmetry groups. 
In classical field theory the jet-bundle extensions
$
j^{r}\varphi(x)
$
do verify Euler-Lagrange equations. To write them we need the formal derivatives defined by
\be
d_{\nu}\phi_{\{\mu_{1},\dots,\mu_{r}\}} \equiv \phi_{\{\nu,\mu_{1},\dots,\mu_{r}\}}.
\ee

We suppose that in the algebra 
$
{\rm Alg}(\Xi_{r})
$
generated by the variables 
$
\xi_{p}
$
there is a natural conjugation
$
A \rightarrow A^{\dagger}.
$
If $A$ is some monomial in these variables, there is a canonical way to associate to $A$ a Wick 
monomial: we associate to every classical field
$
\xi_{a}, a \in {\cal A}
$
a quantum free field denoted by
$
\xi^{\rm quant}_{a}(x), a \in {\cal A}
$
and determined by the $2$-point function
\be
<\Omega, \xi^{\rm quant}_{a}(x), \xi^{\rm quant}_{b}(y) \Omega> = - i~D_{ab}^{(+)}(x - y)\times {\bf 1}.
\label{2-point}
\ee
Here 
\be
D_{ab}(x) = D_{ab}^{(+)}(x) + D_{ab}^{(-)}(x)
\ee
is the causal Pauli-Jordan distribution associated to the two fields; it is (up to some numerical factors) a polynomial
in the derivatives applied to the Pauli-Jordan distribution. We understand by 
$
D^{(\pm)}_{ab}(x)
$
the positive and negative parts of
$
D_{ab}(x)
$.
From (\ref{2-point}) we have
\be
[ \xi_{a}(x), \xi_{b}(y) ] = - i~ D_{ab}(x - y) \times {\bf 1} 
\ee
where by 
$
[\cdot, \cdot ]
$
we mean the graded commutator. 

The $n$-point functions for
$
n \geq 3
$
are obtained assuming that the truncated Wightman functions are null: see \cite{BLOT}, relations (8.74) and (8.75) and proposition 8.8
from there. The definition of these truncated Wightman functions involves the Fermi parities
$
|\xi_{p}|
$
of the fields
$
\xi_{p}, p \in P.
$

Afterwards we define
$$
\xi^{\rm quant}_{a;\mu_{1},\dots,\mu_{n}}(x) \equiv \partial_{\mu_{1}}\dots \partial_{\mu_{n}}\xi^{\rm quant}_{a}(x), a \in {\cal A}
$$
which amounts to
\be
[ \xi_{a;\mu_{1}\dots\mu_{m}}(x), \xi_{b;\nu_{1}\dots\nu_{n}}(y) ] =
(-1)^{n}~i~\partial_{\mu_{1}}\dots \partial_{\mu_{m}}\partial_{\nu_{1}}\dots \partial_{\nu_{n}}D_{ab}(x - y )\times {\bf 1}.
\label{2-point-der}
\ee
More sophisticated ways to define the free fields involve the GNS construction. 

The free quantum fields are generating a Fock space 
$
{\cal F}
$
in the sense of the Borchers algebra: formally it is generated by states of the form
$
\xi^{\rm quant}_{a_{1}}(x_{1})\dots \xi^{\rm quant}_{a_{n}}(x_{n})\Omega
$
where 
$
\Omega
$
the vacuum state.
The scalar product in this Fock space is constructed using the $n$-point distributions and we denote by
$
{\cal F}_{0} \subset {\cal F}
$
the algebraic Fock space.

One can prove that the quantum fields are free, i.e. they verify some free field equation; in particular every field must verify 
Klein Gordon equation for some mass $m$
\be
(\square + m^{2})~\xi^{\rm quant}_{a}(x) = 0
\label{KG}
\ee
and it follows that in momentum space they must have the support on the hyperboloid of mass $m$. This means that 
they can be split in two parts
$
\xi^{\rm quant (\pm)}_{a}
$
with support on the upper (resp. lower) hyperboloid of mass $m$. We convene that 
$
\xi^{\rm quant (+)}_{a} 
$
resp.
$
\xi^{\rm quant (-)}_{a} 
$
correspond to the creation (resp. annihilation) part of the quantum field. The expressions
$
\xi^{\rm quant (+)}_{p} 
$
resp.
$
\xi^{\rm quant (-)}_{p} 
$
for a generic
$
\xi_{p},~ p \in P
$
are obtained in a natural way, applying partial derivatives. For a general discussion of this method of constructing free fields, see 
ref. \cite{BLOT} - especially prop. 8.8.
The Wick monomials are leaving invariant the algebraic Fock space.
The definition for the Wick monomials is contained in the following Proposition.

\begin{prop}
The operator-valued distributions
$
N(\xi_{q_{1}}(x_{1}),\dots,\xi_{q_{n}}(x_{n}))
$
are uniquely defined by:

\be
N(\xi_{q_{1}}(x_{1}),\dots,\xi_{q_{n}}(x_{n}))\Omega = \xi_{q_{1}}^{(+)}(x_{1})\dots  \xi_{q_{n}}^{(+)}(x_{n})\Omega
\ee

\bea
[ \xi_{p}(y), N(\xi_{q_{1}}(x_{1}),\dots,\xi_{q_{n}}(x_{n})) ]
\nonumber\\
= \sum_{m=1}^{n} \prod_{l <m} (-1)^{|\xi_{p}||\xi_{q_{l}}|}~[ \xi_{p}(y), \xi_{q_{m}}(x_{m})] 
~N(\xi_{q_{1}}(x_{1}),\dots,\hat{m},\dots,\xi_{q_{n}}(x_{n}))
\nonumber\\
= - i~\sum_{m=1}^{n} \prod_{l <m} (-1)^{|\xi_{p}||\xi_{q_{l}}|}~D_{pq_{m}}(y - x_{m})
~N(\xi_{q_{1}}(x_{1}),\dots,\hat{m},\dots,\xi_{q_{n}}(x_{n}))
\eea

\be
N(\emptyset) = I.
\ee

The expression
$
N(\xi_{q_{1}}(x_{1}),\dots,\xi_{q_{n}}(x_{n}))
$
is (graded) symmetrical in the arguments.
\end{prop}

The expressions
$
N(\xi_{q_{1}}(x_{1}),\dots,\xi_{q_{n}}(x_{n}))
$
are called {\it Wick monomials}. There is an alternative definition based on the splitting of the fields into the creation and annihilation 
part for which we refer to \cite{algebra}.

It is a non-trivial result of Wightman and G\aa rding \cite{WG} that in
$
N(\xi_{q_{1}}(x_{1}),\dots,\xi_{q_{n}}(x_{n}))
$
one can collapse all variables into a single one and still gets an well-defined expression (see Prop. 2.2 of \cite{wick+hopf}). In 
this way we can associate to every monomial $A$ in the jet variables a quantum operator 
$
N(A)
$
which is called the associated Wick monomial. 

One can prove that 
\be
[ N(A(x)), N(B(y)) ] = 0,\quad (x - y)^{2} < 0
\ee
where by
$
[ \cdot,\cdot]
$
we mean the graded commutator. This is the most simple case of causal support property.
Now we are ready for the most general setting. We define for any monomial
$
A \in {\rm Alg}(\Xi_{r})
$
the derivation
\be
\xi \cdot A \equiv (-1)^{|\xi| |A|}~{\partial \over \partial \xi}A
\label{derivative}
\ee
for all
$
\xi \in \Xi_{r}.
$
Here 
$|A|$ 
is the Fermi parity of $A$ and we consider the left derivative in the Grassmann sense. So, the product $\cdot$ is defined as an map
$
\Xi_{r} \times {\rm Alg}(\Xi_{r}) \rightarrow {\rm Alg}(\Xi_{r}).
$
An expression
$
E(A_{1}(x_{1}),\dots,A_{n}(x_{n}))
$
is called {\it of Wick type} iff verifies:

\bea
[ \xi_{p}(y), E(A_{1}(x_{1}),\dots,A_{n}(x_{n}))  ]
\nonumber\\
= \sum_{m=1}^{n} \prod_{l \leq m} (-1)^{|\xi_{p}||A_{l}|}~\sum_{q}~[ \xi_{p}(y), \xi_{q}(x_{m})]~
E(A_{1}(x_{1}),\dots,\xi_{q}\cdot A_{m}(x_{m}),\dots,A_{n}(x_{n}))
\nonumber\\
= - i~\sum_{m=1}^{n} \prod_{l \leq m} (-1)^{|\xi_{p}||A_{l}|}~\sum_{q}~D_{pq}(y - x_{m})~
E(A_{1}(x_{1}),\dots,\xi_{q}\cdot A_{m}(x_{m}),\dots,A_{n}(x_{n}))
\label{comm-wick}
\eea

\be
E(A_{1}(x_{1}),\dots,A_{n}(x_{n}),{\bf 1}) = E(A_{1}(x_{1}),\dots,A_{n}(x_{n})) 
\ee

\be
E(1) = {\bf 1}.
\ee

\newpage
\subsection{Bogoliubov Axioms \label{Bogoliubov}}
Suppose the monomials
$
A_{1},\dots,A_{n} \in {\rm Alg}(\Xi_{r})
$
are self-adjoint:
$
A_{j}^{\dagger} = A_{j},~\forall j = 1,\dots,n
$
and of Fermi number
$
f_{i}.
$

The chronological products
$$ 
T(A_{1}(x_{1}),\dots,A_{n}(x_{n})) \equiv T^{A_{1},\dots,A_{n}}(x_{1},\dots,x_{n}) \quad n = 1,2,\dots
$$
are some distribution-valued operators leaving invariant the algebraic Fock space and verifying the following set of axioms:
\begin{itemize}
\item
{\bf Skew-symmetry} in all arguments:
\be
T(\dots,A_{i}(x_{i}),A_{i+1}(x_{i+1}),\dots,) =
(-1)^{f_{i} f_{i+1}} T(\dots,A_{i+1}(x_{i+1}),A_{i}(x_{i}),\dots)
\ee

\item
{\bf Poincar\'e invariance}: we have a natural action of the Poincar\'e group in the
space of Wick monomials and we impose that for all 
$g \in inSL(2,\C)$
we have:
\be
U_{g} T(A_{1}(x_{1}),\dots,A_{n}(x_{n})) U^{-1}_{g} =
T(g\cdot A_{1}(x_{1}),\dots,g\cdot A_{n}(x_{n}))
\label{invariance}
\ee
where in the right hand side we have the natural action of the Poincar\'e group on
$
\Xi
$.

Sometimes it is possible to supplement this axiom by other invariance
properties: space and/or time inversion, charge conjugation invariance, global
symmetry invariance with respect to some internal symmetry group, supersymmetry,
etc.
\item
{\bf Causality}: if 
$
y \cap (x + \bar{V}^{+}) = \emptyset
$
then we denote this relation by
$
x \succeq y
$.
Suppose that we have 
$x_{i} \succeq x_{j}, \quad \forall i \leq k, \quad j \geq k+1$;
then we have the factorization property:
\be
T(A_{1}(x_{1}),\dots,A_{n}(x_{n})) =
T(A_{1}(x_{1}),\dots,A_{k}(x_{k}))~~T(A_{k+1}(x_{k+1}),\dots,A_{n}(x_{n}));
\label{causality}
\ee

\item
{\bf Unitarity}: We define the {\it anti-chronological products} using a convenient notation introduced
by Epstein-Glaser, adapted to the Grassmann context. If 
$
X = \{j_{1},\dots,j_{s}\} \subset N \equiv \{1,\dots,n\}
$
is an ordered subset, we define
\be
T(X) \equiv T(A_{j_{1}}(x_{j_{1}}),\dots,A_{j_{s}}(x_{j_{s}})).
\ee
Let us consider some Grassmann variables
$
\theta_{j},
$
of parity
$
f_{j},  j = 1,\dots, n
$
and let us define
\be
\theta_{X} \equiv \theta_{j_{1}} \cdots \theta_{j_{s}}.
\ee
Now let
$
(X_{1},\dots,X_{r})
$
be a partition of
$
N = \{1,\dots,n\}
$
where
$
X_{1},\dots,X_{r}
$
are ordered sets. Then we define the (Koszul) sign
$
\epsilon(X_{1},\dots,X_{r})
$
through the relation
\be
\theta_{1} \cdots \theta_{n} = \epsilon(X_{1}, \dots,X_{r})~\theta_{X_{1}} \dots \theta_{X_{r}}
\ee
and the antichronological products are defined according to
\be
(-1)^{n} \bar{T}(N) \equiv \sum_{r=1}^{n} 
(-1)^{r} \sum_{I_{1},\dots,I_{r} \in Part(N)}
\epsilon(X_{1},\dots,X_{r})~T(X_{1})\dots T(X_{r})
\label{antichrono}
\ee
Then the unitarity axiom is:
\be
\bar{T}(N) = T(N)^{\dagger}.
\label{unitarity}
\ee
\item
{\bf The ``initial condition"}:
\be
T(A(x)) =  N(A(x)).
\ee

\item
{\bf Power counting}: We can also include in the induction hypothesis a limitation on the order of
singularity of the vacuum averages of the chronological products associated to
arbitrary Wick monomials
$A_{1},\dots,A_{n}$;
explicitly:
\be
\omega(<\Omega, T^{A_{1},\dots,A_{n}}(X)\Omega>) \leq
\sum_{l=1}^{n} \omega(A_{l}) - 4(n-1)
\label{power}
\ee
where by
$\omega(d)$
we mean the order of singularity of the (numerical) distribution $d$ and by
$\omega(A)$
we mean the canonical dimension of the Wick monomial $W$.

\item
{\bf Wick expansion property}: In analogy to (\ref{comm-wick}) we require
\bea
[ \xi_{p}(y), T(A_{1}(x_{1}),\dots,A_{n}(x_{n})) ]
\nonumber\\
= \sum_{m=1}^{n}~\prod_{l \leq m} (-1)^{|\xi_{p}||A_{l}|}~\sum_{q}~[\xi_{p}(y), \xi_{q}(x_{m})]~
T(A_{1}(x_{1}),\dots,\xi_{q}\cdot A_{m}(x_{m}),\dots, A_{n}(x_{n}))
\nonumber\\
= - i~\sum_{m=1}^{n}~\prod_{l \leq m} (-1)^{|\xi_{p}||A_{l}|}~\sum_{q}~D_{pq}(y - x_{m} )~
T(A_{1}(x_{1}),\dots,\xi_{q}\cdot A_{m}(x_{m}),\dots, A_{n}(x_{n}))
\nonumber\\
\label{wick}
\eea

In fact we can impose a sharper form:
\bea
[ \xi_{p}^{(\epsilon)}(y), T(A_{1}(x_{1}),\dots,A_{n}(x_{n})) ]
\nonumber\\
= - i~\sum_{m=1}^{n}~\prod_{l \leq m} (-1)^{|\xi_{p}||A_{l}|}~\sum_{q}~D_{pq}^{(-\epsilon)}(y - x_{m} )~
T(A_{1}(x_{1}),\dots,\xi_{q}\cdot A_{m}(x_{m}),\dots, A_{n}(x_{n}))
\nonumber\\
\label{wick-e}
\eea
\end{itemize}

Up to now, we have defined the chronological products only for self-adjoint Wick monomials 
$
W_{1},\dots,W_{n}
$
but we can extend the definition for Wick polynomials by linearity. 

The chronological products
$
T(A_{1}(x_{1}),\dots,A_{n}(x_{n}))
$
are not uniquely defined by the axioms presented above. They can be modified with quasi-local expressions i.e. 
expressions localized on the big diagonal 
$
x_{1} = \cdots = x_{n};
$
such expressions are of the type
\be
N(A_{1}(x_{1}),\dots,A_{n}(x_{n})) = P_{j}(\partial)\delta(X)~W_{j}(X)
\ee
where
$
\delta(X) \equiv \delta(x_{1} - x_{n}) \dots, \delta(x_{n-1} - x_{n}), 
$
the expressions
$
P_{j}(\partial)
$
are polynomials in the partial derivatives and 
$
W_{j}(X)
$
are Wick polynomials. There are some restrictions on these quasi-local expressions such that the Bogoliubov axioms remain true.
One such consistency relations comes from Wick expansion property. 
\newpage
\subsection{Quantum Gravity in the Linear Approximation\label{gr}}

If we consider pure massless gravity then the jet variables are 
$
(H_{\mu\nu}, \Phi, u_{\rho}, \tilde{u}_{\sigma})
$
where 
$
H_{\mu\nu}
$
and 
$
\Phi
$
are Grassmann even and  
$
u_{\rho}, \tilde{u}_{\sigma}
$
are Grassmann odd variables. Also 
$
H_{\mu\nu}
$
is symmetric and traceless.
The interaction Lagrangian is determined by gauge invariance. Namely we define the {\it gauge charge} operator by
\bea
d_{Q} H_{\mu\nu} = -{i\over 2}~\Bigl( d_{\mu}u_{\nu} + d_{\nu}u_{\mu} - {1\over 2} \eta_{\mu\nu}~d_{\rho}u^{\rho}\Bigl),
\qquad
d_{Q}\Phi = {i\over 2}~d_{\rho}u^{\rho}
\nonumber\\
d_{Q} u_{\rho} = 0,
\qquad
d_{Q} \tilde{u}_{\sigma} = i~\Bigl( d^{\lambda}H_{\sigma\lambda} + {1\over 2} d_{\sigma}\Phi\Bigl)
\label{Q-H}
\eea
where 
$
d^{\mu}
$
is the formal derivative. The gauge charge operator squares to zero:
\be
d_{Q}^{2} \simeq  0
\ee
where by
$
\simeq
$
we mean, modulo the equation of motion. 

Then we define the associated Fock space by the non-zero $2$-point distributions are
\bea
<\Omega, H_{\mu\nu}(x_{1}) H_{\rho\sigma}(x_{2})\Omega> = 
- {i\over 2}~\Bigl(\eta_{\mu\rho}~\eta_{\nu\sigma} + \eta_{\mu\sigma}~\eta_{\nu\rho} - {1\over 2}~\eta_{\mu\nu}~\eta_{\rho\sigma}\Bigl)~
D_{0}^{(+)}(x_{1} - x_{2}),
\nonumber\\
<\Omega, \Phi(x_{1}) \Phi(x_{2})\Omega> = i~D_{0}^{(+)}(x_{1} - x_{2})
\nonumber \\
<\Omega, u_{\rho}(x_{1}) \tilde{u}_{\sigma}(x_{2})\Omega> = i~\eta_{\rho\sigma}~D_{0}^{(+)}(x_{1} - x_{2}),
\nonumber\\
<\Omega, \tilde{u}_{\sigma}(x_{1}) u_{\rho}(x_{2})\Omega> = - i~\eta_{\rho\sigma}~D_{0}^{(+)}(x_{1} - x_{2}).
\label{2-H}
\eea

Here we are using the Pauli-Jordan distribution
\be
D_{m}(x) = D_{m}^{(+)}(x) + D_{m}^{(-)}(x)
\ee
where
\be
D_{m}^{(\pm)}(x) =
\pm {i \over (2\pi)^{3}}~\int dp e^{- i p\cdot x} \theta(\pm p_{0}) \delta(p^{2} -
m^{2})
\ee
and
\be
D^{(-)}(x) = - D^{(+)}(- x).
\ee

We stress that in (\ref{Q-H}) we are dealing with the classical jet variables, but in the previous relation we have the 
associated quantum variables; we have omitted the super-script ``quant" for simplicity.

The definitions above are describing a system of massless particles of helicity $2$ as it is proved in \cite{gravity}. 

If we define the derivative on the jet variables by
\be
H_{\mu\nu}\cdot H_{\rho\sigma} \equiv {1\over 2}~\Bigl(\eta_{\mu\rho}~\eta_{\nu\sigma} + \eta_{\mu\sigma}~\eta_{\nu\rho}
- {1\over 2}~\eta_{\mu\nu}~\eta_{\rho\sigma}\Bigl)
\ee
then we have for any monomial $A$
\be
[ H_{\mu\nu}(x), A(y) ] = [ H_{\mu\nu}(x), H_{\rho\sigma}(y) ]~H^{\rho\sigma}\cdot A(y) 
\ee
so we do not have to worry about overcounting. It is more convenient to define
\be
h_{\mu\nu} \equiv H_{\mu\nu} + {1\over 2}~\eta_{\mu\nu}~\Phi  
\ee
and then we have
\bea
<\Omega, h_{\mu\nu}(x_{1}) h_{\rho\sigma}(x_{2})\Omega> = 
- {i\over 2}~\Bigl(\eta_{\mu\rho}~\eta_{\nu\sigma} + \eta_{\mu\sigma}~\eta_{\nu\rho} 
- \eta_{\mu\nu}\eta_{\rho\sigma}\Bigl)~D_{0}^{(+)}(x_{1} - x_{2})
\label{2-h}
\eea
\bea
d_{Q} h_{\mu\nu} = -{i\over 2}~\Bigl( d_{\mu}u_{\nu} + d_{\nu}u_{\mu} - \eta_{\mu\nu}~d_{\rho}u^{\rho}\Bigl),
\nonumber\\
d_{Q} u_{\rho} = 0,
\qquad
d_{Q} \tilde{u}_{\sigma} = i~d^{\lambda}h_{\sigma\lambda}
\label{Q-h}
\eea
and
\be
h_{\mu\nu}\cdot h_{\rho\sigma} \equiv {1\over 2}~(\eta_{\mu\rho}~\eta_{\nu\sigma} + \eta_{\mu\sigma}~\eta_{\nu\rho})
\ee
In \cite{Sc2}, \cite{massive} it is proved that the relation
\be
d_{Q}T \sim {\rm total~divergence}
\ee
fixes uniquely (up to a coboundary) the expression $T$ if we require: (a) Lorentz covariance; (b) $T$ should be tri-linear in the jet 
variables; (c) 
$
\omega(T) \leq 5;
$
(there are no solutions with
$
\omega(T) \leq 4).
$
We can choose, up to a coboundary 
\be
T = \sum_{j=1}^{9}~T_{j}
\label{T}
\ee
where
\bea
T_{1} \equiv - h_{\alpha\beta}~d^{\alpha}h~d^{\beta}h
\nonumber \\
T_{2} \equiv 2~h^{\alpha\beta}~d_{\alpha}h_{\mu\nu}~d_{\beta}h^{\mu\nu}
\nonumber \\
T_{3} \equiv 4~h_{\alpha\beta}~d_{\nu}h^{\beta\mu}~d_{\mu}h^{\alpha\nu}
\nonumber \\
T_{4} \equiv 2~h_{\alpha\beta}~d_{\mu}h^{\alpha\beta}~d^{\mu}h
\nonumber \\
T_{5} \equiv - 4~h_{\alpha\beta}~d_{\nu}h^{\alpha\mu}~d^{\nu}{h^{\beta}}_{\mu}
\nonumber \\
T_{6} \equiv - 4~u^{\mu}~d_{\beta}\tilde{u}_{\nu}~d_{\mu}h^{\nu\beta}
\nonumber \\
T_{7} \equiv 4~d_{\nu}u^{\beta}~d_{\mu}\tilde{u}_{\beta}~h^{\mu\nu}
\nonumber \\
T_{8} \equiv - 4~d^{\nu}u_{\nu}~d_{\alpha}\tilde{u}_{\beta}~h^{\alpha\beta}
\nonumber \\
T_{9} \equiv 4~d_{\nu}u^{\mu}~d_{\mu}\tilde{u}_{\beta}~h^{\nu\beta}
\label{T-int}
\eea

Then the expression gives a Wick polynomial 
$
T^{\rm quant}
$
formally the same, but: 
(a) the jet variables must be replaced by the associated quantum fields; (b) the formal derivative 
$
d^{\mu}
$
goes in the true derivative in the coordinate space; (c) Wick ordering should be done to obtain well-defined operators. We also 
have an associated {\it gauge charge} operator in the Fock space given by the graded commutator
\be
d_{Q} A \equiv [ Q, A ]. 
\ee
One can be proved that
$
Q^{2} = 0
$
and
\be
~[Q, T^{\rm quant}(x) ] = {\rm total~divergence}
\label{gauge1}
\ee
where the equations of motion are automatically used because the quantum fields are on-shell.

From now on we abandon the super-script {\it quant} because it will be obvious from the context if we refer 
to the classical expression  or to its quantum counterpart.

We conclude our presentation with a generalization of (\ref{gauge1}). In fact, it can be proved that (\ref{gauge1}) implies
the existence of Wick polynomials
$
T^{\alpha}, T^{\alpha\beta}, T^{\alpha\beta\gamma} 
$
such that we have:
\be
~[Q, T^{I} ] = i \partial_{\mu}T^{I\mu}
\label{gauge2}
\ee
for any multi-index $I$ with the convention
$
T^{\emptyset} \equiv T.
$
Explicitly:
\be
T^{\alpha} = \sum_{j=1}^{19}~T_{j}^{\alpha}
\label{T1}
\ee

\bea
T_{1}^{\alpha} \equiv 4~u^{\mu}~d_{\mu}h_{\beta\nu}~d^{\beta}h^{\alpha\nu}
\nonumber \\
T_{2}^{\alpha} \equiv - 2~u^{\mu}~d_{\mu}h^{\beta\nu}~d^{\alpha}h_{\beta\nu}
\nonumber \\
T^{\alpha}_{3} \equiv - 2 ~u^{\alpha}~d^{\beta}h^{\mu\nu}~d_{\mu}h_{\beta\nu}
\nonumber \\
T_{4}^{\alpha} \equiv - 4~d_{\nu}u_{\beta}~d_{\mu}h^{\alpha\beta}~h^{\mu\nu}
\nonumber \\
T_{5}^{\alpha} \equiv 4~d_{\nu}u^{\nu}~d^{\mu}h^{\alpha\beta}~h_{\beta\mu}
\nonumber \\
T_{6}^{\alpha} \equiv u^{\alpha}~d_{\beta}h_{\mu\nu}~d^{\beta}h^{\mu\nu} 
\nonumber \\
T_{7}^{\alpha} \equiv - 2~d_{\nu}u^{\nu}~h_{\mu\beta}~d^{\alpha}h^{\mu\beta}
\nonumber \\
T_{8}^{\alpha} \equiv - {1\over 2}~u^{\alpha}~d_{\mu}h~d^{\mu}h
\nonumber \\
T_{9}^{\alpha} \equiv d^{\nu}u_{\nu}~h~d^{\alpha}h
\nonumber \\
T_{10}^{\alpha} \equiv u^{\nu}~d^{\alpha}h~d_{\nu}h
\nonumber \\
T_{11}^{\alpha} \equiv - 2~d_{\nu}u_{\mu}~h^{\mu\nu}~d^{\alpha}h
\nonumber \\
T_{12}^{\alpha} \equiv 4~d^{\nu}u_{\mu}~d^{\alpha}h^{\mu\beta}~h_{\beta\nu}
\nonumber \\
T_{13}^{\alpha} \equiv - 4~d^{\nu}u^{\mu}~d_{\mu}h^{\alpha\beta}~h_{\beta\nu}
\nonumber \\
T_{14}^{\alpha} \equiv - 2~u^{\mu}~d_{\mu}u_{\nu}~d^{\alpha}\tilde{u}^{\nu}
\nonumber \\
T_{15}^{\alpha} \equiv 2~u^{\mu}~d^{\nu}u^{\alpha}~d_{\mu}\tilde{u}_{\nu}
\nonumber \\
T_{16}^{\alpha} \equiv - 2~u^{\alpha} d_{\nu}u_{\mu}~d^{\mu}\tilde{u}^{\nu}
\nonumber \\
T_{17}^{\alpha} \equiv 2~d^{\nu}u_{\nu}~d^{\mu}u^{\alpha}~\tilde{u}_{\mu}
\nonumber \\
T_{18}^{\alpha} \equiv 2~u^{\mu}~d_{\mu}d_{\nu}u^{\nu}~\tilde{u}^{\alpha}
\nonumber \\
T_{19}^{\alpha} \equiv - 2~u^{\alpha}~d_{\mu}d_{\nu}u^{\mu}~\tilde{u}^{\nu}
\label{Tmu-int}
\eea

\be
T^{\alpha\beta} = \sum_{j=1}^{5}~T_{j}^{\alpha\beta}
\label{T2}
\ee

\bea
T^{\alpha\beta}_{1} \equiv 2~u^{\mu}~d_{\mu}u_{\nu}~d^{\beta}h^{\alpha\nu}
- (\alpha \leftrightarrow \beta)
\nonumber \\
T^{\alpha\beta}_{2} \equiv 2~u^{\mu}~d_{\nu}u^{\alpha}~d_{\mu}h^{\beta\nu}
- (\alpha \leftrightarrow \beta)
\nonumber \\
T^{\alpha\beta}_{3} \equiv - 2~u^{\alpha}~d_{\nu}u_{\mu}~d^{\mu}h^{\beta\nu}
- (\alpha \leftrightarrow \beta)
\nonumber \\
T^{\alpha\beta}_{4} \equiv 4~d^{\mu}u^{\alpha}~d^{\nu}u^{\beta}~h_{\mu\nu}
\nonumber \\
T^{\alpha\beta}_{5} \equiv 2~d_{\nu}u^{\nu}~d_{\mu}u^{\alpha}~h^{\mu\beta}
- (\alpha \leftrightarrow \beta)
\label{Tmunu-int}
\eea

\be
T^{\alpha\beta\gamma} = \sum_{j=1}^{6}~T_{j}^{\alpha\beta\gamma}
\label{T3}
\ee

\bea
T^{\alpha\beta\gamma}_{1} \equiv d^{\alpha}u^{\beta}~u_{\mu}~d^{\mu}u^{\gamma}
- (\alpha \leftrightarrow \beta)
\nonumber \\
T^{\alpha\beta\gamma}_{2} \equiv d^{\beta}u^{\gamma}~u_{\mu}~d^{\mu}u^{\alpha}
- (\alpha \leftrightarrow \beta)
\nonumber \\
T^{\alpha\beta\gamma}_{3} \equiv d^{\gamma}u^{\alpha}~u_{\mu}~d^{\mu}u^{\beta}
- (\alpha \leftrightarrow \beta)
\nonumber \\
T^{\alpha\beta\gamma}_{4} \equiv d^{\beta}u^{\mu}~d_{\mu}u^{\alpha}~u^{\gamma}
- (\alpha \leftrightarrow \beta)
\nonumber \\
T^{\alpha\beta\gamma}_{5} \equiv d^{\gamma}u^{\mu}~d_{\mu}u^{\beta}~u^{\alpha}
- (\alpha \leftrightarrow \beta)
\nonumber \\
T^{\alpha\beta\gamma}_{6} \equiv d^{\alpha}u^{\mu}~d_{\mu}u^{\gamma}~u^{\beta}
- (\alpha \leftrightarrow \beta)
\label{Tmunurho-int}
\eea

Finally we give the relation expressing gauge invariance in order $n$ of the perturbation theory. We define the operator 
$
\delta
$
on chronological products by:
\bea
\delta T(T^{I_{1}}(x_{1}),\dots,T^{I_{n}}(x_{n})) \equiv 
\sum_{m=1}^{n}~( -1)^{s_{m}}\partial_{\mu}^{m}T(T^{I_{1}}(x_{1}), \dots,T^{I_{m}\mu}(x_{m}),\dots,T^{I_{n}}(x_{n}))
\label{derT}
\eea
with
\be
s_{m} \equiv \sum_{p=1}^{m-1} |I_{p}|,
\ee
then we define the operator
\be
s \equiv d_{Q} - i \delta.
\label{s-n}
\ee

Gauge invariance in an arbitrary order is then expressed by
\be
sT(T^{I_{1}}(x_{1}),\dots,T^{I_{n}}(x_{n})) = 0.
\label{brst-n}
\ee
This concludes the necessary prerequisites.
\newpage


We notice that in (\ref{Q-h}) and in (\ref{gauge2}) we have a pattern of the type:
\be
d_{Q}A = {\rm total~divergence}.
\label{total-div}
\ee
This pattern remains essentially true for Wick submonomials if we use the definition (\ref{derivative}). We consider the expressions
(\ref{T1}), (\ref{T2}), (\ref{T3}) and define:
\bea
B_{\mu\nu} \equiv \tilde{u}_{\mu,\nu} \cdot T = 
4~( u^{\rho}~d_{\rho}h_{\mu\nu} - d^{\rho}u_{\mu}~h_{\nu\rho} - d^{\rho}u_{\nu}~h_{\mu\rho}
- d^{\rho}u_{\rho}~h_{\mu\nu})
\nonumber\\
C_{\mu\nu} \equiv h_{\mu\nu} \cdot T = - d_{\mu}h~d_{\nu}h + 2 d_{\mu}h_{\rho\sigma}~d_{\nu}h^{\rho\sigma}
+ 4 d^{\sigma}h_{\mu\rho}~d^{\rho}h_{\nu\sigma} + 2 d_{\rho}h_{\mu\nu}~d^{\rho}h
- 4 d^{\sigma}h_{\mu\rho}~d_{\sigma}{h_{\nu}}^{\rho}
\nonumber\\
+ 2 ( d_{\mu}u^{\rho}~d_{\nu}\tilde{u}_{\rho} + d_{\nu}u^{\rho}~d_{\mu}\tilde{u}_{\rho} )
- 2 d^{\rho}u_{\rho}~(~d_{\nu}\tilde{u}_{\mu} + ~d_{\mu}\tilde{u}_{\nu}) 
+ 2 ( d_{\mu}u^{\rho}~d_{\rho}\tilde{u}_{\nu} + d_{\nu}u^{\rho}~d_{\rho}\tilde{u}_{\mu} )
\nonumber\\
D_{\mu} \equiv u_{\mu} \cdot T = - 4 d^{\alpha}\tilde{u}^{\beta}~d_{\mu}h_{\alpha\beta}
\nonumber\\
E_{\mu\nu,\rho} \equiv h_{\mu\nu,\rho} \cdot T = 4 ( h_{\rho\sigma} d^{\sigma}h_{\mu\nu} 
- {h_{\mu}}^{\sigma} d_{\rho}h_{\nu\sigma} -  {h_{\nu}}^{\sigma} d_{\rho}h_{\mu\sigma} )
\nonumber\\
+ 2 h_{\mu\nu}~d_{\rho}h + 4 ( {h_{\mu}}^{\sigma} d_{\nu}h_{\rho\sigma} +  {h_{\nu}}^{\sigma} d_{\mu}h_{\rho\sigma})
- 2 u_{\rho} (d_{\mu}\tilde{u}_{\nu} + d_{\nu}\tilde{u}_{\mu} ) 
+ 2 \eta_{\mu\nu}~( - h_{\rho\sigma} d^{\sigma}h + h^{\alpha\beta} d_{\rho}h_{\alpha\beta} )
\nonumber\\
F_{\mu,\nu} \equiv u_{\mu,\nu} \cdot T_{\mu} = 4 [ ( d^{\rho}\tilde{u}_{\mu} + d_{\mu}\tilde{u}^{\rho})~h_{\nu\rho}
- \eta_{\mu\nu} d_{\sigma}\tilde{u}_{\rho}~h^{\rho\sigma} ]
\nonumber\\
G_{\mu} \equiv \tilde{u}_{\mu} \cdot T = 0
\nonumber\\
H_{\mu;\alpha\beta} \equiv u_{\mu,\alpha\beta} \cdot T = 0.
\label{sub1}
\eea
We also have
\bea
B_{\mu,\nu,\rho} \equiv \tilde{u}_{\mu,\nu} \cdot T_{\rho} = 2 ( - u_{\nu}~d_{\mu}u_{\rho} +  u_{\rho}~d_{\mu}u_{\nu})
+ 2 \eta_{\nu\rho}~u^{\sigma}d_{\sigma}u_{\mu}
\nonumber\\
C_{\mu\nu,\rho} \equiv h_{\mu\nu}\cdot T_{\rho}  = 
2 (- d_{\mu}u^{\sigma}~d_{\nu}h_{\rho\sigma} - d_{\nu}u^{\sigma}~d_{\mu}h_{\rho\sigma} 
+ d_{\mu}u^{\sigma}~d_{\rho}h_{\nu\sigma} + d_{\nu}u^{\sigma}~d_{\rho}h_{\mu\sigma}
\nonumber\\
- d_{\mu}u^{\sigma}~d_{\sigma}h_{\nu\rho} - d_{\nu}u^{\sigma}~d_{\sigma}h_{\mu\rho})
- ( d_{\mu}u_{\nu} + d_{\nu}u_{\mu} ) d_{\rho}h 
\nonumber\\
+ 2 d_{\sigma}u^{\sigma}~( d_{\nu}h_{\rho\mu} + d_{\mu}h_{\rho\nu} - d_{\rho}h_{\mu\nu} )
+ \eta_{\mu\nu}~d_{\sigma}u^{\sigma}~d_{\rho}h
\nonumber\\
D_{\mu,\rho} \equiv u_{\mu}\cdot T_{\rho} = - 4 d_{\mu}h^{\alpha\beta} d_{\beta}h_{\rho\alpha}
+ 2 d_{\mu}h^{\alpha\beta} d_{\rho}h_{\alpha\beta} - d_{\mu}h~d_{\rho}h
\nonumber\\
+ 2 ( d_{\mu}u_{\nu} d_{\rho}\tilde{u}^{\nu} -  d_{\nu}u_{\rho} d_{\mu}\tilde{u}^{\nu}
- d_{\mu}d_{\nu}u^{\nu}~\tilde{u}_{\rho} )
\nonumber\\
+ \eta_{\mu\rho}~\Bigl( 2 d^{\nu}h^{\alpha\beta} d_{\beta}h_{\nu\alpha} - d_{\nu}h_{\alpha\beta} d^{\nu}h^{\alpha\beta}
+ {1 \over 2}~d_{\nu}h~d^{\nu}h + 2 d_{\beta}u_{\alpha}~d^{\alpha}\tilde{u}^{\beta} 
+ 2 d_{\nu}d_{\sigma}u^{\sigma}~\tilde{u}^{\nu} \Bigl)
\nonumber\\
E_{\mu\nu,\lambda,\rho} \equiv h_{\mu\nu,\lambda}\cdot T_{\rho} =
2 u_{\lambda}~( d_{\nu}h_{\rho\mu} + d_{\mu}h_{\rho\nu} - d_{\rho}h_{\mu\nu})
- 2 u_{\rho}~( d_{\nu}h_{\lambda\mu} + d_{\mu}h_{\lambda\nu} - d_{\lambda}h_{\mu\nu})
\nonumber\\
+ 2 \eta_{\mu\rho} ( u^{\sigma}~d_{\sigma}h_{\nu\lambda} - d^{\sigma}u_{\nu}~h_{\sigma\lambda}
+ d_{\sigma}u^{\sigma}~h_{\nu\lambda} - d^{\sigma}u_{\lambda}~h_{\nu\sigma}) + (\mu \leftrightarrow \nu)
\nonumber\\
+ 2 \eta_{\rho\lambda}~( - u^{\sigma}~d_{\sigma}h_{\mu\nu} - d^{\sigma}u_{\sigma}~h_{\mu\nu}
+ d^{\sigma}u_{\mu}~h_{\nu\sigma} + d^{\sigma}u_{\nu}~h_{\mu\sigma})
\nonumber\\
+ \eta_{\mu\nu}~(- u_{\rho} d_{\lambda}h + u_{\lambda} d_{\rho}h )
+ \eta_{\mu\nu}~\eta_{\rho\lambda}~( d_{\sigma}u^{\sigma}~h + u^{\sigma} d_{\sigma}h 
- 2 d_{\beta}u_{\alpha}~h^{\alpha\beta})
\nonumber\\
F_{\mu\nu,\rho} \equiv u_{\mu,\nu}\cdot T_{\rho} =
4 ( d^{\sigma}h_{\rho\mu}~h_{\nu\sigma} - d_{\rho}h_{\mu\sigma}~{h_{\nu}}^{\sigma}
+  d_{\mu}h_{\rho\sigma}~{h_{\nu}}^{\sigma})
\nonumber\\
+ 2 h_{\mu\nu}~d_{\rho}h - 2 ( u_{\nu}~d_{\rho}\tilde{u}_{\mu} + u_{\rho}~d_{\mu}\tilde{u}_{\nu} )
\nonumber\\
+ \eta_{\mu\nu}~( - 4 d_{\beta}h_{\rho\alpha}~h^{\alpha\beta} + 2 d_{\rho}h_{\alpha\beta}~h^{\alpha\beta}
- h~d_{\rho}h - 2 d_{\sigma}u_{\rho}~\tilde{u}^{\sigma})
\nonumber\\
+ 2 \eta_{\mu\rho}~( u^{\sigma}~d_{\sigma}\tilde{u}_{\nu} + d_{\sigma}u^{\sigma}~\tilde{u}_{\nu})
\nonumber\\
G_{\mu,\rho} \equiv \tilde{u}_{\mu}\cdot T_{\rho} = 2 ( - d_{\nu}u^{\nu}~d_{\mu}u_{\rho}
+ u_{\rho}~d_{\mu}d_{\nu}u^{\nu} - \eta_{\mu\rho}~u^{\alpha}~d_{\alpha}d_{\beta}u^{\beta})
\nonumber\\
H_{\mu,\alpha\beta,\rho} \equiv u_{\mu,\alpha\beta} \cdot T_{\rho} = 
\eta_{\mu\beta}~(u_{\alpha}~\tilde{u}_{\rho} - u_{\rho}~\tilde{u}_{\alpha} ) + (\alpha \leftrightarrow \beta) 
\label{sub2}
\eea
\bea
B_{\mu,\nu,\alpha\beta} \equiv \tilde{u}_{\mu,\nu} \cdot T_{\alpha\beta} = 0
\nonumber\\
C_{\mu\nu,\alpha\beta} \cdot h_{\mu\nu}\cdot T_{\alpha\beta}  = 
2 ( d_{\mu}u_{\alpha}~d_{\nu}u_{\beta} + d_{\nu}u_{\alpha}~d_{\mu}u_{\beta} 
+ d_{\rho}u^{\rho}~[ ( \eta_{\beta\nu}~d_{\mu}u_{\alpha} + \eta_{\beta\mu}~d_{\nu}u_{\alpha} ) - (\alpha \leftrightarrow \beta) ]
\nonumber\\
D_{\mu,\alpha\beta} \equiv u_{\mu}\cdot T_{\alpha\beta} = 2 ( d_{\mu}u^{\rho} d_{\beta}h_{\rho\alpha}
+ d^{\rho}u_{\alpha} d_{\mu}h_{\rho\beta} - \eta_{\alpha\mu}~d^{\sigma}u^{\rho}~d_{\rho}h_{\beta\sigma} ) - (\alpha \leftrightarrow \beta)
\nonumber\\
E_{\mu\nu,\lambda,\alpha\beta} \equiv h_{\mu\nu,\lambda}\cdot T_{\alpha\beta} =
[ \eta_{\beta\lambda} ( \eta_{\alpha\mu} u^{\rho}~d_{\rho}u_{\nu} + \eta_{\alpha\nu}~u^{\rho}~d_{\rho}u_{\mu})
\nonumber\\
+ u_{\lambda}~( \eta_{\beta\mu}~d_{\nu}u_{\alpha} + \eta_{\beta\nu} d_{\mu}u_{\alpha} )
- (\eta_{\beta\mu}~u_{\alpha}~d_{\nu}u_{\lambda} + \eta_{\beta\nu}~u_{\alpha}~d_{\mu}u_{\lambda}) ] - (\alpha \leftrightarrow \beta)
\nonumber\\
F_{\mu\nu,\alpha\beta} \equiv u_{\mu,\nu}\cdot T_{\alpha\beta} =
2 ( - u_{\nu}~d_{\beta}h_{\mu\alpha} - \eta_{\mu\alpha}~u^{\rho}~d^{\rho}h_{\beta\nu}
+ u_{\alpha}~d_{\mu}h_{\beta\nu} 
\nonumber\\
+ 2 \eta_{\alpha\mu}~d^{\rho}u_{\beta}~h_{\nu\rho} + + \eta_{\mu\nu}~d^{\rho}u_{\alpha}~h_{\beta\rho} - 
\eta_{\mu\alpha}~d^{\rho}u_{\rho}~h_{\nu\beta} ) - (\alpha \leftrightarrow \beta)
\nonumber\\
G_{\mu,\alpha\beta} \equiv \tilde{u}_{\mu}\cdot T_{\alpha\beta} = 0
\nonumber\\
H_{\mu,\alpha\beta,\rho\sigma} \equiv u_{\mu,\alpha\beta} \cdot T_{\rho\sigma} = 0.
\label{sub3}
\eea
\bea
B_{\mu,\nu,\alpha\beta\gamma} \equiv \tilde{u}_{\mu,\nu} \cdot T_{\alpha\beta\gamma} = 0
\nonumber\\
C_{\mu\nu,\alpha\beta\gamma} \cdot h_{\mu\nu}\cdot T_{\alpha\beta\gamma}  = 0
\nonumber\\
D_{\mu,\alpha\beta\gamma} \equiv u_{\mu}\cdot T_{\alpha\beta\gamma} = ( d_{\alpha}u_{\beta} d_{\mu}u_{\gamma}
+ d_{\beta}u_{\gamma} d_{\mu}u_{\alpha} + d_{\gamma}u_{\alpha} d_{\mu}u_{\beta}
\nonumber\\
- \eta_{\mu\gamma}~d_{\beta}u_{\rho} d^{\rho}u_{\alpha} - \eta_{\mu\alpha}~d_{\gamma}u_{\rho} d^{\rho}u_{\beta}
- \eta_{\mu\beta}~d_{\alpha}u_{\rho} d^{\rho}u_{\gamma} ) - (\alpha \leftrightarrow \beta)
\nonumber\\
E_{\mu\nu,\lambda,\alpha\beta\gamma} \equiv h_{\mu\nu,\lambda}\cdot T_{\alpha\beta\gamma} = 0
\nonumber\\
F_{\mu\nu,\alpha\beta\gamma} \equiv u_{\mu,\nu}\cdot T_{\alpha\beta\gamma} =
[ - u^{\rho} (\eta_{\mu\beta}~\eta_{\nu\alpha}~d_{\rho}u_{\gamma} + \eta_{\mu\gamma}~\eta_{\nu\beta}~d_{\rho}u_{\alpha}
+ \eta_{\mu\alpha}~\eta_{\nu\gamma}~d_{\rho}u_{\beta} )
\nonumber\\
- ( \eta_{\mu\gamma}~d_{\alpha}u_{\beta} + \eta_{\mu\alpha}~d_{\beta}u_{\gamma} + \eta_{\mu\beta}~d_{\gamma}u_{\alpha} ) u_{\nu}
\nonumber\\
- ( \eta_{\nu\beta}~d_{\mu}u_{\alpha}~u_{\gamma} + \eta_{\nu\gamma}~d_{\mu}u_{\beta}~u_{\alpha}
+ \eta_{\nu\alpha}~d_{\mu}u_{\gamma}~u_{\beta} )
\nonumber\\
+ ( \eta_{\mu\alpha}~d_{\beta}u_{\nu}~u_{\gamma} + \eta_{\mu\beta}~d_{\gamma}u_{\nu}~u_{\alpha}
+ \eta_{\mu\gamma}~d_{\alpha}u_{\nu}~u_{\beta} ) ] - (\alpha \leftrightarrow \beta)
\nonumber\\
G_{\mu,\alpha\beta\gamma} \equiv \tilde{u}_{\mu}\cdot T_{\alpha\beta\gamma} = 0
\nonumber\\
H_{\mu,\alpha\beta\gamma,\rho\sigma\lambda} \equiv u_{\mu,\alpha\beta} \cdot T_{\rho\sigma\lambda} = 0
\label{sub4}
\eea

Then we try to extend the structure (\ref{total-div}) to the Wick submonomials defined above. We have the formal derivative
\be
\delta A \equiv d_{\mu}A^{\mu}
\ee
used in the definition of gauge invariance (\ref{derT}) + (\ref{s-n}); we also define the derivative
$
\delta^{\prime}
$
by
\bea
\delta^{\prime}B_{\mu,\nu} = G_{\mu,\nu}
\nonumber\\
\delta^{\prime}E_{\mu\nu,\lambda} = C_{\mu\nu,\lambda}
\nonumber\\
\delta^{\prime}F_{\mu,\nu} = D_{\mu,\nu} - C_{\mu,\nu} + {1\over 2}~\eta_{\mu\nu}~C,\qquad C \equiv \eta^{\rho\sigma}~C_{\rho\sigma}
\nonumber\\
\delta^{\prime}H_{\mu,\alpha\beta} = {1\over 2} \Bigl( F_{\mu,\alpha,\beta} - E_{\mu\alpha,\beta} 
+ {1\over 2}~\eta_{\mu\alpha}~E_{\beta}\Bigl) + (\alpha \leftrightarrow \beta),\qquad E_{\beta} \equiv \eta^{\rho\sigma}~E_{\rho\sigma,\beta}
\nonumber\\
\delta^{\prime}E_{\mu\nu,\lambda,\rho} = - C_{\mu\nu,\lambda,\rho} 
+ {1\over 2}~(\eta_{\mu\lambda}~G_{\nu,\rho} + \eta_{\nu\lambda}~G_{\mu,\rho})
\nonumber\\
\delta^{\prime}F_{\mu,\nu,\rho} = - D_{\mu,\nu\rho} + C_{\mu\nu, \rho} - {1\over 2}~\eta_{\mu\nu}~C_{\rho},
\qquad C_{\rho} \equiv \eta^{\alpha\beta}~C_{\alpha\beta,\rho}
\nonumber\\
\delta^{\prime}H_{\mu,\alpha\beta,\rho} = {1\over 2} \Bigl( - F_{\mu,\alpha,\beta\rho} + E_{\mu\alpha,\beta,\rho} 
- {1\over 2}~\eta_{\mu\alpha}~E_{\beta,\rho}\Bigl) + (\alpha \leftrightarrow \beta),\qquad E_{\beta,\rho} \equiv \eta^{\rho\sigma}~E_{\rho\sigma,\beta,\rho}
\nonumber\\
\delta^{\prime}F_{\mu,\nu,\alpha\beta} = D_{\mu,\alpha\beta\nu} - C_{\mu\nu,\alpha\beta} 
+ {1\over 2}~\eta_{\mu\nu}~\tilde{C}_{\alpha\beta}, \qquad \tilde{C}_{\alpha\beta} \equiv \eta^{\rho\sigma}~C_{\rho\sigma,\alpha\beta}
\label{dprime}
\eea
and $0$ for the other Wick submonomials (\ref{sub1}). Finally
\be
s \equiv d_{Q} - i\delta, \qquad s^{\prime} \equiv s - i\delta^{\prime} = d_{Q} - i(\delta +\delta^{\prime}).
\ee
Then we have the structure
\be
s^{\prime} A = 0
\ee
for all expressions
$
A = T^{I}, 
$
$
B_{a\mu}, C_{a\mu},
$
etc. and also for the basic jet variables
$
v_{a\mu}, u_{a}, \tilde{u}_{a}.
$
\newpage
\section{Tree Contributions\label{tree}}

The Hopf structure of pure gravity is similar to the the Hopf structure of the pure Yang-Mills model given in \cite{wick+hopf}.

We implement the Wick theorem for the pure  massless gravity model. First we use the more precise form of Wick theorem for the expressionsand compute the expressions
$$
T(T^{I}(x_{1})^{(k)}, A_{2},\dots,A_{n}). 
$$
Because loop contributions are trivial, only the case 
$
k = 2
$
i.e. when we ``pull out" two factors from the first entry
$
T^{I}
$
is relevant. We have:
\bea
T(T(x_{1})^{(2)}, A_{2}(x_{2}),\dots,A_{n}(x_{n})) =
\nonumber\\
: C^{\mu\nu}(x_{1})~T(h_{\mu\nu}(x_{1})^{(0)}, A_{2}(x_{2}),\dots,A_{n}(x_{n})):
\nonumber\\
+ : E^{\mu\nu,\lambda}(x_{1})~T(h_{\mu\nu,\lambda}(x_{1})^{(0)}, A_{2}(x_{2}),\dots,A_{n}(x_{n})):
\nonumber\\
- : D^{\mu}(x_{1})~T(u_{\mu}(x_{1})^{(0)}, A_{2}(x_{2}),\dots,A_{n}(x_{n})):
\nonumber\\
- : F^{\mu,\nu}(x_{1})~T(u_{\mu,\nu}(x_{1})^{(0)}, A_{2}(x_{2}),\dots,A_{n}(x_{n})):
\nonumber\\
- : B^{\mu,\nu}(x_{1})~T(\tilde{u}_{\mu,\nu}(x_{1})^{(0)}, A_{2}(x_{2}),\dots,A_{n}(x_{n})):
\label{t2a}
\eea
\bea
T(T^{\rho}(x_{1})^{(2)}, A_{2}(x_{2}),\dots,A_{n}(x_{n})) =
\nonumber\\
: C^{\mu\nu,\rho}(x_{1})~T(h_{\mu\nu}(x_{1})^{(0)}, A_{2}(x_{2}),\dots,A_{n}(x_{n})):
\nonumber\\
+ : E^{\mu\nu,\lambda,\rho}(x_{1})~T(h_{\mu\nu,\lambda}(x_{1})^{(0)}, A_{2}(x_{2}),\dots,A_{n}(x_{n})):
\nonumber\\
- : D^{\mu,\rho}(x_{1})~T(u_{\mu}(x_{1})^{(0)}, A_{2}(x_{2}),\dots,A_{n}(x_{n})):
\nonumber\\
- : F^{\mu,\nu,\rho}(x_{1})~T(u_{\mu,\nu}(x_{1})^{(0)}, A_{2}(x_{2}),\dots,A_{n}(x_{n})):
\nonumber\\
- : H^{\mu,\alpha\beta,\rho}(x_{1})~T(u_{\mu,\alpha\beta}(x_{1})^{(0)}, A_{2}(x_{2}),\dots,A_{n}(x_{n})):
\nonumber\\
- : G^{\mu,\rho}(x_{1})~T(\tilde{u}_{\mu}(x_{1})^{(0)}, A_{2}(x_{2}),\dots,A_{n}(x_{n})):
\nonumber\\
- : B^{\mu,\nu,\rho}(x_{1})~T(\tilde{u}_{\mu,\nu}(x_{1})^{(0)}, A_{2}(x_{2}),\dots,A_{n}(x_{n})):
\label{t2b}
\eea
\bea
T(T^{\rho\sigma}(x_{1})^{(2)}, A_{2}(x_{2}),\dots,A_{n}(x_{n})) =
\nonumber\\
: C^{\mu\nu,\rho\sigma}(x_{1})~T(h_{\mu\nu}(x_{1})^{(0)}, A_{2}(x_{2}),\dots,A_{n}(x_{n})):
\nonumber\\
+ E^{\mu\nu,\lambda,\rho\sigma}(x_{1})~T(h_{\mu\nu,\lambda}(x_{1})^{(0)}, A_{2}(x_{2}),\dots,A_{n}(x_{n})):
\nonumber\\
- : D^{\mu,\rho\sigma}(x_{1})~T(u_{\mu}(x_{1})^{(0)}, A_{2}(x_{2}),\dots,A_{n}(x_{n})):
\nonumber\\
- : F^{\mu,\nu,\rho\sigma}(x_{1})~T(u_{\mu,\nu}(x_{1})^{(0)}, A_{2}(x_{2}),\dots,A_{n}(x_{n})):
\label{t2c}
\eea
and
\bea
T(T^{\rho\sigma\tau}(x_{1})^{(2)}, A_{2}(x_{2}),\dots,A_{n}(x_{n})) =
\nonumber\\
: D^{\mu,\rho\sigma\tau}(x_{1})~T(u_{\mu}(x_{1})^{(0)}, A_{2}(x_{2}),\dots,A_{n}(x_{n})):
\nonumber\\
+ F^{\mu,\nu,\rho\sigma\tau}(x_{1})~T(u_{\mu,\nu,}(x_{1})^{(0)}, A_{2}(x_{2}),\dots,A_{n}(x_{n})):
\label{t2d}
\eea

Here the expressions
$
T(A_{1}(x_{1})^{(0)}, A_{2}(x_{2}),\dots,A_{n}(x_{n}))
$
are of Wick type only in 
$
A_{2},\cdots,A_{n}.
$
There are various signs $-$ because we have a true Grassmann structure:
the variables
$
u_{\mu}, \tilde{u}_{\nu}
$
are of ghost number $1$ (resp. $-1$); the Wick submonomials
$
B_{\mu\nu}, C_{\mu\nu,\rho}, E_{\mu\nu,\lambda,\rho}, D_{\mu,\alpha\beta}, F_{\mu,\nu,\alpha\beta}
$
are of ghost number $1$ and
$
D_{\mu}, F_{\mu,\nu}
$
are of ghost number $-1$.

Now we study tree contributions. If we iterate the preceding formulas we get:
\bea
T(T(x_{1})^{(2)}, T(x_{2})^{(2)}) =
: C^{\mu_{1}\nu_{1}}(x_{1})~C^{\mu_{2}\nu_{2}}(x_{2}):~T(h_{\mu_{1}\nu_{1}}(x_{1})^{(0)}, h_{\mu_{2}\nu_{2}}(x_{2})^{(0)})
\nonumber\\
+ : E^{\mu_{1}\nu_{1},\lambda_{1}}(x_{1})~E^{\mu_{2}\nu_{2},\lambda_{2}}(x_{2}):~
T(h_{\mu_{1}\nu_{1},\lambda_{1}}(x_{1})^{(0)}, h_{\mu_{1}\nu_{2},\lambda_{2}}(x_{2})^{(0)})
\nonumber\\
+ [ : C^{\mu_{1}\nu_{1}}(x_{1})~E^{\mu_{2}\nu_{2},\lambda}(x_{2}):~T(h_{\mu_{1}\nu_{1}}(x_{1})^{(0)},h_{\mu_{2}\nu_{2},\lambda}(x_{2})^{(0)})
+ ( x_{1}  \leftrightarrow x_{2}) ]
\nonumber\\
- [ : D^{\mu_{1}}(x_{1})~B^{\mu_{2},\nu}(x_{2}):~T(u_{\mu_{1}}(x_{1})^{(0)}, \tilde{u}_{\mu_{2},\nu}(x_{2})^{(0)}) 
+ ( x_{1}  \leftrightarrow x_{2}) ]
\nonumber\\
- [ : F^{\mu_{1},\nu_{1}}(x_{1})~B^{\mu_{2},\nu_{2}}(x_{2}):~T(u_{\mu_{1},\nu_{1}}(x_{1})^{(0)}, \tilde{u}_{\mu_{2},\nu_{2}}(x_{2})^{(0)}) 
+ ( x_{1}  \leftrightarrow x_{2}) ]
\label{t2-2-1}
\eea
\bea
T(T^{\rho}(x_{1})^{(2)}, T(x_{2})^{(2)}) =
 : C^{\mu_{1}\nu_{1},\rho}(x_{1})~C^{\mu_{2}\nu_{2}}(x_{2}):~T(h_{\mu_{1}\nu_{1}}(x_{1})^{(0)}, h_{\mu_{2}\nu_{2}}(x_{2})^{(0)})
\nonumber\\
+ : C^{\mu_{1}\nu_{1},\rho}(x_{1})~E^{\mu_{2}\nu_{2},\lambda}(x_{2}):~
T(h_{\mu_{1}\nu_{1}}(x_{1})^{(0)},h_{\mu_{2}\nu_{2},\lambda}(x_{2})^{(0)})
\nonumber\\
+ : E^{\mu_{1}\nu_{1},\lambda,\rho}(x_{1})~C^{\mu_{2}\nu_{2}}(x_{2}):~
T(h_{\mu_{1}\nu_{1},\lambda}(x_{1})^{(0)}, h_{\mu_{2}\nu_{2}}(x_{2})^{(0)})
\nonumber\\
+ : E^{\mu_{1}\nu_{1},\lambda_{1},\rho}(x_{1})~E^{\mu_{2}\nu_{2},\lambda_{2}}(x_{2}):~
T(h_{\mu_{1}\nu_{1},\lambda_{1}}(x_{1})^{(0)}, h_{\mu_{2}\nu_{2},\lambda_{2}}(x_{2})^{(0)})
\nonumber\\
- : D^{\mu_{1},\rho}(x_{1})~B^{\mu_{2},\nu}(x_{2}):~T(u_{\mu_{1}}(x_{1})^{(0)}, \tilde{u}_{\mu_{2},\nu}(x_{2})^{(0)})
\nonumber\\
- : F^{\mu_{1},\nu_{1},\rho}(x_{1})~B^{\mu_{2},\nu_{2}}(x_{2}):~T(u_{\mu_{1},\nu_{1}}(x_{1})^{(0)}, \tilde{u}_{\mu_{2},\nu_{2}}(x_{2})^{(0)})
\nonumber\\
- : H^{\mu_{1},\alpha\beta,\rho}(x_{1})~B^{\mu_{2},\nu}(x_{2}):~T(u_{\mu_{1},\alpha\beta}(x_{1})^{(0)}, \tilde{u}_{\mu_{2},\nu}(x_{2})^{(0)})
\nonumber\\
- : G^{\mu_{1},\rho}(x_{1})~D^{\mu_{2}}(x_{2}):~T(\tilde{u}_{\mu_{1}}(x_{1})^{(0)}, u_{\mu_{2}}(x_{2})^{(0)})
\nonumber\\
- : G^{\mu_{1},\rho}(x_{1})~F^{\mu_{2},\nu}(x_{2}):~T(\tilde{u}_{\mu_{1}}(x_{1})^{(0)}, u_{\mu_{2},\nu}(x_{2})^{(0)})
\nonumber\\
- : B^{\mu_{1},\nu,\rho}(x_{1})~D^{\mu_{2}}(x_{2}):~T(\tilde{u}_{\mu_{1},\nu}(x_{1})^{(0)}, u_{\mu_{2}}(x_{2})^{(0)})
\nonumber\\
- : B^{\mu_{1},\nu_{1},\rho}(x_{1})~F^{\mu_{2},\nu_{2}}(x_{2}):~
T(\tilde{u}_{\mu_{1},\nu_{1}}(x_{1})^{(0)}, u_{\mu_{2},\nu_{2}}(x_{2})^{(0)})
\label{t2-2-2}
\eea
\bea
T(T^{\rho}(x_{1})^{(2)}, T^{\sigma}(x_{2})^{(2)}) =
 : C^{\mu_{1}\nu_{1},\rho}(x_{1})~C^{\mu_{2}\nu_{2},\sigma}(x_{2}):~T(h_{\mu_{1}\nu_{1}}(x_{1})^{(0)}, h_{\mu_{2}\nu_{2}}(x_{2})^{(0)})
\nonumber\\
+ : E^{\mu_{1}\nu_{1},\lambda_{1},\rho}(x_{1})~E^{\mu_{2}\nu_{2},\lambda_{2},\sigma}(x_{2}):~
T(h_{\mu_{1}\nu_{1},\lambda_{1}}(x_{1})^{(0)},h_{\mu_{2}\nu_{2},\lambda_{2}}(x_{2})^{(0)})
\nonumber\\
+ [ : C^{\mu_{1}\nu_{1},\rho}(x_{1})~E^{\mu_{2}\nu_{2},\lambda,\sigma}(x_{2}):~
T(h_{\mu_{1}\nu_{1}}(x_{1})^{(0)}, h_{\mu_{2}\nu_{2},\lambda}(x_{2})^{(0)}) 
- ( x_{1} \leftrightarrow x_{2}, \rho \leftrightarrow \sigma ) ]
\nonumber\\
+ [ : D^{\mu_{1},\rho}(x_{1})~G^{\mu_{2},\sigma}(x_{2}):~T(u_{\mu_{1}}(x_{1})^{(0)}, \tilde{u}_{\mu_{2}}(x_{2})^{(0)})
- ( x_{1} \leftrightarrow x_{2}, \rho \leftrightarrow \sigma ) ]
\nonumber\\
- [ : D^{\mu_{1},\rho}(x_{1})~B^{\mu_{2},\nu,\sigma}(x_{2}):~T(u_{\mu_{1}}(x_{1})^{(0)}, \tilde{u}_{\mu_{2},\nu}(x_{2})^{(0)})
- ( x_{1} \leftrightarrow x_{2}, \rho \leftrightarrow \sigma ) ]
\nonumber\\
- [ : F^{\mu_{1},\nu,\rho}(x_{1})~G^{\mu_{2},\sigma}(x_{2}):~T(u_{\mu_{1},\nu}(x_{1})^{(0)}, \tilde{u}_{\mu_{2}}(x_{2})^{(0)})
- ( x_{1} \leftrightarrow x_{2}, \rho \leftrightarrow \sigma ) ]
\nonumber\\
- [ : F^{\mu_{1},\nu_{1},\rho}(x_{1})~B^{\mu_{2},\nu_{2},\sigma}(x_{2}):~
T(u_{\mu_{1},\nu_{1}}(x_{1})^{(0)}, \tilde{u}_{\mu_{2},\nu_{2}}(x_{2})^{(0)})
- ( x_{1} \leftrightarrow x_{2}, \rho \leftrightarrow \sigma ) ]
\nonumber\\
- [ : H^{\mu_{1},\alpha\beta,\rho}(x_{1})~G^{\mu_{2},\sigma}(x_{2}):~
T(u_{\mu_{1},\alpha\beta}(x_{1})^{(0)}, \tilde{u}_{\mu_{2}}(x_{2})^{(0)})
- ( x_{1} \leftrightarrow x_{2}, \rho \leftrightarrow \sigma ) ]
\nonumber\\
- [ : H^{\mu_{1},\alpha\beta,\rho}(x_{1})~B^{\mu_{2},\nu,\sigma}(x_{2}):~
T(u_{\mu_{1},\alpha\beta}(x_{1})^{(0)}, \tilde{u}_{\mu_{2},\nu}(x_{2})^{(0)})
- ( x_{1} \leftrightarrow x_{2}, \rho \leftrightarrow \sigma ) ]
\label{t2-2-3}
\eea
\newpage
\bea
T(T^{\rho\sigma}(x_{1})^{(2)}, T(x_{2})^{(2)}) =
 : C^{\mu_{1}\nu_{1},\rho\sigma}(x_{1})~C^{\mu_{2}\nu_{2}}(x_{2}):~T(h_{\mu_{1}\nu_{1}}(x_{1})^{(0)}, h_{\mu_{2}\nu_{2}}(x_{2})^{(0)})
\nonumber\\
+ : C^{\mu_{1}\nu_{1},\rho\sigma}(x_{1})~E^{\mu_{2}\nu_{2},\lambda}(x_{2}):~
T(h_{\mu_{1}\nu_{1}}(x_{1})^{(0)},h_{\mu_{2}\nu_{2},\lambda}(x_{2})^{(0)})
\nonumber\\
+ : E^{\mu_{1}\nu_{1},\lambda,\rho\sigma}(x_{1})~C^{\mu_{2}\nu_{2}}(x_{2}):~
T(h_{\mu_{1}\nu_{1},\lambda}(x_{1})^{(0)}, h_{\mu_{2}\nu_{2}}(x_{2})^{(0)})
\nonumber\\
+ : E^{\mu_{1}\nu_{1},\lambda_{1},\rho\sigma}(x_{1})~E^{\mu_{2}\nu_{2},\lambda_{2}}(x_{2}):~
T(h_{\mu_{1}\nu_{1},\lambda_{1}}(x_{1})^{(0)}, h_{\mu_{2}\nu_{2},\lambda_{2}}(x_{2})^{(0)})
\nonumber\\
- : D^{\mu_{1},\rho\sigma}(x_{1})~B^{\mu_{2},\nu}(x_{2}):~T(u_{\mu_{1}}(x_{1})^{(0)}, \tilde{u}_{\mu_{2},\nu}(x_{2})^{(0)})
\nonumber\\
- : F^{\mu_{1},\nu_{1},\rho\sigma}(x_{1})~B^{\mu_{2},\nu_{2}}(x_{2}):~
T(u_{\mu_{1},\nu_{1}}(x_{1})^{(0)}, \tilde{u}_{\mu_{2},\nu_{2}}(x_{2})^{(0)})
\label{t2-2-4}
\eea
\bea
T(T^{\rho\sigma}(x_{1})^{(2)}, T^{\tau}(x_{2})^{(2)}) =
 : C^{\mu_{1}\nu_{1},\rho\sigma}(x_{1})~C^{\mu_{2}\nu_{2},\tau}(x_{2}):~
 T(h_{\mu_{1}\nu_{1}}(x_{1})^{(0)}, h_{\mu_{2}\nu_{2}}(x_{2})^{(0)})
\nonumber\\
+ : C^{\mu_{1}\nu_{1},\rho\sigma}(x_{1})~E^{\mu_{2}\nu_{2},\lambda,\tau}(x_{2}):~
T(h_{\mu_{1}\nu_{1}}(x_{1})^{(0)},h_{\mu_{2}\nu_{2},\lambda}(x_{2})^{(0)})
\nonumber\\
+ : E^{\mu_{1}\nu_{1},\lambda,\rho\sigma}(x_{1})~C^{\mu_{2}\nu_{2},\tau}(x_{2}):~
T(h_{\mu_{1}\nu_{1},\lambda}(x_{1})^{(0)}, h_{\mu_{2}\nu_{2}}(x_{2})^{(0)})
\nonumber\\
+ : E^{\mu_{1}\nu_{1},\lambda_{1},\rho\sigma}(x_{1})~E^{\mu_{2}\nu_{2},\lambda_{2},\tau}(x_{2}):~
T(h_{\mu_{1}\nu_{1},\lambda_{1}}(x_{1})^{(0)}, h_{\mu_{2}\nu_{2},\lambda_{2}}(x_{2})^{(0)})
\nonumber\\
+ : D^{\mu_{1},\rho\sigma}(x_{1})~G^{\mu_{2},\tau}(x_{2}):~T(u_{\mu_{1}}(x_{1})^{(0)}, \tilde{u}_{\mu_{2}}(x_{2})^{(0)})
\nonumber\\
+ : D^{\mu_{1},\rho\sigma}(x_{1})~B^{\mu_{2},\nu,\tau}(x_{2}):~
T(u_{\mu_{1},\nu_{1}}(x_{1})^{(0)}, \tilde{u}_{\mu_{2},\nu}(x_{2})^{(0)})
\nonumber\\
+ : F^{\mu_{1},\nu,\rho\sigma}(x_{1})~G^{\mu_{2},\tau}(x_{2}):~
T(u_{\mu_{1},\nu}(x_{1})^{(0)}, \tilde{u}_{\mu_{2}}(x_{2})^{(0)})
\nonumber\\
+ : F^{\mu_{1},\nu_{1},\rho\sigma}(x_{1})~B^{\mu_{2},\nu_{2},\tau}(x_{2}):~
T(u_{\mu_{1},\nu_{1}}(x_{1})^{(0)}, \tilde{u}_{\mu_{2},\nu_{2}}(x_{2})^{(0)})
\label{t2-2-5}
\eea
\bea
T(T^{\rho\sigma\tau}(x_{1})^{(2)}, T(x_{2})^{(2)}) =
- : D^{\mu_{1},\rho\sigma\tau}(x_{1})~B^{\mu_{2},\nu}(x_{2}):~T(u_{\mu_{1}}(x_{1})^{(0)}, \tilde{u}_{\mu_{2},\nu}(x_{2})^{(0)})
\nonumber\\
- : F^{\mu_{1},\nu_{1},\rho\sigma\tau}(x_{1})~B^{\mu_{2},\nu_{2}}(x_{2}):~
T(u_{\mu_{1},\nu_{1}}(x_{1})^{(0)}, \tilde{u}_{\mu_{2},\nu_{2}}(x_{2})^{(0)})
\label{t2-2-6}
\eea
\bea
T(T^{\rho_{1}\sigma_{1}}(x_{1})^{(2)}, T^{\rho_{2}\sigma_{2}}(x_{2})^{(2)}) =
\nonumber\\
 : C^{\mu_{1}\nu_{1},\rho_{1}\sigma_{1}}(x_{1})~C^{\mu_{2}\nu_{2},\rho_{2}\sigma_{2}}(x_{2}):~
 T(h_{\mu_{1}\nu_{1}}(x_{1})^{(0)}, h_{\mu_{2}\nu_{2}}(x_{2})^{(0)})
\nonumber\\
+ : E^{\mu_{1}\nu_{1},\lambda_{1},\rho_{1}\sigma_{1}}(x_{1})~E^{\mu_{2}\nu_{2},\lambda_{2},\rho_{2}\sigma_{2}}(x_{2}):~
T(h_{\mu_{1}\nu_{1},\lambda_{1}}(x_{1})^{(0)}, h_{\mu_{2}\nu_{2},\lambda_{2}}(x_{2})^{(0)})
\nonumber\\
+ [ : C^{\mu_{1}\nu_{1},\rho_{1}\sigma_{1}}(x_{1})~E^{\mu_{2}\nu_{2},\lambda,\rho_{2}\sigma_{2}}(x_{2}):~
 T(h_{\mu_{1}\nu_{1}}(x_{1})^{(0)}, h_{\mu_{2}\nu_{2},\lambda}(x_{2})^{(0)})
\nonumber\\
 + ( x_{1} \leftrightarrow x_{2}, \rho_{1} \leftrightarrow \rho_{2}, \sigma_{1} \leftrightarrow \sigma_{2} ) ]
\label{t2-2-7}
\eea
\bea
T(T^{\rho\sigma\tau}(x_{1})^{(2)}, T^{\nu}(x_{2})^{(2)}) =
: D^{\mu_{1},\rho\sigma\tau}(x_{1})~G^{\mu_{2},\nu}(x_{2}):~T(u_{\mu_{1}}(x_{1})^{(0)}, \tilde{u}_{\mu_{2}}(x_{2})^{(0)})
\nonumber\\
+ : D^{\mu_{1},\rho\sigma\tau}(x_{1})~B^{\mu_{2},\lambda,\nu}(x_{2}):~
T(u_{\mu_{1}}(x_{1})^{(0)}, \tilde{u}_{\mu_{2},\lambda}(x_{2})^{(0)})
\nonumber\\
+ : F^{\mu_{1},\lambda,\rho\sigma\tau}(x_{1})~G^{\mu_{2},\nu}(x_{2}):~
T(u_{\mu_{1},\lambda}(x_{1})^{(0)}, \tilde{u}_{\mu_{2}}(x_{2})^{(0)})
\nonumber\\
+ : F^{\mu_{1},\lambda_{1},\rho\sigma\tau}(x_{1})~B^{\mu_{2},\lambda_{2},\nu}(x_{2}):~
T(u_{\mu_{1},\lambda_{1}}(x_{1})^{(0)}, \tilde{u}_{\mu_{2},\lambda_{2}}(x_{2})^{(0)})
\label{t2-2-8}
\eea
\newpage
We must give the values of the chronological products
$
T(\xi^{(0)}_{p}(x_{1}), \xi^{(0)}_{q}(x_{2}))
$
which are not unique. For the pure massless gravity model we consider the causal commutators of the basic fields and perform the 
causal splitting; we obtain:
\bea
T(h_{\mu\nu}(x_{1})^{0)}, h_{\rho\sigma}(x_{2})^{0)}) = 
- {i\over 2}~\Bigl(\eta_{\mu\rho}~\eta_{\nu\sigma} + \eta_{\mu\sigma}~\eta_{\nu\rho} 
- {1\over 2}~\eta_{\mu\nu}\eta_{\rho\sigma}\Bigl)~D_{0}^{F}(x_{1} - x_{2})
\nonumber \\
T(u_{\mu}(x_{1})^{0)}, \tilde{u}_{\nu}(x_{2})^{0)}) = i~\eta_{\mu\nu}~D_{0}^{F}(x_{1} - x_{2}),
\nonumber\\
T(\tilde{u}_{\mu}(x_{1})^{0)}, u_{\nu}(x_{2})^{0)}) = - i~\eta_{\mu\nu}~D_{0}(x_{1} - x_{2}).
\label{comm-2}
\eea
and, according to (\ref{2-point-der}):
\be
T(\xi_{a,\mu}(x_{1})^{(0)}, \xi_{b,\nu}(x_{2})^{(0)}) = 
i \partial_{\mu}^{1}\partial_{\nu}^{2}T(\xi_{a}(x_{1})^{(0)}, \xi_{b}(x_{2})^{(0)})
\label{chr-3}
\ee
etc. This gives:
\bea
T(T(x_{1})^{(2)}, T(x_{2})^{(2)}) =
- i~D^{F}_{0}( x_{1} - x_{2})~ \Bigl[ : C^{\mu\nu}(x_{1})~C_{\mu\nu}(x_{2}): - {1\over 2}~: C(x_{1})~C(x_{2}): \Bigl]
\nonumber\\
+ i~\partial_{\lambda}D^{F}_{0}( x_{1} - x_{2})~
\Bigl\{ \Bigl[ : C_{\mu\nu}(x_{1})~E^{\mu\nu,\lambda}(x_{2}): - {1\over 2}~: C(x_{1})~E^{\lambda}(x_{2}):
\nonumber\\
+ : D_{\mu}(x_{1})~B^{\mu,\lambda}(x_{2}): \Bigl] + ( x_{1} \leftrightarrow x_{2}) \Bigl\}
\nonumber\\
+ i~\partial_{\lambda_{1}}\partial_{\lambda_{2}}D^{F}_{0}( x_{1} - x_{2})~
\Bigl[ : E^{\mu\nu,\lambda_{1}}(x_{1})~{E_{\mu\nu,}}^{\lambda_{2}}(x_{2}):
- {1\over 2}~: E^{\lambda_{1}}(x_{1})~E^{\lambda_{2}}(x_{2}):
\nonumber\\
+ : F^{\mu,\lambda_{1}}(x_{1})~{B_{\mu,}}^{\lambda_{2}}(x_{2})
- {B_{\mu,}}^{\lambda_{1}}(x_{1})~F^{\mu,\lambda_{2}}(x_{2}) \Bigl]
\label{t2-2-1a}
\eea
\bea
T(T^{\rho}(x_{1})^{(2)}, T(x_{2})^{(2)}) =
\nonumber\\
- i~D^{F}_{0}( x_{1} - x_{2})~\Bigl[ : C^{\mu\nu,\rho}(x_{1})~C_{\mu\nu}(x_{2}): - {1\over 2}~: C^{\rho}(x_{1})~C(x_{2}): 
- : G^{\mu,\rho}(x_{1})~D_{\mu}(x_{2}): \Bigl]
\nonumber\\
+ i~\partial_{\lambda}D^{F}_{0}( x_{1} - x_{2})~\Bigl[ : C^{\mu\nu,\rho}(x_{1})~{E_{\mu\nu,}}^{\lambda}(x_{2}): 
- {1\over 2}~: C^{\rho}(x_{1})~E^{\lambda}(x_{2}):
\nonumber\\
- : E^{\mu\nu,\lambda,\rho}(x_{1})~C_{\mu\nu}(x_{2}): 
+ {1\over 2}~: E^{\lambda,\rho}(x_{1})~C(x_{2}): 
\nonumber\\
+ : D^{\mu,\rho}(x_{1})~{B_{\mu,}}^{\lambda}(x_{2}): - : G^{\mu,\rho}(x_{1})~{F_{\mu,}}^{\lambda}(x_{2}): 
+ : B^{\mu,\lambda,\rho}(x_{1})~D_{\mu}(x_{2}): 
\Bigl]
\nonumber\\
+ i~\partial_{\lambda_{1}}\partial_{\lambda_{2}}D^{F}_{0}( x_{1} - x_{2})~
\Bigl[ : E^{\mu\nu,\lambda_{1},\rho}(x_{1})~{E_{\mu\nu,}}^{\lambda_{2}}(x_{2}):
- {1\over 2}~: E^{\lambda_{1},\rho}(x_{1})~E^{\lambda_{2}}(x_{2}):
\nonumber\\
+ : F^{\mu,\lambda_{1},\rho}(x_{1})~{B_{\mu,}}^{\lambda_{2}}(x_{2})
- {B_{\mu,}}^{\lambda_{1},\rho}(x_{1})~F^{\mu,\lambda_{2}}(x_{2}) \Bigl]
\nonumber\\
+ i~\partial_{\lambda_{1}}\partial_{\lambda_{2}}\partial_{\lambda_{3}}D^{F}_{0}( x_{1} - x_{2})~
: H^{\mu,\lambda_{1}\lambda_{2},\rho}(x_{1})~{B_{\mu,}}^{\lambda_{3}}(x_{2}):
\label{t2-2-2a}
\eea
\newpage
\bea
T(T^{\rho}(x_{1})^{(2)}, T^{\sigma}(x_{2})^{(2)}) =
\nonumber\\
- i~D^{F}_{0}( x_{1} - x_{2})~\Bigl[ : C^{\mu\nu,\rho}(x_{1})~{C_{\mu\nu,}}^{\sigma}(x_{2}): 
- {1\over 2}~: C^{\rho}(x_{1})~C^{\sigma}(x_{2}): 
\nonumber\\
- : D^{\mu,\rho}(x_{1})~{G_{\mu,}}^{\sigma}(x_{2}): + {G_{\mu,}}^{\rho}(x_{1})~D^{\mu,\sigma}(x_{1}):\Bigl]
\nonumber\\
+ i~\partial_{\lambda}D^{F}_{0}( x_{1} - x_{2})~\Bigl[ : C^{\mu\nu,\rho}(x_{1})~{E_{\mu\nu,}}^{\lambda,\sigma}(x_{2}): 
- {1\over 2}~: C^{\rho}(x_{1})~E^{\lambda,\sigma}(x_{2}):
\nonumber\\
- : D^{\mu,\rho}(x_{1})~{B_{\mu,}}^{\lambda,\sigma}(x_{2}): 
+ : F^{\mu,\lambda,\rho}(x_{1})~{G_{\mu,}}^{\sigma}(x_{2}): 
\nonumber\\
- : {E^{\mu\nu,}}^{\lambda,\rho}(x_{1})~{C_{\mu\nu}}^{\sigma}(x_{2}): + {1\over 2}~: E^{\lambda,\rho}(x_{1})~C^{\sigma}(x_{2}): 
\nonumber\\
- : {B^{\mu,}}^{\lambda,\rho}(x_{1})~D^{\mu,\sigma}(x_{2}): + : {G_{\mu,}}^{\rho}(x_{2})~F^{\mu,\lambda,\sigma}(x_{1}): \Bigl]
\nonumber\\
+ i~\partial_{\lambda_{1}}\partial_{\lambda_{2}}D^{F}_{0}( x_{1} - x_{2})~
\Bigl[ : E^{\mu\nu,\lambda_{1},\rho}(x_{1})~{E_{\mu\nu,}}^{\lambda_{2},\sigma}(x_{2}):
- {1\over 2}~: E^{\lambda_{1},\rho}(x_{1})~E^{\lambda_{2},\sigma}(x_{2}):
\nonumber\\
- : F^{\mu,\lambda_{1},\rho}(x_{1})~{B_{\mu,}}^{\lambda_{2},\sigma}(x_{2})
+ {B_{\mu,}}^{\lambda_{1},\rho}(x_{1})~F^{\mu,\lambda_{2},\sigma}(x_{2}) 
\nonumber\\
+ : H^{\mu,\lambda_{1}\lambda_{2},\rho}(x_{1})~{G_{\mu,}}^{\sigma}(x_{2}):
- : {G_{\mu,}}^{\rho}(x_{1})~ H^{\mu,\lambda_{1}\lambda_{2},\sigma}(x_{1}): \Bigl]
\nonumber\\
- i~\partial_{\lambda_{1}}\partial_{\lambda_{2}}\partial_{\lambda_{3}}D^{F}_{0}( x_{1} - x_{2})~
[ : H^{\mu,\lambda_{1}\lambda_{2},\rho}(x_{1})~{B_{\mu,}}^{\lambda_{3},\sigma}(x_{2}):
- :{B_{\mu,}}^{\lambda_{3},\rho}(x_{1})~ H^{\mu,\lambda_{1}\lambda_{2},\sigma}(x_{1}): ]
\label{t2-2-3a}
\eea
\bea
T(T^{\rho\sigma}(x_{1})^{(2)}, T(x_{2})^{(2)}) =
\nonumber\\
- i~D^{F}_{0}( x_{1} - x_{2})~\Bigl[ : C^{\mu\nu,\rho\sigma}(x_{1})~C_{\mu\nu}(x_{2}): 
- {1\over 2}~: C^{\rho\sigma}(x_{1})~C(x_{2}): \Bigl]
\nonumber\\
+ i~\partial_{\lambda}D^{F}_{0}( x_{1} - x_{2})~\Bigl[ : C^{\mu\nu,\rho\sigma}(x_{1})~{E_{\mu\nu,}}^{\lambda}(x_{2}): 
- {1\over 2}~: C^{\rho\sigma}(x_{1})~E^{\lambda}(x_{2}):
\nonumber\\
- : E^{\mu\nu,\lambda,\rho\sigma}(x_{1})~C_{\mu\nu}(x_{2}): + {1\over 2}~: E^{\lambda,\rho\sigma}(x_{1})~C(x_{2}): 
\nonumber\\
+ : D^{\mu,\rho\sigma}(x_{1})~{B^{\mu,}}^{\lambda}(x_{2}):  \Bigl]
\nonumber\\
+ i~\partial_{\lambda_{1}}\partial_{\lambda_{2}}D^{F}_{0}( x_{1} - x_{2})~
\Bigl[ : E^{\mu\nu,\lambda_{1},\rho\sigma}(x_{1})~{E_{\mu\nu,}}^{\lambda_{2}}(x_{2}):
- {1\over 2}~: E^{\lambda_{1},\rho\sigma}(x_{1})~E^{\lambda_{2}}(x_{2}):
\nonumber\\
+ : F^{\mu,\lambda_{1},\rho\sigma}(x_{1})~{B_{\mu,}}^{\lambda_{2}}(x_{2})
\label{t2-2-4a}
\eea
\newpage
\bea
T(T^{\rho\sigma}(x_{1})^{(2)}, T^{\tau}(x_{2})^{(2)}) =
\nonumber\\
- i~D^{F}_{0}( x_{1} - x_{2})~\Bigl[ : C^{\mu\nu,\rho\sigma}(x_{1})~{C_{\mu\nu,}}^{\tau}(x_{2}): 
- {1\over 2}~: C^{\rho\sigma}(x_{1})~C^{\tau}(x_{2}): 
\nonumber\\
- : D^{\mu,\rho\sigma}(x_{1})~{G_{\mu,}}^{\tau}(x_{2}): \Bigl]
\nonumber\\
+ i~\partial_{\lambda}D^{F}_{0}( x_{1} - x_{2})~\Bigl[ : C^{\mu\nu,\rho\sigma}(x_{1})~{E_{\mu\nu,}}^{\lambda,\tau}(x_{2}): 
- {1\over 2}~: C^{\rho\sigma}(x_{1})~E^{\lambda,\tau}(x_{2}):
\nonumber\\
- : E^{\mu\nu,\lambda,\rho\sigma}(x_{1})~{C_{\mu\nu,}}^{\tau}(x_{2}): + {1\over 2}~: E^{\lambda,\rho\sigma}(x_{1})~C^{\tau}(x_{2}): 
\nonumber\\
- : D^{\mu,\rho\sigma}(x_{1})~{B^{\mu,}}^{\lambda,\tau}(x_{2}):  
+ : F^{\mu,\lambda,\rho\sigma}(x_{1})~{G^{\mu,}}^{\tau}(x_{2}): \Bigl]
\nonumber\\
+ i~\partial_{\lambda_{1}}\partial_{\lambda_{2}}D^{F}_{0}( x_{1} - x_{2})~
\Bigl[ : E^{\mu\nu,\lambda_{1},\rho\sigma}(x_{1})~{E_{\mu\nu,}}^{\lambda_{2},\tau}(x_{2}):
- {1\over 2}~: E^{\lambda_{1},\rho\sigma}(x_{1})~E^{\lambda_{2},\tau}(x_{2}):
\nonumber\\
- : F^{\mu,\lambda_{1},\rho\sigma}(x_{1})~{B_{\mu,}}^{\lambda_{2},\tau}(x_{2}) \Bigl]
\label{t2-2-5a}
\eea
\bea
T(T^{\rho\sigma\tau}(x_{1})^{(2)}, T(x_{2})^{(2)}) =
i~\partial_{\lambda}D^{F}_{0}( x_{1} - x_{2})~: D^{\mu,\rho\sigma\tau}(x_{1})~{B_{\mu,}}^{\lambda}(x_{2}): 
\nonumber\\
+ i~\partial_{\lambda_{1}}\partial_{\lambda_{2}}D^{F}_{0}( x_{1} - x_{2})~
: F^{\mu,\lambda_{1},\rho\sigma\tau}(x_{1})~{B_{\mu,}}^{\lambda_{2}}(x_{2}):
\label{t2-2-6a}
\eea
\bea
T(T^{\rho_{1}\sigma_{1}}(x_{1})^{(2)}, T^{\rho_{2}\sigma_{2}}(x_{2})^{(2)}) =
\nonumber\\
- i~D^{F}_{0}( x_{1} - x_{2})~\Bigl[ : C^{\mu\nu,\rho_{1}\sigma_{1}}(x_{1})~{C_{\mu\nu,}}^{\rho_{2}\sigma_{2}}(x_{2}): 
- {1\over 2}~: C^{\rho_{1}\sigma_{1}}(x_{1})~C^{\rho_{2}\sigma_{2}}(x_{2}):  \Bigl]
\nonumber\\
+ i~\partial_{\lambda}D^{F}_{0}( x_{1} - x_{2})~
\Bigl[ : C^{\mu\nu,\rho_{1}\sigma_{1}}(x_{1})~{E_{\mu\nu,}}^{\lambda,\rho_{2}\sigma_{2}}(x_{2}):
- {1\over 2}~: C^{\rho_{1}\sigma_{1}}(x_{1})~E^{\lambda,\rho_{2}\sigma_{2}}(x_{2}):
\nonumber\\
- : {E^{\mu\nu,}}^{\lambda,\rho_{1}\sigma_{1}}(x_{1})~C^{\mu\nu,\rho_{2}\sigma_{2}}(x_{2}): 
+ {1\over 2}~: E^{\lambda,\rho_{1}\sigma_{1}}(x_{1})~C^{\rho_{2}\sigma_{2}}(x_{2}):  \Bigl]
\nonumber\\
+ i~\partial_{\lambda_{1}}\partial_{\lambda_{2}}D^{F}_{0}( x_{1} - x_{2})~
\Bigl[ : E^{\mu\nu,\lambda_{1},\rho_{1}\sigma_{1}}(x_{1})~{E_{\mu\nu,}}^{\lambda_{2},\rho_{2}\sigma_{2}}(x_{2}):
- {1\over 2}~: E^{\lambda_{1},\rho_{1}\sigma_{1}}(x_{1})~E^{\lambda_{2},\rho_{2}\sigma_{2}}(x_{2}): \Bigl]
\nonumber\\
\label{t2-2-7a}
\eea
\bea
T(T^{\rho\sigma\tau}(x_{1})^{(2)}, T^{\nu}(x_{2})^{(2)}) =
i~D^{F}_{0}( x_{1} - x_{2})~: D^{\mu,\rho\sigma\tau}(x_{1})~{G_{\mu,}}^{\nu}(x_{2}): 
\nonumber\\
+ i~\partial_{\lambda}D^{F}_{0}( x_{1} - x_{2})~
[ - : D^{\mu,\rho\sigma\tau}(x_{1})~{B_{\mu,}}^{\lambda,\nu}(x_{2}):
+ : F^{\mu,\lambda,\rho\sigma\tau}(x_{1})~{G_{\mu,}}^{\nu}(x_{2}): ]
\nonumber\\
- i~\partial_{\lambda_{1}}\partial_{\lambda_{2}}D^{F}_{0}( x_{1} - x_{2})~
: F^{\mu,\lambda_{1},\rho\sigma\tau}(x_{1})~{B_{\mu,}}^{\lambda_{2},\nu}(x_{2}):
\label{t2-2-8a}
\eea
Here
\bea
C \equiv \eta^{\mu\nu}~C_{\mu\nu}, \qquad C_{\rho} \equiv \eta^{\mu\nu}~C_{\mu\nu,\rho}, \qquad 
C_{\rho\sigma} \equiv \eta^{\mu\nu}~C_{\mu\nu,\rho\sigma},
\nonumber\\
E_{\lambda} \equiv \eta^{\mu\nu}~E_{\mu\nu,\lambda}, \qquad E_{\lambda,\rho} \equiv \eta^{\mu\nu}~E_{\mu\nu,\lambda,\rho}, 
\qquad E_{\lambda,\rho\sigma} \equiv \eta^{\mu\nu}~C_{\mu\nu,\lambda,\rho\sigma}.
\eea
\newpage
From these formulas we can determine now if gauge invariance is true; in fact, we have anomalies, as it is well known.
The cancelations are more subtle than in the case of pure Yang-Mills case. One has to decompose the expresions from 
(\ref{sub1}) -(\ref{sub4}) according to they tensorial character; we need:
\bea
E^{\mu\nu,\rho} = E^{\mu\nu,\rho}_{1} + \eta^{\mu\nu}~E_{2}^{\rho}
\nonumber\\
B^{\mu,\nu,\rho} = B^{\mu,\nu,\rho}_{1} + \eta^{\nu\rho}~B_{1}^{\mu}
\nonumber\\
E^{\mu\nu,\lambda,\rho} = E^{\mu\nu,\lambda,\rho}_{1} 
+ \eta^{\mu\rho}~E^{\lambda\nu}_{2} + \eta^{\nu\rho}~E^{\lambda\mu}_{2} 
+ \eta^{\lambda\rho}~E^{\mu\nu}_{3} + \eta^{\mu\nu}~E^{\lambda\rho}_{4}
\nonumber\\
E^{\lambda,\rho} \equiv \eta_{\mu\nu}~E^{\mu\nu,\lambda,\rho} 
= E^{\lambda,\rho}_{1} + 2~E^{\lambda\nu}_{2} + \eta^{\lambda\rho}~E_{3} + 4~E^{\lambda\rho}_{4}
\nonumber\\
H^{\mu,\alpha\beta,\rho} = \eta_{\mu\beta}~H^{\alpha\rho} +  (\alpha \leftrightarrow \beta) 
\nonumber\\
E^{\mu\nu,\lambda,\alpha\beta} = 
( \eta^{\mu\beta}~\tilde{E}^{\nu,\lambda,\alpha}_{1} + \eta^{\nu\beta}~\tilde{E}^{\mu,\lambda,\alpha}_{1} 
+ \eta^{\mu\alpha}~\eta^{\beta\lambda}~\tilde{E}^{\nu}_{2} + \eta^{\nu\alpha}~\eta^{\beta\lambda}~\tilde{E}^{\mu}_{2}) 
- (\alpha \leftrightarrow \beta)
\nonumber\\
F^{\mu\nu,\alpha\beta\gamma} = 
\eta^{\mu\alpha}~F^{\nu,\beta\gamma}_{1} + \eta^{\mu\beta}~F^{\nu,\gamma\alpha}_{1}
+ \eta^{\mu\gamma}~F^{\nu,\alpha\beta}_{1}
\nonumber\\
+ \eta^{\nu\alpha}~F^{\mu,\beta\gamma}_{2} + \eta^{\nu\beta}~F^{\mu,\gamma\alpha}_{2}
+ \eta^{\nu\gamma}~F^{\mu,\alpha\beta}_{2}
\nonumber\\
+ (\eta^{\mu\alpha}~\eta^{\nu\beta} - \eta^{\mu\beta}~\eta^{\nu\alpha})~F^{\gamma}_{3}
+ (\eta^{\mu\beta}~\eta^{\nu\gamma} - \eta^{\mu\gamma}~\eta^{\nu\beta})~F^{\alpha}_{3}
+ (\eta^{\mu\gamma}~\eta^{\nu\alpha} - \eta^{\mu\alpha}~\eta^{\nu\gamma})~F^{\beta}_{3}
\eea
where
\bea
E^{\mu\nu,\rho}_{1} = 4 ( h^{\rho\sigma} d_{\sigma}h^{\mu\nu} 
- {h^{\mu}}_{\sigma} d^{\rho}h^{\nu\sigma} -  {h^{\nu}}_{\sigma} d^{\rho}h^{\mu\sigma} )
\nonumber\\
+ 2 h^{\mu\nu}~d^{\rho}h + 4 ( {h^{\mu}}_{\sigma} d^{\nu}h^{\rho\sigma} +  {h^{\nu}}_{\sigma} d^{\mu}h^{\rho\sigma})
- 2 u^{\rho} (d^{\mu}\tilde{u}^{\nu} + d^{\nu}\tilde{u}^{\mu} ) 
\nonumber\\
E_{2}^{\rho} = 2~( - h^{\rho\sigma} d_{\sigma}h + h_{\alpha\beta} d^{\rho}h^{\alpha\beta} )
\nonumber\\
\nonumber\\
B^{\mu,\nu,\rho}_{1} = 2 ( - u^{\nu}~d^{\mu}u^{\rho} +  u^{\rho}~d^{\mu}u^{\nu})
\nonumber\\
B_{1}^{\mu} = 2~u_{\sigma}d^{\sigma}u^{\mu}
\nonumber\\
\nonumber\\
E^{\mu\nu,\lambda,\rho}_{1} =
2 [ u^{\lambda}~( d^{\nu}h^{\rho\mu} + d^{\mu}h^{\rho\nu} - d^{\rho}h^{\mu\nu})
- u^{\rho}~( d^{\nu}h^{\lambda\mu} + d^{\mu}h^{\lambda\nu} - d^{\lambda}h^{\mu\nu}) ]
\nonumber\\
E^{\lambda\nu}_{2}= 2 ( u_{\sigma}~d^{\sigma}h^{\nu\lambda} - d_{\sigma}u^{\nu}~h^{\sigma\lambda}
+ d^{\sigma}u_{\sigma}~h^{\nu\lambda} - d_{\sigma}u^{\lambda}~h^{\nu\sigma})
\nonumber\\
E^{\mu\nu}_{3} = 2~( - u_{\sigma}~d^{\sigma}h^{\mu\nu} - d_{\sigma}u^{\sigma}~h^{\mu\nu}
+ d_{\sigma}u^{\mu}~h^{\nu\sigma} + d_{\sigma}u^{\nu}~h^{\mu\sigma})
\nonumber\\
+ \eta^{\mu\nu}~( d_{\sigma}u^{\sigma}~h + u^{\sigma} d_{\sigma}h 
- 2 d_{\beta}u_{\alpha}~h^{\alpha\beta})
\nonumber\\
E^{\lambda\rho}_{4} = - u^{\rho} d^{\lambda}h + u^{\lambda} d^{\rho}h 
\nonumber\\
\nonumber\\
E^{\lambda,\rho}_{1} \equiv \eta_{\mu\nu}~E^{\mu\nu,\lambda,\rho}_{1} ,\qquad
E_{3} = \eta_{\mu\nu}~E^{\mu\nu}_{3}  
\nonumber\\
\nonumber\\
H^{\alpha\rho} = u^{\alpha}~\tilde{u}^{\rho} - u^{\rho}~\tilde{u}^{\alpha} 
\nonumber\\
\nonumber\\
\tilde{E}^{\nu,\lambda,\alpha}_{1} = u^{\lambda}~d^{\nu}u^{\alpha} - u^{\alpha}~d^{\nu}u^{\lambda} 
\nonumber\\
\tilde{E}^{\nu}_{2} = u_{\rho}~d^{\rho}u^{\nu}
\nonumber\\
\nonumber\\
F^{\nu,\beta\gamma}_{1} = d^{\alpha}u^{\nu} u^{\beta} - d^{\alpha}u^{\beta} u^{\nu} - (\alpha \leftrightarrow \beta)
\nonumber\\
F^{\mu,\beta\gamma}_{2} = d^{\mu}u^{\beta}~u^{\alpha} - (\alpha \leftrightarrow \beta)
\nonumber\\
F^{\alpha}_{3} = u_{\rho}~d^{\rho}u^{\alpha}
\eea

Then we have, with the notations introduced above:
\begin{thm}
The following formulas are true
\bea
sT(T(x_{1})^{(2)}, T(x_{2})^{(2)}) =
\nonumber\\
\delta( x_{1} - x_{2})~ (2 : E_{3}^{\mu\nu}~C_{\mu\nu}: - : E_{3}~C: + 2 :B_{1}^{\mu}~D_{\mu}:)(x_{2})
\nonumber\\
+ \partial_{\lambda}\delta( x_{1} - x_{2})~
\Bigl\{ \Bigl[ - : E_{3\mu\nu}(x_{1})~E^{\mu\nu,\lambda}(x_{2}): + {1\over 2}~: E_{3}(x_{1})~E^{\lambda}(x_{2}): 
- B_{1\mu}(x_{1})~F^{\mu,\lambda}(x_{2}): \Bigl] 
\nonumber\\
- [x_{1} \leftrightarrow x_{2}] \Bigl\}
\label{st2-2-1}
\eea
\bea
sT(T^{\rho}(x_{1})^{(2)}, T(x_{2})^{(2)}) =
\delta( x_{1} - x_{2})~ \Bigl( - : B_{1\sigma}~C_{\rho\sigma}: +  {1\over 2}~: B_{1}^{\rho}~C: 
\nonumber\\
+ :C^{\mu\nu,\rho}~E_{3\mu\nu}: - {1\over 2}~: C^{\rho}~E_{3}: - :D^{\mu,\rho}~B_{1\mu}\Bigl)(x_{2}): 
\nonumber\\
+ \partial_{\lambda}\delta( x_{1} - x_{2})~
\Bigl[ : B_{1\sigma}(x_{1})~E^{\rho\sigma,\lambda}(x_{2}): - {1\over 2}~: B_{1}^{\rho}(x_{1})~E^{\lambda}(x_{2}):
\nonumber\\
+ : E^{\mu\nu,\lambda,\rho}(x_{1})~E_{3\mu\nu}(x_{2}):  - {1\over 2}~: E^{\lambda,\rho}(x_{1})~E_{3}(x_{2}): \Bigl]
\nonumber\\
- \partial_{\lambda_{1}}\partial_{\lambda_{2}}\delta( x_{1} - x_{2})~
:H^{\mu,\lambda_{1}\lambda_{2},\rho}(x_{1})~B_{1\mu}(x_{2}):
\label{st2-2-2}
\eea
\bea
sT(T^{\rho}(x_{1})^{(2)}, T^{\sigma}(x_{2})^{(2)}) =
\delta( x_{1} - x_{2})~\Bigl[ \Bigl(: C^{\mu\sigma,\rho}~B_{1\mu}: - {1\over 2}~: C^{\rho}~B_{1}^{\sigma}:\Bigl) (x_{2})
- (\rho \leftrightarrow \sigma) \Bigl]
\nonumber\\
+ \partial_{\lambda}\delta( x_{1} - x_{2})~
\Bigl\{ \Bigl[ : E^{\sigma\mu,\lambda,\rho}(x_{1})~B_{1\mu}(x_{2}): 
- {1\over 2}~: E^{\lambda,\rho}(x_{1})~B_{1}^{\sigma}(x_{2}): \Bigl] + 
(x_{1} \leftrightarrow x_{2}, \rho \leftrightarrow \sigma) \Bigl\}
\label{st2-2-3}
\eea
\bea
sT(T^{\rho\sigma}(x_{1})^{(2)}, T(x_{2})^{(2)}) =
\delta( x_{1} - x_{2})~\Bigl( - : C^{\mu\nu,\rho\sigma}~E_{3\mu\nu}: +  {1\over 2}~: C^{\rho\sigma}~E_{3}:\Bigl)(x_{2})
\nonumber\\
+ \partial_{\lambda}\delta( x_{1} - x_{2})~
\Bigl[ - : E^{\mu\nu,\lambda,\rho\sigma}(x_{1})~E_{3\mu\nu}(x_{2}): 
+ {1\over 2}~: E^{\lambda,\rho\sigma}(x_{1})~E_{3}(x_{2}): 
\nonumber\\
+ : \tilde{E}_{1}^{\mu\nu,\rho\sigma}(x_{1})~{B_{\mu,}}^{\lambda}(x_{2}): 
+ : \tilde{E}_{2}^{\rho}(x_{1})~B^{\sigma,\lambda}(x_{2}):
- : \tilde{E}_{2}^{\sigma}(x_{1})~B^{\rho,\lambda}(x_{2}): \Bigl]
\label{st2-2-4}
\eea
\bea
sT(T^{\rho\sigma}(x_{1})^{(2)}, T^{\tau}(x_{2})^{(2)}) =
{1\over 2}~ \delta( x_{1} - x_{2})~( : B_{1}^{\mu,\rho\sigma}~{G_{\mu,}}^{\tau}: 
+ : B_{1}^{\rho}~G^{\sigma,\tau}: - : B_{1}^{\sigma}~G^{\rho,\tau}:
\nonumber\\
- 2 : C^{\mu\tau,\rho\sigma,}~B_{1\mu}: +  : C^{\rho\sigma}~B_{1}^{\tau}:)(x_{2})
\label{st2-2-5}
\eea
\bea
sT(T^{\rho\sigma\tau}(x_{1})^{(2)}, T(x_{2})^{(2)}) =
- \delta( x_{1} - x_{2})~: D^{\mu,\rho\sigma\tau}~B_{1\mu}: (x_{2})
\nonumber\\
- \partial_{\lambda}\delta( x_{1} - x_{2})~: F^{\mu,\lambda,\rho\sigma\tau}(x_{1})~B_{1\mu}(x_{2}): 
\label{st2-2-6}
\eea
\bea
sT(T^{\rho_{1}\sigma_{1}}(x_{1})^{(2)}, T^{\rho_{2}\sigma_{2}}(x_{2})^{(2)}) = 0
\label{st2-2-7}
\eea
\bea
sT(T^{\rho\sigma\tau}(x_{1})^{(2)}, T^{\nu}(x_{2})^{(2)}) = 0
\label{st2-2-8}
\eea
\label{sTT}
\end{thm}
{\bf Proof:} We consider for illustration only the first relation. We have to compute the expressions
$
A = d_{Q}T(T(x_{1})^{(2)}, T(x_{2})^{(2)}), B = - i\partial_{\mu}^{1}T(T^{\mu}(x_{1})^{(2)}, T(x_{2})^{(2)}),
C = - i\partial_{\mu}^{2}T(T(x_{1})^{(2)}, T^{\mu}(x_{2})^{(2)})
$
using the relations . The expression in the left hand side of is
$
A + B + C.
$
There are a lot of cancelations and only the terms with 
$
\Box
$
acting on 
$
D^{F}_{0}( x_{1} - x_{2}), \partial_{\rho}D^{F}_{0}( x_{1} - x_{2})
$
survive. 

Some identities are valid and must be used to obtain the compensations:
\bea
E^{\mu\nu,\lambda,\rho}_{1} = - ( \lambda \leftrightarrow \rho), \qquad 
E^{\lambda,\rho}_{4} = - ( \lambda \leftrightarrow \rho), \qquad 
E^{\lambda,\rho}_{1} = - ( \lambda \leftrightarrow \rho)
\nonumber\\
E^{\lambda,\nu}_{2} = {1\over 2}~B^{\lambda,\nu}
\nonumber\\
{\cal S}_{\lambda\rho}(E^{\lambda\mu,\rho}_{1}) = {\cal S}_{\lambda\rho}(F^{\mu,\lambda,\rho}_{1}) 
\nonumber\\
\tilde{E}^{\rho,\lambda,\sigma}_{1} = - ( \lambda \leftrightarrow \sigma)
\nonumber\\
B^{\mu,\lambda,\rho}_{1} = - 2~\tilde{E}^{\mu,\lambda,\rho}_{1}
\nonumber\\
B^{\mu}_{1} = 2~\tilde{E}^{\mu}_{2}
\nonumber\\
F^{\lambda,\rho\sigma}_{1} = \tilde{E}^{\sigma,\rho,\lambda}_{1} - \tilde{E}^{\rho,\sigma,\lambda}_{1}
\nonumber\\
F^{\mu,\rho\sigma}_{2} = \tilde{E}^{\mu,\rho,\sigma}_{1}
\nonumber\\
F^{\mu}_{3} = \tilde{E}^{\mu}_{2}.
\eea
$\qed$

From  (\ref{st2-2-1}) we can obtain a simple form of the anomaly:
\be
sT(T(x_{1})^{(2)}, T(x_{2})^{(2)}) = \delta( x_{1} - x_{2})~{\cal A}(T,T)(x_{2})
\ee
where
\bea
{\cal A}(T,T) \equiv 2 : E_{3}^{\mu\nu}~C_{\mu\nu}: - : E_{3}~C: + 2 :B_{1}^{\mu}~D_{\mu}: 
- :\partial_{\lambda}B_{1\mu}~F^{\mu,\lambda}: - :B_{1\mu}~\partial_{\lambda}F^{\mu,\lambda}:
\nonumber\\
+ :\partial_{\lambda}E_{3\mu\nu}~E_{1}^{\mu\nu,\lambda}: - {1\over 2}~: \partial_{\lambda}E_{3}~E_{2}^{\lambda}:
- :E_{3\mu\nu}~\partial_{\lambda}E_{1}^{\mu\nu,\lambda}: + {1\over 2}~: E_{3}~\partial_{\lambda}E_{2}^{\lambda}:
\nonumber\\
\label{st2-2-1a}
\eea
The previous anomaly can be written in a simpler form:
\bea
{\cal A}(T,T) = : B^{\mu,\nu}~C_{\mu\nu}: + 2 :B_{1}^{\mu}~D_{\mu}: 
- : B_{\mu,\nu}~\partial_{\lambda}E^{\mu\nu,\lambda}: - 2~:B_{1}^{\mu}~\partial^{\lambda}F_{\mu,\lambda}:
\nonumber\\
- \partial_{\lambda} \Bigl( {1\over 2}~: B_{\mu,\nu}~E^{\mu\nu,\lambda}: + : B_{1\mu}~F^{\mu,\lambda}: \Bigl)
\label{st2-2-1aa}
\eea

We consider only the first contribution
\bea
{\cal A}^{\prime}(T,T) = : B^{\mu,\nu}~( C_{\mu\nu} - \partial_{\lambda}E^{\mu\nu,\lambda}):~ 
+ 2 :B_{1}^{\mu}~( D_{\mu} - \partial^{\lambda}F_{\mu,\lambda}):
\label{st2-2-1aaa}
\eea

It is useful to consider separately the contributions trilinear and linear in $h$ namely 
$
{\cal A}_{uhhh}
$
and
$
{\cal A}_{uu\tilde{u}h}
$
respectively. After some calculations one get:
\be
{\cal A}_{uhhh} = u_{\rho}~{\cal A}^{\rho} + {u_{\rho,}}^{\rho}~{\cal A} + u_{\mu,\nu}~{\cal A}^{\mu\nu}
\label{Auhhh}
\ee
where
\bea
{\cal A}_{\rho} \equiv 4~h_{,\rho}~{\cal D} + 4~h_{\mu\nu,\rho}~{\cal D}^{\mu\nu}
\nonumber\\
{\cal A} \equiv 4~h~{\cal D} + 4~h_{\mu\nu}~{\cal D}^{\mu\nu}
\nonumber\\
{\cal A}_{\mu\nu} = - 8~h_{\mu\nu}~{\cal D} - 8~{h_{\nu}}^{\lambda}~{\cal D}_{\mu\lambda}
\eea
where
\be
{\cal D} \equiv 2 ( {h^{\rho\sigma}}_{,\sigma}~h_{,\rho} + h^{\rho\sigma}~h_{,\rho\sigma}
- h^{\rho\sigma,\lambda}~h_{\rho\sigma,\lambda})
\ee
and
\bea
{\cal D}_{\mu\nu} \equiv - h_{,\mu}~h_{,\nu} + 2~{h^{\rho\sigma}}_{,\mu}~h_{\rho\sigma,\nu} 
+ 4~( h_{\mu\rho,\sigma}~{h_{\nu}}^{\sigma,\rho} + h_{\mu\rho,\sigma}~{h_{\nu}}^{\rho,\sigma})
\nonumber\\
- 4~( {h^{\rho\sigma}}_{,\rho}~h_{\mu\nu,\sigma} + h^{\rho\sigma}~h_{\mu\nu,\rho\sigma})
- 4~( h_{\mu\rho,\sigma}~{h^{\rho\sigma}}_{,\nu} + h_{\nu\rho,\sigma}~{h^{\rho\sigma}}_{,\mu}
+ h_{\mu\rho}~{h^{\rho\sigma}}_{,\nu\sigma} + h_{\nu\rho}~{h^{\rho\sigma}}_{,\mu\sigma}).
\eea

To obtain a simpler form for 
$
{\cal A}_{uu\tilde{u}h}
$
it is useful to eliminate total derivatives and exhibit it in the form:
$
{\cal A}_{uu\tilde{u}h} = {\rm total~divergence} + h_{\mu\nu}~{\cal B}^{\mu\nu}
$
i.e. to eliminate the derivatives on 
$
h_{\mu\nu}.
$
The end result is very nice i.e.
$
{\cal B}^{\mu\nu} = 0
$
so 
$
{\cal A}_{uu\tilde{u}h}
$
is a total derivative:
\be
{\cal A}_{uu\tilde{u}h} = d_{\mu}~[ 4~u^{\mu}~h_{\rho\sigma}~{\cal C}^{\rho\sigma} 
- 16~u_{\sigma}u^{\rho,\sigma}~(\tilde{u}_{\rho,\lambda} + \tilde{u}_{\lambda,\rho})~h^{\mu\lambda} ]
\label{Ahuuu}
\ee
where
\be
{\cal C}_{\rho\sigma} = 2~[ {u^{\lambda}}_{,\rho}~(\tilde{u}_{\lambda,\sigma} +  \tilde{u}_{\sigma,\lambda})
+ u^{\lambda}~\tilde{u}_{\rho,\sigma\lambda} ] + (\rho \leftrightarrow \sigma).
\ee

Now we can try to eliminate the anomaly 
$
{\cal A}_{uhhh}
$
using finite renormalizations:
\be
T^{\rm ren}(T^{I}(x_{1}),T^{J}(x_{2})) = T(T^{I}(x_{1}),T^{J}(x_{2})) + \delta(x_{1} - x_{2})~N(T^{I},T^{J})(x_{2})
\ee
with the polynomials
$
N(T^{I},T^{J})
$
verify the symmetry property:
\be
N(T^{I},T^{J}) = (-1)^{|I||J|}~N(T^{J},T^{I})
\ee
and
\be
gh(N(T^{I},T^{J})) = |I| + |J|, \qquad \omega(N(T^{I},T^{J})) \leq 6.
\ee
Because the anomaly 
$
{\cal A}_{uhhh}
$
is quadri-linear, we are looking for an expression 
$
N(T,T)
$
which is also quadri-linear. The result is \cite{massive}:
\begin{thm}
The anomaly
$
{\cal A}_{uhhh}
$
can be eliminated if one takes 
\bea
N(T,T) = i~( - 16~h^{\mu\nu}~h^{\rho\sigma}~d_{\rho}h_{\mu\nu}~d_{\sigma}h
+ 8~h^{\mu\nu}~h^{\rho\sigma}~d_{\lambda}h_{\mu\nu}~d^{\lambda}h_{\rho\sigma}
\nonumber\\
+ 32~h^{\mu\nu}~h_{\nu\rho}~d^{\alpha}h^{\rho\beta}~d_{\beta}h_{\mu\alpha}
+ 32~h^{\mu\nu}~h^{\rho\sigma}~d_{\mu}h_{\rho\alpha}~d_{\nu}{h_{\sigma}}^{\alpha}
\nonumber\\
- 32~h^{\mu\nu}~h_{\nu\rho}~d^{\alpha}h_{\mu\beta}~d_{\alpha}h^{\rho\beta}
\nonumber\\
- 16 h^{\mu\nu}~h^{\rho\sigma}~d_{\lambda}h_{\mu\rho}~d^{\lambda}h_{\nu\sigma}
+ 16 h^{\mu\nu}~{h_{\nu}}^{\rho}~d^{\lambda}h_{\mu\rho}~d_{\lambda}h)
\label{second-order}
\eea
\end{thm}
{\bf Proof:}
We consider finite renormalizations of the type
\be
R(T^{I}(x_{1}),T^{J}(x_{2})) = \delta(x_{1} - x_{2})~N(T^{I},T^{J})(x_{2}) 
\label{R}
\ee
where the polynomials
$
N(T^{I},T^{J})
$
verify the symmetry property:
\be
N(T^{I},T^{J}) = (-1)^{|I||J|}~N(T^{J},T^{I}).
\label{symN}
\ee
Then we have by direct computation that the expression
\bea
sR(T^{I}(x_{1}),T^{J}(x_{2})) \equiv d_{Q}R(T^{I}(x_{1}),T^{J}(x_{2}))
\nonumber\\
- i~\partial_{\mu}^{1}R(T^{I\mu}(x_{1}),T^{J}(x_{2}))
- i~(-1)^{|I|}~\partial_{\mu}^{2}R(T^{I}(x_{1}),T^{J\mu}(x_{2}))
\label{sR}
\eea
is:
\bea
sR(T^{I}(x_{1}),T^{J}(x_{2})) = \delta(x_{1} - x_{2})~[ d_{Q}N(T^{I},T^{J}) - i~(-1)^{|I|}~\partial_{\mu}N(T^{I},T^{J\mu})](x_{2})
\nonumber\\
- i~\partial_{\mu}\delta(x_{1} - x_{2})~[ N(T^{I\mu},T^{J}) - (-1)^{|I|}~N(T^{I},T^{J\mu})](x_{2}) 
\eea
In particular, due to the symmetry property
\bea
sR(T(x_{1}),T(x_{2})) = \delta(x_{1} - x_{2})~[ d_{Q}N(T,T) - i~\partial_{\mu}N(T,T^{\mu})](x_{2})
\eea
Taking into account the form of the anomaly (\ref{st2-2-1aa}) it follows that we must have 
\be
{\cal A}(T,T) + d_{Q}N(T,T) - i~\partial_{\mu}N(T,T^{\mu}) = 0
\ee
and from here with (\ref{st2-2-1aaa})
\be
{\cal A}^{\prime}(T,T) + d_{Q}N(T,T) - i~\partial_{\mu}N^{\prime}(T,T^{\mu}) = 0
\ee
and finally with (\ref{Ahuuu})
\be
{\cal A}_{uhhh} + d_{Q}N(T,T) - i~\partial_{\mu}N^{\prime\prime}(T,T^{\mu}) = 0.
\ee

It is useful to transform the expression (\ref{Auhhh}) of
$
{\cal A}_{uhhh}
$
such that we eliminate, up to a total derivative, all terms with derivatives on the ghost field i.e. we write
\be
{\cal A}_{uhhh} = u_{\rho}~\tilde{\cal A}^{\rho} + d_{\rho}{\cal B}^{\rho}.
\ee
Indeed, we obtain from (\ref{Auhhh})
\be
\tilde{\cal A}_{\rho} = {\cal A}_{\rho} - d_{\rho}{\cal A} - d^{\sigma}{\cal A}_{\rho\sigma},
\qquad
{\cal B}^{\rho} = u^{\rho}~{\cal A} + u_{\sigma}~{\cal A}^{\sigma\rho}.
\ee

The strategy is to find an expression $N$ such that
\be
d_{Q}N = i~(u_{\rho}~\tilde{\cal A}_{\rho} + d_{\rho}N^{\rho}).
\ee
For this purpose, we make an ansatz for $N$, as an expresssion quadri-linear in $h$; it is sufficient to consider 
terms of the type
$
h h dh dh
$.
There are $43$ terms of this type so we have an ansatz
\be
N = \sum_{j=1}^{43}~a_{j}N_{j}
\ee
We do not provide the complete list, but only give
\bea
N_{1} = h^{\mu\nu}~h^{\rho\sigma}~d_{\rho}h_{\mu\nu}~d_{\sigma}h
\nonumber\\
N_{2} = h^{\mu\nu}~h^{\rho\sigma}~d_{\lambda}h_{\mu\nu}~d^{\lambda}h_{\rho\sigma}
\nonumber\\
N_{3} = h^{\mu\nu}~h_{\nu\rho}~d^{\alpha}h^{\rho\beta}~d_{\beta}h_{\mu\alpha}
\nonumber\\
N_{4} = h^{\mu\nu}~h^{\rho\sigma}~d_{\mu}h_{\rho\alpha}~d_{\nu}{h_{\sigma}}^{\alpha}
\nonumber\\
N_{5} = h^{\mu\nu}~h_{\nu\rho}~d^{\alpha}h_{\mu\beta}~d_{\alpha}h^{\rho\beta}
\nonumber\\
N_{6} = h^{\mu\nu}~h^{\rho\sigma}~d_{\lambda}h_{\mu\rho}~d^{\lambda}h_{\nu\sigma}
\nonumber\\
N_{7} = h^{\mu\nu}~{h_{\nu}}^{\rho}~d^{\lambda}h_{\mu\rho}~d_{\lambda}h
\eea
By direct computation we can exhibit the expresions
$
d_{Q}N_{j}
$
in the form
\be
d_{Q}N_{j} = i(u_{\rho}~{\cal A}_{j}^{\rho} + d_{\rho}N_{j}^{\rho}) 
\ee
and we impose the equation
\be
\sum a_{j}~{\cal A}_{j}^{\rho} = \tilde{\cal A}^{\rho}.
\ee
This equation is equivalent to a system of linear equations for the coefficients 
$
a_{j};
$
the solution is not unique. If we consider our lucky guess i.e. only the coefficients
$
a_{j}\quad j = 1,\dots,7
$
then we get an unique solution, namely the expression (\ref{second-order}) from the statement.
$\qed$

\newpage
\section{Finite Renormalizations\label{finite}}

We have mentioned in the preceding Section that the solution for the finite renormalization $N$ is not unique.
Here we investgate how big is this non-uniqueness. In other words, we suppose that we have a solution
$
T(T^{I}(x_{1}),T^{j}(x_{2}))
$
such that Bogoliubov axioms and gauge invariance i.e. the cocyle relations
\be
sT(T^{I}(x_{1}),T^{j}(x_{2})) = 0
\ee
are fulfiled. Then the arbitrariness is given by finite renormalizations of the type:
\bea
R(T^{I}(x_{1}),T^{J}(x_{2})) = \delta(x_{1} - x_{2})~N(T^{I},T^{J})(x_{2})
+ \partial_{\mu}\delta(x_{1} - x_{2})~N(T^{I},T^{J})^{\mu}(x_{2})
\nonumber\\
+ \partial_{\mu}\partial_{\nu}(x_{1} - x_{2})~N(T^{I},T^{J})^{\mu\nu}(x_{2})
\label{R-IJ}
\eea
where the polynomials $N$ are Lorentz covariant, from power counting we have
\be
\omega(N(T^{I},T^{J})) \leq 6, \quad \omega(N(T^{I},T^{J})^{\mu}) \leq 5,\quad
\omega(N(T^{I},T^{J})^{\mu\nu}) \leq 4
\ee
and we also have
\be
gh(N(T^{I},T^{J}))^{\dots} = |I| + |J|
\ee
and we can suppose that
$
N(T^{I},T^{J})^{\mu\nu}
$
is symmetric in 
$
\mu \leftrightarrow \nu.
$

If we want to preserve the symmetry properties of the chronological products (see Section \ref{Bogoliubov}) we must impose
\be
R(T^{I}(x_{1}),T^{J}(x_{2})) = (-1)^{|I||J|}~R(T^{J}(x_{2}),T^{I}(x_{1}))
\label{S-IJ}
\ee
which is equivalent to the following system of equations:
\bea
N(T^{I},T^{J}) = (-1)^{|I||J|}~[ N(T^{J},T^{I}) + d_{\mu}N(T^{J},T^{I})^{\mu}
+ d_{\mu} d_{\nu}N(T^{J},T^{I})^{\mu\nu} ]
\nonumber\\
N(T^{I},T^{J})^{\mu} = - (-1)^{|I||J|}~[ N(T^{J},T^{I})^{\mu} + 2 d_{\nu}N(T^{J},T^{I})^{\mu\nu} ]
\nonumber\\
N(T^{I},T^{J})^{\mu\nu} = (-1)^{|I||J|}~N(T^{J},T^{I})^{\mu\nu}.
\label{symR}
\eea

More important, if we want that gauge invariance 
$
sT(T^{I}(x_{1}),T^{J}(x_{2})) = 0
$
is preserved we must also have the cocyle relations
\be
sR(T^{I}(x_{1}),T^{J}(x_{2})) = 0
\label{sR-IJ}
\ee
which is equivalent to the following system of equations:
\bea
d_{Q}N(T^{I},T^{J}) = i~(-1)^{|I|}~d_{\mu}N(T^{I},T^{J\mu})
\nonumber\\
d_{Q}N(T^{I},T^{J})^{\mu} = i~\{ N(T^{I\mu},T^{J}) - (-1)^{|I|}~[ N(T^{I},T^{J\mu}) - d_{\nu}N(T^{I},T^{J\nu})^{\mu} ]\}
\nonumber\\
d_{Q}N(T^{I},T^{J})^{\rho\sigma} = i~{\cal S}_{\rho\sigma} \{ N(T^{I\rho},T^{J})^{\sigma} 
- (-1)^{|I|}~[ N(T^{I},T^{J\rho})^{\sigma} - d_{\mu}N(T^{I},T^{J\mu})^{\rho\sigma} ]\}
\nonumber\\
{\cal S}_{\mu\rho\sigma} [ N(T^{I\mu},T^{J})^{\rho\sigma} - (-1)^{|I|}~N(T^{I},T^{J\mu})^{\rho\sigma} ] = 0.
\label{renR}
\eea

The restrictions from the preceding relations are rather severe and we start to analyse this cohomology problem. 
\begin{thm}
Suppose we have a cocyle 
$
R(T^{I}(x_{1}),T^{j}(x_{2}))
$
of the type (\ref{R-IJ}) and verifying the cocyle equation (\ref{sR-IJ}). Then $R$ is cohomologous with an cocyle
verifying
\be
N(T^{I},T^{J})^{\mu\nu} = 0.
\ee
\label{NIJ2}
\end{thm}
{\bf Proof:} (i) We first define the cochain
\be
S(T^{\mu},T) = \partial_{\nu}\delta(x_{1} - x_{2})~S^{\mu\nu}(x_{2})
\ee
and compute the coboundary
\be
\delta S(T,T) = \delta(x_{1} - x_{2})~d_{\mu}d_{\nu}S^{\mu\nu}(x_{2}) 
- 2 \partial_{\mu}\delta(x_{1} - x_{2})~d_{\nu}S^{\mu\nu}(x_{2})  
+ 2 \partial_{\mu}\partial_{\nu}\delta(x_{1} - x_{2})~S^{\mu\nu}(x_{2}).  
\ee

So, if we choose
\be
S^{\mu\nu} = {1\over 2}~N(T,T)^{\mu\nu}
\ee
the expression 
\be
R^{\prime}(T,T) \equiv R(T,T) - \delta S(T,T)
\ee
which is cohomologous to 
$
R(T,T)
$
has a simpler form i.e.
\be
R^{\prime}(T,T) = \delta(x_{1} - x_{2})~R(x_{2}) + \partial_{\nu}\delta(x_{1} - x_{2})~R^{\mu}(x_{2})
\ee
and this means that we can take 
\be
N(T,T)^{\mu\nu} = 0.
\label{N1}
\ee

(ii) We now define the cochain
\bea
S(T^{\mu},T^{\nu}) = \partial_{\rho}\delta(x_{1} - x_{2})~S^{\{\mu\nu\},\rho}(x_{2}) 
- 2~\delta(x_{1} - x_{2})~\partial_{\rho}S^{\{\mu\nu\},\rho}(x_{2})
\nonumber\\
S(T^{\mu\nu},T) = \partial_{\rho}\delta(x_{1} - x_{2})~S^{[\mu\nu],\rho}(x_{2})
\eea
such that we have the symmetry property
\be
S(T^{\mu},T^{\nu}) = - S(T^{\nu},T^{\mu}).
\ee

Now we compute the coboundary
\be
\delta S(T^{\mu},T) = \partial_{\nu}^{1}S(T^{\mu\nu},T) - \partial_{\nu}^{2}S(T^{\mu},T^{\nu})
\ee
and we find
\be
\delta S(T^{\mu},T) = \partial_{\rho}\partial_{\sigma}S^{\mu\rho,\sigma}(x_{2}) + \cdots 
\ee
where
\be
S^{\mu\nu,\rho} \equiv S^{\{\mu\nu\},\rho} + S^{[\mu\nu],\rho}
\ee
and $\cdots$ are terms with one and no derivatives on $\delta$. If we choose
\be
S^{\mu\nu,\rho} = N(T^{\mu},T)^{\nu\rho}
\ee
then it follows that the cocyle
$
R(T^{\mu},T)
$
is cohomologous with a cocyle having
\be
N(T^{\mu},T)^{\rho\sigma} = 0.
\label{N2}
\ee

(iii) Let us define the cochain
\be
S(T^{\mu\nu},T^{\rho}) = \partial_{\sigma}\delta(x_{1} - x_{2})~S^{[\mu\nu][\rho]\sigma}(x_{2})
\ee
and compute the coboundary
\be
\delta S(T^{\mu},T^{\nu}) = \partial_{\rho}^{1}S(T^{\mu\rho},T^{\nu}) - \partial_{\rho}^{2}S(T^{\mu},T^{\nu\rho}).
\ee
After some computations we find
\be
\delta S(T^{\mu},T^{\nu}) = \partial_{\rho}\partial_{\sigma}\delta(x_{1} - x_{2})~
(S^{[\mu\rho][\nu]\sigma} - S^{[\nu\rho][\mu]\sigma})(x_{2}) + \cdots
\ee
where $\cdots$ are terms with one and no derivatives on $\delta$. We choose
\be
S^{[\mu\nu][\rho]\sigma} \equiv {1\over 2}~[ N(T^{\mu},T^{\nu})^{\rho\sigma} 
+ N(T^{\mu},T^{\rho})^{\nu\sigma} + N(T^{\nu},T^{\rho})^{\mu\sigma} ].
\ee
From the last relation (\ref{symR}) we find out that 
\be
N(T^{\mu},T^{\nu})^{\rho\sigma} = - N(T^{\nu},T^{\mu})^{\rho\sigma} 
\ee
so the preceding definition is consistent: we have the anti-symmetry property in
$
\mu \leftrightarrow \nu.
$
Moreover, we easily obtain:
\be
S^{[\mu\rho][\nu]\sigma} - S^{[\nu\rho][\mu]\sigma} = N(T^{\mu},T^{\nu})^{\rho\sigma}
\ee
so
\be
\delta S(T^{\mu},T^{\nu}) = 
\partial_{\rho}\partial_{\sigma}\delta(x_{1} - x_{2})~N(T^{\mu},T^{\nu})^{\rho\sigma}(x_{2}) + \cdots
\ee
It follows that the cocyle
$
R(T^{\mu},T^{\nu})
$
is cohomologous with a cocyle having
\be
N(T^{\mu},T^{\nu})^{\rho\sigma} = 0.
\label{N3}
\ee

(iv) We choose the cochain
\bea
S(T^{\mu\nu\rho},T) = \partial_{\sigma}\delta(x_{1} - x_{2})~S^{[\mu\nu\rho]\sigma}(x_{2})
\nonumber\\
S(T^{\mu\nu},T^{\rho}) = 0
\eea
so in this way we preserve the preceding relation (\ref{N3}). We compute the coboundary
\be
\delta S(T^{\mu\nu},T) = \partial_{\rho}^{1}S(T^{\mu\nu\rho},T) +  \partial_{\rho}^{2}S(T^{\mu\nu},T^{\rho})
\ee
and we find  immediately
\be
\delta S(T^{\mu\nu},T) = \partial_{\rho}\partial_{\sigma}\delta(x_{1} - x_{2})~
S^{[\mu\nu\rho]\sigma}(x_{2})
\ee
Let us take
\be
S^{[\mu\nu\rho]\sigma} \equiv N(T^{\mu\nu},T)^{\rho\sigma} + N(T^{\nu\rho},T)^{\mu\sigma} +
N(T^{\rho\mu},T)^{\nu\sigma}
\ee
such that we have complete anti-symmetry in the indexes
$
\mu,\nu,\rho.
$
From the last relation (\ref{renR}) we find out that 
\be
{\cal S}_{\nu\rho\sigma} [ N(T^{\mu\nu},T)^{\rho\sigma} + N(T^{\mu},T^{\nu})^{\rho\sigma} ] = 0 
\ee
If we use (\ref{N3}) we are left with
\be
{\cal S}_{\nu\rho\sigma} N(T^{\mu\nu},T)^{\rho\sigma} = 
N(T^{\mu\nu},T)^{\rho\sigma} + N(T^{\mu\rho},T)^{\nu\sigma} + N(T^{\mu\sigma},T)^{\nu\rho} = 0. 
\ee
If we use this relation we end up with
\be
S^{[\mu\nu\rho]\sigma} = N(T^{\mu\nu},T)^{\rho\sigma} + N(T^{\rho\sigma},T)^{\mu\nu}
\ee
which implies
\be
\delta S(T^{\mu\nu},T) = \partial_{\rho}\partial_{\sigma}\delta(x_{1} - x_{2})~
N(T^{\mu\nu},T^{\rho\sigma})(x_{2}). 
\ee
It follows that 
$
R(T^{\mu\nu},T)
$
is cohomologous with a cocyle having
\be
N(T^{\mu\nu},T)^{\rho\sigma} = 0.
\label{N4}
\ee

(v) We define the cochain
\bea
S(T^{\mu\nu\rho},T^{\sigma}) = \partial_{\lambda}\delta(x_{1} - x_{2})~S^{[\mu\nu\rho][\sigma]\lambda}(x_{2})
\nonumber\\
S(T^{\mu\nu\rho\sigma},T) = 0
\eea
and compute the coboundary
\be
\delta S(T^{\mu\nu\rho},T) = \partial_{\sigma}^{1}S(T^{\mu\nu\rho\sigma},T) 
-  \partial_{\sigma}^{2}S(T^{\mu\nu\rho},T^{\sigma}).
\ee
We immediately get:
\be
\delta S(T^{\mu\nu\rho},T) = \partial_{\lambda}\partial_{\sigma}\delta(x_{1} - x_{2})~
S^{[\mu\nu\rho][\sigma]\lambda}(x_{2}) + \cdots
\ee
We choose
\be
S^{[\mu\nu\rho][\sigma]\lambda} \equiv N(T^{\mu\nu\rho},T)^{\lambda\sigma}
\ee
and get 
\be
\delta S(T^{\mu\nu\rho},T) = \partial_{\lambda}\partial_{\sigma}\delta(x_{1} - x_{2})~
N(T^{\mu\nu\rho},T)^{\lambda\sigma}(x_{2}) + \cdots
\ee
so 
$
R(T^{\mu\nu\rho},T)
$
is cohomologous with a cocyle having
\be
N(T^{\mu\nu\rho},T)^{\alpha\beta} = 0.
\label{N5}
\ee

(vi) We use the cochain
\bea
S(T^{\mu\nu},T^{\rho\sigma}) = \partial_{\lambda}\delta(x_{1} - x_{2})~S^{[\mu\nu][\rho\sigma]\lambda}(x_{2})
\nonumber\\
S(T^{\mu\nu\rho},T^{\sigma}) = 0
\eea
and this preserves (\ref{N5}). We compute the coboundary
\be
\delta S(T^{\mu\nu},T^{\rho}) = \partial_{\sigma}^{1}S(T^{\mu\nu\sigma},T^{\rho}) 
+  \partial_{\sigma}^{2}S(T^{\mu\nu},T^{\rho\sigma})
\ee
and we immediately obtain
\be
\delta S(T^{\mu\nu},T^{\rho}) = - \partial_{\lambda}\partial_{\sigma}\delta(x_{1} - x_{2})~
S^{[\mu\nu][\rho\sigma]\lambda}(x_{2}) + \cdots
\ee
Let us choose
\be
S^{[\mu\nu][\rho\sigma]\lambda} \equiv - {1\over 2}~
[ N(T^{\mu\nu},T^{\rho})^{\sigma\lambda} - N(T^{\mu\nu},T^{\sigma})^{\rho\lambda}]
\ee
From the last relation (\ref{renR}) we find out that 
\be
{\cal S}_{\rho\sigma\lambda} [ N(T^{\mu\nu\rho},T)^{\sigma\lambda} - N(T^{\mu\nu},T^{\rho})^{\sigma\lambda} ] = 0 
\ee
and if we use (\ref{N5}) we are left with
\be
{\cal S}_{\rho\sigma\lambda} N(T^{\mu\nu},T^{\rho})^{\sigma\lambda} = 0 \qquad
\Leftrightarrow \qquad
{\cal S}_{\sigma\lambda} N(T^{\mu\nu},T^{\sigma})^{\rho\lambda} = - N(T^{\mu\nu},T^{\rho})^{\sigma\lambda}.
\ee
Then we have
\be
\delta S(T^{\mu\nu},T^{\rho}) = \partial_{\lambda}\partial_{\sigma}\delta(x_{1} - x_{2})~
N(T^{\mu\nu},T^{\rho})^{\sigma\lambda}(x_{2}) + \cdots
\ee
so 
$
R(T^{\mu\nu},T^{\rho})
$
is cohomologous with a cocyle having
\be
N(T^{\mu\nu},T^{\rho})^{\alpha\beta} = 0.
\label{N6}
\ee

(vii) We consider the cochain
\be
S(T^{\mu\nu\rho\sigma},T^{\lambda}) = 
\partial_{\alpha}\delta(x_{1} - x_{2})~S^{[\mu\nu\rho\sigma][\lambda]\alpha}(x_{2})
\ee
and compute the coboundary
\be
\delta S(T^{\mu\nu\rho\sigma},T) = \partial_{\lambda}^{2}S(T^{\mu\nu\rho\sigma},T^{\lambda}).
\ee
The result is
\be
\delta S(T^{\mu\nu\rho\sigma},T) = - \partial_{\lambda}\partial_{\alpha}\delta(x_{1} - x_{2})~
S^{[\mu\nu\rho\sigma][\lambda]\alpha}(x_{2}) + \cdots
\ee
If we choose
\be
S^{[\mu\nu\rho\sigma][\lambda]\alpha} \equiv - N(T^{\mu\nu\rho\sigma},T)^{\lambda\alpha}
\ee
then we have
\be
\delta S(T^{\mu\nu\rho\sigma},T) = \partial_{\lambda}\partial_{\alpha}\delta(x_{1} - x_{2})~
N(T^{\mu\nu\rho\sigma},T)^{\lambda\alpha}(x_{2}) + \cdots
\ee
so 
$
R(T^{\mu\nu\rho\sigma},T)
$
is cohomologous with a cocyle having
\be
N(T^{\mu\nu\rho\sigma},T)^{\alpha\beta} = 0.
\label{N7}
\ee

(viii) We start from the cochain
\bea
S(T^{\mu\nu\rho},T^{\sigma\lambda}) = 
\partial_{\alpha}\delta(x_{1} - x_{2})~S^{[\mu\nu\rho][\sigma\lambda]\alpha}(x_{2})
\nonumber\\
S(T^{\mu\nu\rho\sigma},T^{\lambda}) = 0
\eea
so (\ref{N7}) is preserved. We compute the coboundary
\be
\delta S(T^{\mu\nu\rho},T^{\sigma}) = \partial_{\lambda}^{1}S(T^{\mu\nu\rho\lambda},T^{\sigma})
- \partial_{\lambda}^{2}S(T^{\mu\nu\rho},T^{\sigma\lambda})
\ee
and the result is:
\be
\delta S(T^{\mu\nu\rho},T^{\sigma}) = \partial_{\lambda}\partial_{\alpha}\delta(x_{1} - x_{2})~
S^{[\mu\nu\rho][\sigma\lambda]\alpha}(x_{2}) + \cdots
\ee
We choose
\be
S^{[\mu\nu\rho][\sigma\lambda]} \equiv {a\over 2}~
[ N(T^{\mu\nu\rho},T^{\sigma})^{\alpha\lambda} - N(T^{\mu\nu\rho},T^{\lambda})^{\sigma\alpha}]
\ee
From the last relation (\ref{renR}) we find out that 
\be
{\cal S}_{\sigma\alpha\beta} [ N(T^{\mu\nu\rho\sigma},T)^{\alpha\beta} 
+ N(T^{\mu\nu\rho},T^{\sigma})^{\alpha\beta} ] = 0 
\ee
and if we use (\ref{N7}) we are left with
\be
{\cal S}_{\sigma\alpha\beta} N(T^{\mu\nu\rho},T^{\sigma})^{\alpha\beta} = 0 
\qquad \Leftrightarrow \qquad
N(T^{\mu\nu\rho},T^{\sigma})^{\alpha\beta} + N(T^{\mu\nu\rho},T^{\alpha})^{\sigma\beta}
+ N(T^{\mu\nu\rho},T^{\beta})^{\sigma\alpha} = 0.
\ee
It follows that:
\be
\delta S(T^{\mu\nu\rho},T^{\sigma}) = {3a \over 4}~\partial_{\lambda}\partial_{\alpha}\delta(x_{1} - x_{2})~
N(T^{\mu\nu\rho},T^{\sigma})^{\lambda\alpha}(x_{2}) + \cdots
\ee
If we choose 
$
a = {4\over 3}
$
it follows that 
$
R(T^{\mu\nu\rho},T^{\sigma})
$
is cohomologous with a cocyle having
\be
N(T^{\mu\nu\rho},T^{\sigma})^{\alpha\beta} = 0.
\label{N8}
\ee

(ix) We are left with the last step, namely the consideration of
$
N(T^{\mu\nu},T^{\rho\sigma})^{\alpha\beta}. 
$
To prove that one can fix this expression to zero, is not as easy as in the previous steps. We need an
ansatz for this expresion. The most general quadri-linear expresion of ghost number $4$ and with the required symmetry properties is:
\bea
N(T^{\mu\nu},T^{\rho\sigma})^{\alpha\beta} = a_{1}~\eta^{\alpha\beta}~u^{\mu}u^{\nu}u^{\rho}u^{\sigma}
\nonumber\\
+ a_{2}~{\cal S}_{\alpha\beta}~( \eta^{\mu\alpha}~u^{\nu}u^{\rho}u^{\sigma}u^{\beta}
- \eta^{\nu\alpha}~u^{\mu}u^{\rho}u^{\sigma}u^{\beta}
+ \eta^{\rho\alpha}~u^{\mu}u^{\nu}u^{\sigma}u^{\beta}
- \eta^{\sigma\alpha}~u^{\mu}u^{\nu}u^{\rho}u^{\beta})
\label{N9}
\eea
From the last relation (\ref{renR}) we find out that 
\be
{\cal S}_{\rho\alpha\beta} [ N(T^{\mu\nu\rho},T^{\sigma})^{\alpha\beta} 
- N(T^{\mu\nu},T^{\sigma\rho})^{\alpha\beta} ] = 0 
\ee
and if use (\ref{N8}) we are left with
\be
{\cal S}_{\rho\alpha\beta} N(T^{\mu\nu},T^{\rho\sigma})^{\alpha\beta} = 0. 
\ee
If we introduce here (\ref{N9}) we find out
$
a_{1} = a_{2}
$
so the generic form becomes:
\bea
N(T^{\mu\nu},T^{\rho\sigma})^{\alpha\beta} = 
a~{\cal S}_{\alpha\beta}~(\eta^{\alpha\beta}~u^{\mu}u^{\nu}u^{\rho}u^{\sigma}
\nonumber\\
+ \eta^{\mu\alpha}~u^{\nu}u^{\rho}u^{\sigma}u^{\beta}
- \eta^{\nu\alpha}~u^{\mu}u^{\rho}u^{\sigma}u^{\beta}
+ \eta^{\rho\alpha}~u^{\mu}u^{\nu}u^{\sigma}u^{\beta}
- \eta^{\sigma\alpha}~u^{\mu}u^{\nu}u^{\rho}u^{\beta})
\label{N9a}
\eea
We try to compensate it using a cochain of the form
\be
S(T^{\mu\nu\lambda},T^{\rho\sigma}) = 
\partial_{\alpha}\delta(x_{1} - x_{2})~S^{[\mu\nu\lambda][\rho\sigma]\alpha}(x_{2})
\ee
and the generic form of $S$ is
\bea
S^{[\mu\nu\lambda][\rho\sigma]\alpha}
= b_{1}~(\eta^{\mu\alpha}~u^{\nu}u^{\rho}u^{\sigma}u^{\lambda}
+ \eta^{\nu\alpha}~u^{\lambda}u^{\rho}u^{\sigma}u^{\mu}
+ \eta^{\lambda\alpha}~u^{\mu}u^{\rho}u^{\sigma}u^{\nu})
\nonumber\\
+ b_{2}~(\eta^{\rho\alpha}~u^{\mu}u^{\nu}u^{\sigma}u^{\lambda}
- \eta^{\sigma\alpha}~u^{\mu}u^{\nu}u^{\rho}u^{\lambda})
\nonumber\\
+ b_{3}~[ (\eta^{\mu\rho}~u^{\nu}u^{\lambda}u^{\sigma}u^{\alpha}
+ \eta^{\nu\rho}~u^{\lambda}u^{\mu}u^{\sigma}u^{\alpha}
+ \eta^{\lambda\rho}~u^{\mu}u^{\nu}u^{\sigma}u^{\alpha})
- (\rho \leftrightarrow \sigma) ]
\eea
We want to preserve (\ref{N8}) so we must have
\be
{\cal S}_{\alpha\beta} S^{[\mu\nu\rho][\sigma\alpha]\beta} = 0. 
\ee
If we insert the previous expresion we get 
$
b_{1} = b_{3} \equiv b, \quad b_{2} = 0.
$
Now we compute the coboundary
\be
\delta S(T^{\mu\nu},T^{\rho\sigma}) = \partial_{\lambda}^{1}S(T^{\mu\nu\lambda},T^{\rho\sigma})
+ \partial_{\lambda}^{2}S(T^{\mu\nu},T^{\rho\sigma\lambda})
\ee
After some standard work we get 
\be
\delta S(T^{\mu\nu},T^{\rho\sigma}) = 
2~\partial_{\alpha}\partial_{\beta}\delta(x_{1} - x_{2})~
S^{[\mu\nu\alpha][\rho\sigma]\beta}(x_{2}) + \cdots
\ee
Using the previous expresion for 
$
S^{[\mu\nu\lambda][\rho\sigma]\alpha}
$
with 
$
b = {a \over 2}
$
we get 
\be
\delta S(T^{\mu\nu},T^{\rho\sigma}) = 
\partial_{\alpha}\partial_{\beta}\delta(x_{1} - x_{2})~N(T^{\mu\nu},T^{\rho\sigma})^{\alpha\beta}(x_{2})
 + \cdots
\ee
It follows that 
$
R(T^{\mu\nu},T^{\rho\sigma})
$
is cohomologous with a cocyle having
\be
N(T^{\mu\nu},T^{\rho\sigma})^{\alpha\beta} = 0.
\label{N9b}
\ee
This finishes the proof.
$\qed$
\begin{rem}
From the previous proof it follows that we do not need all relations (\ref{renR}) to obtain the result, but only the last relation
of (\ref{renR}).
\end{rem}

As a consequence, we have:
\begin{thm}
Suppose the cochain $R$ - see (\ref{R-IJ}) verifies
\be
N(T^{I},T^{J})^{\mu\nu} = 0
\ee
i.e.
\bea
R(T^{I}(x_{1}),T^{J}(x_{2})) = \delta(x_{1} - x_{2})~N(T^{I},T^{J})(x_{2})
+ \partial_{\mu}\delta(x_{1} - x_{2})~N(T^{I},T^{J})^{\mu}(x_{2})
\label{R-IJ-a}
\eea
Then the symmetry properties (\ref{symR}) are equivalent to
\bea
N(T^{I},T^{J}) = (-1)^{|I||J|}~[ N(T^{J},T^{I}) + d_{\mu}N(T^{J},T^{I})^{\mu} ]
\nonumber\\
N(T^{I},T^{J})^{\mu} = - (-1)^{|I||J|}~N(T^{J},T^{I})^{\mu}
\label{symR-a}
\eea

The gauge invariance condition
\be
sR(T^{I}(x_{1}),T^{J}(x_{2})) = 0
\label{sR-IJ-a}
\ee
is equivalent to the following system of equations:
\bea
d_{Q}N(T^{I},T^{J}) = i~(-1)^{|I|}~d_{\mu}N(T^{I},T^{J\mu})
\nonumber\\
d_{Q}N(T^{I},T^{J})^{\mu} = i~[ N(T^{I\mu},T^{J}) - (-1)^{|I|}~[ N(T^{I},T^{J\mu}) - d_{\nu}N(T^{I},T^{J\nu})^{\mu} ]
\nonumber\\
{\cal S}_{\rho\sigma} [ N(T^{I\rho},T^{J})^{\sigma} - (-1)^{|I|}~N(T^{I},T^{J\rho})^{\sigma} ] = 0.
\label{renR-a}
\eea
\end{thm}

The next step is:
\begin{thm}
Suppose we have a cocyle of the type (\ref{R-IJ-a}). Then $R$ is cohomologous to a cocyle verifying
\be
N(T^{I},T^{J})^{\mu} = 0.
\ee
\label{NIJ3}
\end{thm}
{\bf Proof:} (i) We have from the last equation (\ref{renR-a}) for 
$
I = J = \emptyset:
$
\be
N(T,T)^{\mu} = - N(T,T)^{\mu}
\ee
i.e.
\be
N(T,T)^{\mu} = 0.
\label{N10}
\ee

(ii) Let us define the cochain
\bea
S(T^{\mu\nu},T) = \delta(x_{1} - x_{2})~S^{[\mu\nu]}(x_{2})
\nonumber\\
S(T^{\mu},T^{\nu}) = 0
\eea
and compute the coboundary
\be
\delta S(T^{\mu},T) = \partial_{\nu}^{1}S(T^{\mu\nu},T) - \partial_{\nu}^{2}S(T^{\mu},T^{\nu}).
\ee
We obtain:
\be
\delta S(T^{\mu},T) = \partial_{\nu}\delta(x_{1} - x_{2})~S^{[\mu\nu]}(x_{2}).
\ee
We have from the last equation (\ref{renR-a}):
\be
{\cal S}_{\rho\sigma} [ N(T^{\rho},T)^{\sigma} - N(T,T^{\rho})^{\sigma} ] = 0
\ee
and from the last relation (\ref{symR-a})
\be
N(T^{\mu},T)^{\nu} = - N(T,T^{\mu})^{\nu}
\ee
so we have in fact
\be
{\cal S}_{\rho\sigma}  N(T^{\rho},T)^{\sigma} = 0 \qquad \Leftrightarrow \qquad
N(T^{\rho},T)^{\sigma} = - N(T^{\sigma},T)^{\rho}
\ee
It follows that the choice
\be
S^{[\mu\nu]} = N(T^{\mu},T)^{\nu}
\ee
is consistent. It leads to
\be
\delta S(T^{\mu},T) = \partial_{\nu}\delta(x_{1} - x_{2})~N(T^{\mu},T)^{\nu}(x_{2})
\ee
and it follows that 
$
R(T^{\mu},T)
$
is cohomologous with a cocyle having
\be
N(T^{\mu},T)^{\nu} = 0.
\label{N11}
\ee

(iii) We define the cochain
\be
S(T^{\mu\nu},T^{\rho}) = \delta(x_{1} - x_{2})~S^{[\mu\nu][\rho]}(x_{2})
\ee
and compute the coboundary
\be
\delta S(T^{\mu},T^{\nu}) = \partial_{\rho}^{1}S(T^{\mu\rho},T^{\nu}) - \partial_{\rho}^{2}S(T^{\mu},T^{\nu\rho}).
\ee
We obtain:
\be
\delta S(T^{\mu},T^{\nu}) = \partial_{\rho}\delta(x_{1} - x_{2})~(S^{[\mu\rho][\nu]} + S^{[\nu\rho][\mu]})(x_{2})
+ \cdots
\ee
where by $\cdots$ we mean an expresion 
$
\sim \delta(x_{1} - x_{2}).
$
We take
\be
S^{[\mu\nu][\rho]} = {1\over 2}~[ N(T^{\mu},T^{\rho})^{\nu} - N(T^{\nu},T^{\rho})^{\mu}
+ N(T^{\mu\nu},T)^{\rho} ]
\ee
such that the anti-symmetry property in 
$
\mu \leftrightarrow \nu
$
is true. From the last relation (\ref{symR-a}) we have
\be
N(T^{\mu},T^{\nu})^{\rho} = N(T^{\nu},T^{\mu})^{\rho}
\ee
and from the last relation (\ref{renR-a}) we have:
\be
{\cal S}_{\rho\sigma} [ N(T^{\mu\rho},T)^{\sigma} - N(T^{\mu},T^{\rho})^{\sigma} ] = 0.
\ee
If we use these last two relations we obtain
\be
S^{[\mu\rho][\nu]} + S^{[\nu\rho][\mu]} = N(T^{\mu},T^{\nu})^{\rho}
\ee
so 
\be
\delta S(T^{\mu},T^{\nu}) = \partial_{\rho}\delta(x_{1} - x_{2})~N(T^{\mu},T^{\nu})^{\rho}(x_{2}) + \cdots
\ee
It follows that 
$
R(T^{\mu},T^{\nu})
$
is cohomologous with a cocyle having
\be
N(T^{\mu},T^{\nu})^{\rho} = 0.
\label{N12}
\ee

(iv) We define the cochain
\bea
S(T^{\mu\nu\rho},T) = \delta(x_{1} - x_{2})~S^{[\mu\nu\rho]}(x_{2})
\nonumber\\
S(T^{\mu\nu},T^{\rho}) = 0
\eea
such that we preserve (\ref{N12}). Next, we compute the coboundary
\be
\delta S(T^{\mu\nu},T) = \partial_{\rho}^{1}S(T^{\mu\nu\rho},T) + \partial_{\rho}^{2}S(T^{\mu\nu},T^{\rho}).
\ee
and get:
\be
\delta S(T^{\mu\nu},T) = \partial_{\rho}\delta(x_{1} - x_{2})~S^{[\mu\nu\rho]}(x_{2}).
\ee
We take
\be
S^{[\mu\nu\rho]} = {1\over 3}~[ N(T^{\mu\nu},T)^{\rho} + N(T^{\nu\rho},T)^{\mu} + N(T^{\rho\mu},T)^{\nu} ]
\ee
such that we have complete anti-symmetry in the indexes 
$
\mu, \nu, \rho.
$
We use now the last formula (\ref{renR-a}) with the choice (\ref{N12}) and have:
\be
{\cal S}_{\rho\sigma} N(T^{\mu\rho},T)^{\sigma} = 0.
\ee
If we use the preceding formula we obtain that all three terms in the expresion of 
$
S^{[\mu\nu\rho]} 
$
are equal, so we have
\be
S^{[\mu\nu\rho]} = N(T^{\mu\nu},T)^{\rho}
\ee
and from here
\be
\delta S(T^{\mu},T^{\nu}) = \partial_{\rho}\delta(x_{1} - x_{2})~N(T^{\mu\nu},T)^{\rho}(x_{2}).
\ee
It follows that 
$
R(T^{\mu\nu},T)
$
is cohomologous with a cocyle having
\be
N(T^{\mu\nu},T)^{\rho} = 0.
\label{N13}
\ee

(v) We define the cochain
\be
S(T^{\mu\nu\rho},T^{\sigma}) = \delta(x_{1} - x_{2})~S^{[\mu\nu\rho][\sigma]}(x_{2})
\ee
and compute the coboundary
\be
\delta S(T^{\mu\nu\rho},T) = - \partial_{\sigma}^{2}S(T^{\mu\nu\rho},T^{\sigma}).
\ee
We get:
\be
\delta S(T^{\mu\nu\rho},T) = \partial_{\sigma}\delta(x_{1} - x_{2})~S^{[\mu\nu\rho][\sigma]}(x_{2}) + \cdots
\ee
so if we choose
\be
S^{[\mu\nu\rho][\sigma]} = N(T^{\mu\nu\rho},T)^{\sigma}
\ee
we have
\be
\delta S(T^{\mu\nu\rho},T) = \partial_{\sigma}\delta(x_{1} - x_{2})~N(T^{\mu\nu\rho},T)^{\sigma}(x_{2}) + \cdots
\ee
It follows that 
$
R(T^{\mu\nu\rho},T)
$
is cohomologous with a cocyle having
\be
N(T^{\mu\nu\rho},T)^{\sigma} = 0.
\label{N14}
\ee

(vi) We define the cochain
\bea
S(T^{\mu\nu},T^{\rho\sigma}) = \delta(x_{1} - x_{2})~S^{[\mu\nu][\rho\sigma]}(x_{2})
\nonumber\\
S(T^{\mu\nu\rho},T^{\sigma}) = 0
\eea
such that we preserve (\ref{N14}). We compute the coboundary
\be
\delta S(T^{\mu\nu},T^{\rho}) = \partial_{\sigma}^{1}S(T^{\mu\nu\sigma},T^{\rho}) 
+ \partial_{\sigma}^{2}S(T^{\mu\nu},T^{\rho\sigma})
\ee
and we get:
\be
\delta S(T^{\mu\nu},T^{\rho}) = - \partial_{\sigma}\delta(x_{1} - x_{2})~S^{[\mu\nu][\rho\sigma]}(x_{2}) + \cdots
\ee
We take
\be
S^{[\mu\nu][\rho\sigma]} = N(T^{\mu\nu},T^{\rho})^{\sigma}.
\ee
From the last relation (\ref{symR-a}) we have:
\be
N(T^{\mu\nu},T^{\rho})^{\sigma} =  - N(T^{\rho},T^{\mu\nu})^{\sigma}
\ee
and from the last relation (\ref{renR-a}) we get 
\be
{\cal S}_{\rho\sigma} [ N(T^{\mu\nu\rho},T)^{\sigma} - N(T^{\mu\nu},T^{\rho})^{\sigma} ] = 0.
\ee
Taking into account (\ref{N14}) we are left with:
\be
{\cal S}_{\rho\sigma} N(T^{\mu\nu},T^{\rho})^{\sigma} = 0
\ee
and this ensures that the expresion 
$
S^{[\mu\nu][\rho\sigma]}
$
defined above is anti-symmetric to
$
\rho \leftrightarrow \sigma.
$
Also from the last relation (\ref{renR-a}) we get 
\be
{\cal S}_{\rho\sigma} [ N(T^{\mu\rho},T^{\nu})^{\sigma} + N(T^{\mu},T^{\nu\rho})^{\sigma} ] = 0.
\ee
Using this relation we obtain after some computations that 
\be
N(T^{\mu\nu},T^{\rho})^{\sigma} = N(T^{\rho\sigma},T^{\mu})^{\nu}
\ee
so the expresion 
$
S^{[\mu\nu][\rho\sigma]}
$
defined above is symmetric to
$
\mu \leftrightarrow \rho, \nu \leftrightarrow \sigma.
$
So the definition of 
$
S^{[\mu\nu][\rho\sigma]}
$
is consistent and we also have
\be
\delta S(T^{\mu\nu},T^{\rho}) = - \partial_{\sigma}\delta(x_{1} - x_{2})~N(T^{\mu\nu},T^{\rho})^{\sigma}(x_{2}) + \cdots
\ee
It follows that 
$
R(T^{\mu\nu},T^{\rho})
$
is cohomologous with a cocyle having
\be
N(T^{\mu\nu},T^{\rho})^{\sigma} = 0.
\label{N15}
\ee

(vii) We define the cochain
\be
S(T^{\mu\nu\rho\sigma},T^{\lambda}) = \delta(x_{1} - x_{2})~S^{[\mu\nu\rho\sigma][\lambda]}(x_{2})
\ee
and compute the coboundary
\be
\delta S(T^{\mu\nu\rho\sigma},T) = \partial_{\lambda}^{2}S(T^{\mu\nu\rho\sigma},T^{\lambda}).
\ee
We get:
\be
\delta S(T^{\mu\nu\rho\sigma},T) = - \partial_{\lambda}\delta(x_{1} - x_{2})~S^{[\mu\nu\rho\sigma][\lambda]}(x_{2}) 
+ \cdots
\ee
If we take
\be
S^{[\mu\nu\rho\sigma][\lambda]} = N(T^{\mu\nu\rho\sigma},T)^{\lambda}
\ee
we prove that 
$
R(T^{\mu\nu\rho\sigma},T)
$
is cohomologous with a cocyle having
\be
N(T^{\mu\nu\rho\sigma},T)^{\lambda} = 0.
\label{N16}
\ee

(viii) We define the cochain
\bea
S(T^{\mu\nu\rho},T^{\sigma\lambda}) = \delta(x_{1} - x_{2})~S^{[\mu\nu\rho][\sigma\lambda]}(x_{2})
\nonumber\\
S(T^{\mu\nu\rho\sigma},T^{\lambda}) = 0
\eea
such that we preserve (\ref{N16}). We compute the coboundary
\be
\delta S(T^{\mu\nu\rho},T^{\sigma}) = \partial_{\lambda}^{1}S(T^{\mu\nu\rho\lambda},T^{\sigma}) 
- \partial_{\lambda}^{2}S(T^{\mu\nu\rho},T^{\sigma\lambda})
\ee
and we get:
\be
\delta S(T^{\mu\nu\rho},T^{\sigma}) = \partial_{\lambda}\delta(x_{1} - x_{2})~S^{[\mu\nu\rho][\sigma\lambda]}(x_{2}) 
+ \cdots
\ee
We take
\be
S^{[\mu\nu\rho][\sigma\lambda]} = N(T^{\mu\nu\rho},T^{\sigma})^{\lambda}
\ee
and must prove the consistency of this definition. From the last relation (\ref{renR-a}) we have:
\be
{\cal S}_{\sigma\lambda} [ N(T^{\mu\nu\rho\sigma},T)^{\lambda} + N(T^{\mu\nu\rho},T^{\sigma})^{\lambda} ] = 0
\ee
and if we use (\ref{N16}) we are left with
\be
{\cal S}_{\sigma\lambda} N(T^{\mu\nu\rho},T^{\sigma})^{\lambda} = 0
\ee
and this gives the desired consistency of the definition of 
$
S^{[\mu\nu\rho][\sigma\lambda]}.
$
Because 
\be
\delta S(T^{\mu\nu\rho},T^{\sigma}) = \partial_{\lambda}\delta(x_{1} - x_{2})
N(T^{\mu\nu\rho},T^{\sigma})^{\lambda}(x_{2}) + \cdots
\ee
it follows that 
$
R(T^{\mu\nu\rho},T^{\sigma})
$
is cohomologous with a cocyle having
\be
N(T^{\mu\nu\rho},T^{\sigma})^{\lambda} = 0.
\label{N17}
\ee

(ix) We still have to analyse the expresion
$
N(T^{\mu\nu},T^{\rho\sigma})^{\lambda}
$
and as in theorem \ref{NIJ2} we need a general ansatz. Beside the properties of anti-symmetry in
$
\mu \leftrightarrow \nu
$
and
$
\rho \leftrightarrow \sigma
$
we must use the last relation (\ref{symR-a}) to get 
\be
N(T^{\mu\nu},T^{\rho\sigma})^{\lambda} = - N(T^{\rho\sigma},T^{\mu\nu})^{\lambda} 
\ee
so the general ansatz is:
\bea
N(T^{\mu\nu},T^{\rho\sigma})^{\lambda} =
a_{1} ( d^{\lambda}u^{\mu} u^{\nu} u^{\rho} u^{\sigma} - d^{\lambda}u^{\nu} u^{\mu} u^{\rho} u^{\sigma}
- d^{\lambda}u^{\rho} u^{\sigma} u^{\mu} u^{\nu} + d^{\lambda}u^{\sigma} u^{\rho} u^{\mu} u^{\nu})
\nonumber\\
+ a_{2} ( d^{\mu}u^{\lambda} u^{\nu} u^{\rho} u^{\sigma} - d^{\nu}u^{\lambda} u^{\mu} u^{\rho} u^{\sigma}
- d^{\rho}u^{\lambda} u^{\sigma} u^{\mu} u^{\nu} + d^{\sigma}u^{\lambda} u^{\rho} u^{\mu} u^{\nu})
\nonumber\\
+ a_{3} ( d^{\mu}u^{\nu} u^{\rho} u^{\sigma} u^{\lambda} - d^{\nu}u^{\mu} u^{\rho} u^{\sigma} u^{\lambda}
- d^{\rho}u^{\sigma} u^{\mu} u^{\nu} u^{\lambda} + d^{\sigma}u^{\rho} u^{\mu} u^{\nu} u^{\lambda})
\nonumber\\
+ a_{4} ( d^{\mu}u^{\rho} u^{\nu} u^{\sigma} u^{\lambda} - d^{\nu}u^{\rho} u^{\mu} u^{\sigma} u^{\lambda}
- d^{\mu}u^{\sigma} u^{\nu} u^{\rho} u^{\lambda} + d^{\nu}u^{\sigma} u^{\mu} u^{\rho} u^{\lambda}
\nonumber\\
- d^{\rho}u^{\mu} u^{\sigma} u^{\nu} u^{\lambda} + d^{\sigma}u^{\mu} u^{\rho} u^{\nu} u^{\lambda}
+ d^{\rho}u^{\nu} u^{\sigma} u^{\mu} u^{\lambda} - d^{\sigma}u^{\nu} u^{\rho} u^{\mu} u^{\lambda})
\nonumber\\
+ a_{5} ( \eta^{\mu\lambda} u^{\nu} u^{\rho} u^{\sigma} - \eta^{\nu\lambda} u^{\mu} u^{\rho} u^{\sigma}
- \eta^{\rho\lambda} u^{\sigma} u^{\mu} u^{\nu} +  \eta^{\sigma\lambda} u^{\rho} u^{\mu} u^{\nu}) d_{\alpha}u^{\alpha}
\nonumber\\
+ a_{6} ( \eta^{\mu\rho} u^{\nu} u^{\sigma} u^{\lambda} - \eta^{\nu\rho} u^{\mu} u^{\sigma} u^{\lambda}
- \eta^{\mu\sigma} u^{\nu} u^{\rho} u^{\lambda} +  \eta^{\nu\sigma} u^{\mu} u^{\rho} u^{\lambda}) d_{\alpha}u^{\alpha}.
\eea
Form the last relation (\ref{renR-a}) we also have
\be
{\cal S}_{\sigma\lambda} [ N(T^{\mu\nu\sigma},T^{\rho})^{\lambda} - N(T^{\mu\nu},T^{\rho\sigma})^{\lambda} ] = 0
\ee
and if we use (\ref{N17}) we are left with
\be
{\cal S}_{\sigma\lambda} N(T^{\mu\nu},T^{\rho\sigma})^{\lambda} = 0.
\ee
If we substitute the preceding ansatz we get after some computations that 
$
a_{j} = 0,~j = 1,\dots,6
$
so in fact we have 
\be
N(T^{\mu\nu},T^{\rho\sigma})^{\lambda} = 0
\label{N18}
\ee
and this finishes the proof.
$\qed$

As a consequence, we have:
\begin{thm}
Suppose the cochain $R$ - see (\ref{R-IJ}) verifies
\be
N(T^{I},T^{J})^{\mu\nu} = 0, \qquad N(T^{I},T^{J})^{\mu} = 0.
\ee
i.e.
\be
R(T^{I}(x_{1}),T^{J}(x_{2})) = \delta(x_{1} - x_{2})~N(T^{I},T^{J})(x_{2})
\label{R-IJ-b}
\ee
Then the symmetry properties (\ref{symR}) are equivalent to
\bea
N(T^{I},T^{J}) = (-1)^{|I||J|}~N(T^{J},T^{I})
\label{symR-b}
\eea
and gauge invariance condition
\be
sR(T^{I}(x_{1}),T^{J}(x_{2})) = 0
\label{sR-IJ-b}
\ee
is equivalent to the following system of equations:
\bea
d_{Q}N(T^{I},T^{J}) = i~(-1)^{|I|}~d_{\mu}N(T^{I},T^{J\mu})
\nonumber\\
N(T^{I\mu},T^{J}) = (-1)^{|I|}~N(T^{I},T^{J\mu}).
\label{renR-b}
\eea
\end{thm}

We have proved in \cite{wick+hopf} that in this case we have
\begin{thm}
These finite renormalizations do not produce anomalies iff there exists expresions
$
N^{I}
$
such that
\be
N(T^{I},T^{J}) = N^{JI}
\label{NI}
\ee
and
\be
d_{Q}N^{I} = i~d_{\mu}N^{I\mu}.
\label{sN}
\ee
\label{R-finite}
\end{thm}

\section{The Renormalizablity of Quantum Gravity}

It is a comon lore in the literature that the gravity is not renormalizable i.e. the arbitrariness of the 
chronological products increases with the order of the perturbation theory. We prove here that this 
assertion is wrong, at least in the second order of the perturbation theory. As we have proved in the 
preceding Section, the arbitrariness is severely constricted by the gauge invariance condition. In fact,
gauge invariance is a cocyle condition, and we have proved that such a cocyle is cohomologous with one of the 
form
\be
R(T^{I}(x_{1}),T^{J}(x_{2})) = \delta(x_{1} - x_{2})~N^{JI}(x_{2})
\label{R-IJ-c}
\ee
and we must have:
\be
\omega(N) = 6, \qquad gh(N^{I}) = |I|
\ee
and the gauge invariance condition
\be
d_{Q}N^{I} = i~d_{\mu}N^{I\mu}.
\label{sN-a}
\ee
Here we prove that such a cocyle is trivial {\bf in the quadri-linear sector}  i.e. it is a coboundary
$
N = d_{Q}B + d_{\mu}B^{\mu}.
$
First we must give a list of all possible terms. We start with $N$ which is of ghost number $0$. 
We need the list of all terms up to a total derivatives. We have three sectors:

(a) Terms quadri-linear in 
$
h_{\mu\nu}:
$
it is sufficient to consider terms of the type
$
h h dh dh;
$
there are $43$ terms of this type. Terms of the type
$
h h h ddh
$
can be written, up to a total derivative, as a linear combination of these $43$ terms. 

\bea
N_{1} = h^{2}~d_{\rho}h d^{\rho}h
\nonumber\\
\nonumber\\
N_{2,1} = h~d_{\rho}h d_{\sigma}h h^{\rho\sigma} \qquad 
N_{2,2} = h^{2} ~d_{\rho}h d_{\sigma}h^{\rho\sigma}
\nonumber\\
\nonumber\\
N_{3,1} = d_{\rho}h d^{\rho}h h^{\alpha\beta} h_{\alpha\beta} \qquad
N_{3,2} = d_{\rho}h d_{\sigma}h h^{\lambda\rho} {h_{\lambda}}^{\sigma}
\nonumber\\
N_{3,3} = h d_{\rho}h h_{\alpha\beta} d^{\rho}h^{\alpha\beta} \qquad
N_{3,4} = h d_{\rho}h h_{\alpha\beta} d^{\beta}h^{\alpha\rho}
\nonumber\\
N_{3,5} = h d_{\rho}h h^{\rho\sigma} d^{\lambda}h_{\sigma\lambda} \qquad
N_{3,6} = h^{2} d_{\rho}h_{\alpha\beta} d^{\rho}h^{\alpha\beta}
\nonumber\\
N_{3,7} = h^{2} d_{\beta}h_{\rho\alpha} d^{\alpha}h^{\rho\beta} \qquad
N_{3,8} = h^{2} d^{\alpha}h_{\alpha\mu} d_{\beta}h^{\mu\beta}
\nonumber\\
\nonumber\\
N_{4,1} = d_{\rho}h h^{\rho\sigma} h^{\alpha\beta} d_{\sigma}h_{\alpha\beta} \qquad
N_{4,2} = d_{\rho}h h^{\rho\sigma} h^{\alpha\beta} d_{\beta}h_{\alpha\sigma}
\nonumber\\
N_{4,3} = d_{\rho}h h^{\rho\sigma} h_{\sigma\lambda} d_{\alpha}h^{\alpha\lambda} \qquad
N_{4,4} = d_{\rho}h h_{\lambda\alpha} {h^{\lambda}}_{\beta} d^{\rho}h^{\alpha\beta}
\nonumber\\
N_{4,5} = d_{\rho}h h_{\lambda\alpha} {h^{\lambda}}_{\beta} d^{\rho}h^{\alpha\beta} \qquad
N_{4,6} = h h^{\alpha\beta} d_{\alpha}h_{\rho\sigma} d_{\beta}h^{\rho\sigma}
\nonumber\\
N_{4,7} = h h^{\alpha\beta} d_{\sigma}h_{\rho\alpha} d_{\beta}h^{\rho\sigma} \qquad
N_{4,8} = h h^{\alpha\beta} d_{\sigma}h_{\alpha\rho} d^{\sigma}{h_{\beta}}^{\rho}
\nonumber\\
N_{4,9} = h h^{\alpha\beta} d_{\sigma}h_{\rho\alpha} d^{\rho}{h_{\beta}}^{\sigma} \qquad
N_{4,10} = h h^{\alpha\beta} d^{\mu}h_{\alpha\mu} d^{\nu}h_{\beta\nu}
\nonumber\\
N_{4,11} = h h^{\alpha\beta} d_{\mu}h_{\alpha\beta} d_{\nu}h^{\mu\nu} \qquad
N_{4,12} = h h^{\alpha\beta} d_{\alpha}h_{\beta\mu} d_{\nu}h^{\mu\nu}
\nonumber\\
N_{4,13} = d_{\mu}h d_{\nu}h^{\mu\nu} h^{\alpha\beta} h_{\alpha\beta}
\nonumber\\
\nonumber\\
N_{5,1} = h_{\lambda\rho} {h^{\lambda}}_{\sigma} d^{\rho}h^{\alpha\beta} d^{\sigma}h_{\alpha\beta} \qquad
N_{5,2} = h_{\lambda\rho} {h^{\lambda}}_{\sigma} d^{\beta}h^{\rho\alpha} d_{\beta}{h^{\sigma}}_{\alpha}
\nonumber\\
N_{5,3} = h_{\lambda\rho} {h^{\lambda}}_{\sigma} d^{\beta}h^{\rho\alpha} d_{\alpha}{h^{\sigma}}_{\beta} \qquad
N_{5,4} = h_{\lambda\rho} {h^{\lambda}}_{\sigma} d^{\beta}h^{\rho\alpha} d^{\sigma}h_{\alpha\beta}
\nonumber\\
N_{5,5} = h^{\rho\sigma} h^{\alpha\beta} d_{\rho}h_{\lambda\alpha} d_{\sigma}{h^{\lambda}}_{\beta} \qquad
N_{5,6} = h^{\rho\sigma} h^{\alpha\beta} d_{\alpha}h_{\lambda\rho} d^{\sigma}{h^{\lambda}}_{\beta}
\nonumber\\
N_{5,7} = h^{\rho\sigma} h^{\alpha\beta} d_{\sigma}h_{\lambda\rho} d_{\beta}{h^{\lambda}}_{\alpha} \qquad
N_{5,8} = h^{\rho\sigma} h^{\alpha\beta} d_{\lambda}h_{\alpha\rho} d^{\lambda}h_{\beta\sigma}
\nonumber\\
N_{5,9} = h^{\rho\sigma} h^{\alpha\beta} d_{\lambda}h_{\rho\sigma} d^{\lambda}h_{\alpha\beta} \qquad
N_{5,10} = h^{\rho\sigma} h^{\alpha\beta} d_{\rho}h_{\lambda\alpha} d^{\lambda}h_{\beta\sigma}
\nonumber\\
N_{5,11} = h^{\rho\sigma} h^{\alpha\beta} d_{\sigma}h_{\lambda\rho} d^{\lambda}h_{\alpha\beta} \qquad
N_{5,12} = h^{\rho\sigma} h_{\rho\sigma} d^{\lambda}h^{\alpha\beta} d_{\lambda}h_{\alpha\beta} 
\nonumber\\
N_{5,13} = h^{\rho\sigma} h_{\rho\sigma} d^{\lambda}h^{\alpha\beta} d_{\beta}h_{\lambda\alpha} \qquad
N_{5,14} = h_{\lambda\rho} {h^{\lambda}}_{\sigma} d_{\mu}h^{\mu\rho} d_{\nu}h^{\nu\sigma}
\nonumber\\
N_{5,15} = h_{\lambda\rho} {h^{\lambda}}_{\sigma} d^{\mu}h^{\rho\sigma} d^{\nu}h_{\mu\nu} \qquad
N_{5,16} = h_{\lambda\rho} {h^{\lambda}}_{\sigma} d^{\sigma}h^{\mu\rho} d^{\nu}h_{\mu\nu}
\nonumber\\
N_{5,17} = h^{\rho\sigma} h^{\alpha\beta} d^{\lambda}h_{\rho\lambda} d_{\sigma}h_{\alpha\beta} \qquad
N_{5,18} = h^{\rho\sigma} h^{\alpha\beta} d^{\lambda}h_{\rho\lambda} d_{\alpha}h_{\sigma\beta}
\nonumber\\
N_{5,19} = h^{\rho\sigma} h_{\rho\sigma} d^{\alpha}h_{\mu\alpha} d_{\beta}h^{\mu\beta} 
\label{N-1-5}
\eea

(b) Terms of the type
$
u \tilde{u} h h:
$
we have term of the type
$
u \tilde{u} dh dh,~u \tilde{u} h ddh, u d\tilde{u} h dh,~u dd\tilde{u} h h;
$
there are 83 terms of these types. The terms with derivatives on $u$ can be written, up to a total derivative,
as linear combination of these 83 terms.

\bea
N_{6,1} = u^{\mu} \tilde{u}_{\mu} d_{\rho}h d^{\rho}h \qquad
N_{6,2} = u^{\mu} \tilde{u}_{\mu} d_{\rho}h_{\alpha\beta} d^{\rho}h^{\alpha\beta}
\nonumber\\
N_{6,3} = u^{\mu} \tilde{u}_{\mu} d_{\rho}h_{\alpha\beta} d^{\alpha}h^{\rho\beta} \qquad
N_{6,4} = u^{\mu} \tilde{u}_{\mu} d_{\rho}h^{\rho\sigma} d^{\lambda}h_{\lambda\sigma}
\nonumber\\
N_{6,5} = u^{\mu} \tilde{u}_{\mu} d_{\rho}h^{\rho\sigma} d_{\sigma}h \qquad
N_{6,6} = u_{\mu} \tilde{u}_{\nu} d^{\mu}h d^{\nu}h
\nonumber\\
N_{6,7} = u_{\mu} \tilde{u}_{\nu} d^{\mu}h^{\alpha\beta} d^{\nu}h_{\alpha\beta} \qquad
N_{6,8} = u_{\mu} \tilde{u}_{\nu} d^{\mu}h^{\nu\rho} d_{\rho}h
\nonumber\\
N_{6,9} = u_{\mu} \tilde{u}_{\nu} d^{\mu}h^{\nu\rho} d^{\sigma}h_{\rho\sigma} \qquad
N_{6,10} = u_{\mu} \tilde{u}_{\nu} d^{\mu}h d_{\rho}h^{\nu\rho}
\nonumber\\
N_{6,11} = u_{\mu} \tilde{u}_{\nu} d^{\mu}h_{\alpha\beta} d^{\alpha}h^{\nu\beta} \qquad
N_{6,12} = u_{\mu} \tilde{u}_{\nu} d^{\nu}h^{\mu\rho} d_{\rho}h
\nonumber\\
N_{6,13} = u_{\mu} \tilde{u}_{\nu} d^{\nu}h^{\mu\rho} d^{\sigma}h_{\rho\sigma} \qquad
N_{6,14} = u_{\mu} \tilde{u}_{\nu} d^{\nu}h d_{\rho}h^{\mu\rho}
\nonumber\\
N_{6,15} = u_{\mu} \tilde{u}_{\nu} d^{\nu}h_{\alpha\beta} d^{\alpha}h^{\mu\beta} \qquad
N_{6,16} = u_{\mu} \tilde{u}_{\nu} d^{\rho}h^{\mu\nu} d_{\rho}h
\nonumber\\
N_{6,17} = u_{\mu} \tilde{u}_{\nu} d^{\rho}h^{\mu\nu} d^{\sigma}h_{\rho\sigma} \qquad
N_{6,18} = u_{\mu} \tilde{u}_{\nu} d^{\alpha}h^{\mu\beta} d_{\alpha}{h^{\nu}}_{\beta}
\nonumber\\
N_{6,19} = u_{\mu} \tilde{u}_{\nu} d^{\alpha}h^{\mu\beta} d_{\beta}{h^{\nu}}_{\alpha} \qquad
N_{6,20} = u_{\mu} \tilde{u}_{\nu} d_{\rho}h^{\mu\rho} d_{\sigma}h^{\nu\sigma}
\nonumber\\
\nonumber\\
N_{7,1} = u^{\mu} \tilde{u}_{\mu} h d_{\rho}d_{\sigma}h^{\rho\sigma} \qquad
N_{7,2} = u^{\mu} \tilde{u}_{\mu} h_{\alpha\beta} d^{\alpha}d^{\beta}h
\nonumber\\
N_{7,3} = u^{\mu} \tilde{u}_{\mu} h_{\alpha\beta} d^{\alpha}d_{\lambda}h^{\beta\lambda} \qquad
N_{7,4} = u_{\mu} \tilde{u}_{\nu} h^{\mu\nu} d_{\rho}d_{\sigma}h^{\rho\sigma}
\nonumber\\
N_{7,5} = u_{\mu} \tilde{u}_{\nu} h^{\mu\rho} d^{\nu}d_{\rho}h \qquad
N_{7,6} = u_{\mu} \tilde{u}_{\nu} h^{\mu\rho} d_{\rho}d_{\sigma}h^{\nu\sigma}
\nonumber\\
N_{7,7} = u_{\mu} \tilde{u}_{\nu} h^{\nu\rho} d^{\mu}d_{\rho}h \qquad
N_{7,8} = u_{\mu} \tilde{u}_{\nu} h^{\mu\rho} d_{\rho}d_{\sigma}h^{\mu\sigma}
\nonumber\\
N_{7,9} = u_{\mu} \tilde{u}_{\nu} h d^{\mu}d^{\nu}h \qquad
N_{7,10} = u_{\mu} \tilde{u}_{\nu} h_{\alpha\beta} d^{\mu}d^{\nu}h^{\alpha\beta}
\nonumber\\
N_{7,11} = u_{\mu} \tilde{u}_{\nu} h d^{\mu}d_{\rho}h^{\nu\rho} \qquad
N_{7,12} = u_{\mu} \tilde{u}_{\nu} h d^{\nu}d_{\rho}h^{\mu\rho}
\nonumber\\
N_{7,13} = u_{\mu} \tilde{u}_{\nu} h_{\alpha\beta} d^{\mu}d^{\alpha}h^{\nu\beta} \qquad
N_{7,14} = u_{\mu} \tilde{u}_{\nu} h_{\alpha\beta} d^{\nu}d^{\alpha}h^{\mu\beta}
\nonumber\\
N_{7,15} = u_{\mu} \tilde{u}_{\nu} h_{\rho\sigma} d^{\rho}d^{\sigma}h^{\mu\nu} \qquad
N_{7,16} = u_{\mu} \tilde{u}_{\nu} h^{\mu\rho} d^{\nu}d^{\sigma}h_{\rho\sigma}
\nonumber\\
N_{7,17} = u_{\mu} \tilde{u}_{\nu} h^{\nu\rho} d^{\mu}d^{\sigma}h_{\rho\sigma} 
\nonumber\\
\nonumber\\
N_{8,1} = u_{\mu} d^{\mu}\tilde{u}^{\nu} h_{\nu\rho} d^{\rho}h \qquad
N_{8,2} = u_{\mu} d^{\mu}\tilde{u}^{\nu} h_{\nu\rho} d_{\sigma}h^{\rho\sigma}
\nonumber\\
N_{8,3} = u_{\mu} d^{\mu}\tilde{u}^{\nu} h d_{\nu}h \qquad
N_{8,4} = u_{\mu} d^{\mu}\tilde{u}^{\nu} h^{\alpha\beta} d_{\nu}h_{\alpha\beta}
\nonumber\\
N_{8,5} = u_{\mu} d^{\mu}\tilde{u}^{\nu} h d^{\rho}h_{\nu\rho} \qquad
N_{8,6} = u_{\mu} d^{\mu}\tilde{u}^{\nu} h^{\alpha\beta} d_{\alpha}h_{\nu\beta}
\nonumber\\
N_{8,7} = u_{\mu} d^{\nu}\tilde{u}^{\mu} h_{\nu\rho} d^{\rho}h \qquad
N_{8,8} = u_{\mu} d^{\nu}\tilde{u}^{\mu} h_{\nu\rho} d_{\sigma}h^{\rho\sigma}
\nonumber\\
N_{8,9} = u_{\mu} d^{\nu}\tilde{u}^{\mu} h d_{\nu}h \qquad
N_{8,10} = u_{\mu} d^{\nu}\tilde{u}^{\mu} h^{\alpha\beta} d_{\nu}h_{\alpha\beta}
\nonumber\\
N_{8,11} = u_{\mu} d^{\nu}\tilde{u}^{\mu} h d^{\rho}h_{\nu\rho} \qquad
N_{8,12} = u_{\mu} d^{\nu}\tilde{u}^{\mu} h^{\alpha\beta} d_{\alpha}h_{\nu\beta}
\nonumber\\
N_{8,13} = u_{\mu} d\cdot \tilde{u} h^{\mu\nu} d_{\nu}h \qquad
N_{8,14} = u_{\mu} d\cdot \tilde{u} h^{\mu\nu} d^{\rho}h_{\nu\rho}
\nonumber\\
N_{8,15} = u_{\mu} d_{\alpha}\tilde{u}_{\beta} h^{\mu\alpha} d^{\beta}h \qquad
N_{8,16} = u_{\mu} d_{\alpha}\tilde{u}_{\beta} h^{\mu\alpha} d_{\rho}h^{\beta\rho}
\nonumber\\
N_{8,17} = u_{\mu} d_{\alpha}\tilde{u}_{\beta} h^{\mu\beta} d^{\alpha}h \qquad
N_{8,18} = u_{\mu} d_{\alpha}\tilde{u}_{\beta} h^{\mu\beta} d_{\rho}h^{\alpha\rho}
\nonumber\\
N_{8,19} = u_{\mu} d\cdot \tilde{u} h d^{\mu}h \qquad
N_{8,20} = u_{\mu} d\cdot \tilde{u} h_{\alpha\beta} d^{\mu}h^{\alpha\beta}
\nonumber\\
N_{8,21} = u_{\mu} d_{\alpha}\tilde{u}_{\beta} h^{\alpha\beta} d^{\mu}h \qquad
N_{8,22} = u_{\mu} d_{\alpha}\tilde{u}_{\beta} h d^{\mu}h^{\alpha\beta}
\nonumber\\
N_{8,23} = u_{\mu} d_{\alpha}\tilde{u}_{\beta} {h^{\alpha}}_{\lambda} d^{\mu}h^{\beta\lambda} \qquad
N_{8,24} = u_{\mu} d_{\alpha}\tilde{u}_{\beta} {h^{\beta}}_{\lambda} d^{\mu}h^{\alpha\lambda}
\nonumber\\
N_{8,25} = u_{\mu} d\cdot \tilde{u} h_{\alpha\beta} d^{\alpha}h^{\beta\mu} \qquad
N_{8,26} = u_{\mu} d\cdot \tilde{u} h d_{\nu}h^{\mu\nu}
\nonumber\\
N_{8,27} = u_{\mu} d_{\alpha}\tilde{u}_{\beta} h d^{\alpha}h^{\mu\beta} \qquad
N_{8,28} = u_{\mu} d_{\alpha}\tilde{u}_{\beta} h d^{\beta}h^{\mu\alpha}
\nonumber\\
N_{8,29} = u_{\mu} d_{\alpha}\tilde{u}_{\beta} h^{\alpha\beta} d_{\nu}h^{\mu\nu} \qquad
N_{8,30} = u_{\mu} d_{\alpha}\tilde{u}_{\beta} h^{\alpha\lambda} d_{\lambda}h^{\mu\beta}
\nonumber\\
N_{8,31} = u_{\mu} d_{\alpha}\tilde{u}_{\beta} h^{\beta\lambda} d_{\lambda}h^{\mu\alpha} \qquad
N_{8,32} = u_{\mu} d_{\alpha}\tilde{u}_{\beta} h^{\alpha\lambda} d^{\beta}{h^{\mu}}_{\lambda}
\nonumber\\
N_{8,33} = u_{\mu} d_{\alpha}\tilde{u}_{\beta} h^{\beta\lambda} d^{\alpha}{h^{\mu}}_{\lambda} \qquad
N_{8,34} = u_{\mu} d_{\alpha}\tilde{u}_{\beta} h^{\mu\nu} d_{\nu}h^{\alpha\beta}
\nonumber\\
N_{8,35} = u_{\mu} d_{\alpha}\tilde{u}_{\beta} h^{\mu\nu} d^{\alpha}{h_{\nu}}^{\beta} \qquad
N_{8,36} = u_{\mu} d_{\alpha}\tilde{u}_{\beta} h^{\mu\nu} d^{\beta}{h_{\nu}}^{\alpha}
\nonumber\\
\nonumber\\
N_{9,1} = u_{\mu} d^{\mu}d\cdot\tilde{u}^{\nu} h^{2} \qquad
N_{9,2} = u_{\mu} d^{\mu}d\cdot\tilde{u} h_{\alpha\beta} h^{\alpha\beta}
\nonumber\\
N_{9,3} = u_{\mu} d^{\mu}d^{\alpha}\tilde{u}^{\beta} h_{\alpha\beta} h \qquad
N_{9,4} = u_{\mu} d^{\mu}d^{\alpha}\tilde{u}^{\beta} h_{\alpha\lambda} {h_{\beta}}^{\lambda}
\nonumber\\
N_{9,5} = u_{\mu} d^{\alpha}d^{\beta}\tilde{u}^{\mu} h_{\alpha\beta} h \qquad
N_{9,6} = u_{\mu} d^{\alpha}d^{\beta}\tilde{u}^{\mu} h_{\alpha\lambda} {h_{\beta}}^{\lambda}
\nonumber\\
N_{9,7} = u_{\mu} d_{\nu}d\cdot\tilde{u} h^{\mu\nu} h \qquad
N_{9,8} = u_{\mu} d_{\nu}d_{\alpha}\tilde{u}_{\beta} h^{\mu\nu} h^{\alpha\beta}
\nonumber\\
N_{9,9} = u_{\mu} d_{\alpha}d_{\beta}\tilde{u}_{\nu} h^{\mu\nu} h^{\alpha\beta} \qquad
N_{9,10} = u_{\mu} d^{\rho}d\cdot\tilde{u} h^{\mu\nu} h_{\nu\rho}
\label{N-6-9}
\eea

(c) Terms of the type
$
u u \tilde{u} \tilde{u}:
$
we have terms of the type
$
u u d\tilde{u} d\tilde{u}, u u \tilde{u} dd\tilde{u}, u du \tilde{u} d\tilde{u}, u ddu \tilde{u} \tilde{u};
$
there are $23$ terms of these types. The terms of the type
$
du du \tilde{u} \tilde{u}
$
can be be written, up to a total derivative, as linear combination of these 23 terms.

\bea
N_{10,1} = u_{\mu} u_{\nu} d^{\mu}\tilde{u}^{\nu} d\cdot\tilde{u} \qquad
N_{10,2} = u_{\mu} u_{\nu} d^{\mu}\tilde{u}^{\rho} d^{\nu}\tilde{u}_{\rho} 
\nonumber\\
N_{10,3} = u_{\mu} u_{\nu} d^{\mu}\tilde{u}^{\rho} d_{\rho}\tilde{u}^{\nu} \qquad
N_{10,4} = u_{\mu} u_{\nu} d^{\rho}\tilde{u}^{\mu} d_{\rho}\tilde{u}^{\nu} 
\nonumber\\
\nonumber\\
N_{11,1} = u_{\mu} u_{\nu} \tilde{u}^{\mu} d^{\nu}d\cdot\tilde{u} \qquad
N_{11,2} = u_{\mu} u_{\nu} \tilde{u}_{\rho} d^{\rho}d^{\mu}\tilde{u}_{\nu} 
\nonumber\\
\nonumber\\
N_{12,1} = u_{\mu} d^{\mu}u^{\nu} \tilde{u}_{\nu} d\cdot\tilde{u} \qquad
N_{12,2} = u_{\mu} d^{\mu}u^{\nu} \tilde{u}_{\rho} d_{\nu}\tilde{u}^{\rho} 
\nonumber\\
N_{12,3} = u_{\mu} d^{\mu}u^{\nu} \tilde{u}_{\rho} d^{\rho}\tilde{u}_{\nu} \qquad
N_{12,4} = u_{\mu} d^{\nu}u^{\mu} \tilde{u}_{\nu} d\cdot\tilde{u} 
\nonumber\\
N_{12,5} = u_{\mu} d^{\nu}u^{\mu} \tilde{u}_{\rho} d_{\nu}\tilde{u}^{\rho} \qquad
N_{12,6} = u_{\mu} d^{\nu}u^{\mu} \tilde{u}_{\rho} d^{\rho}\tilde{u}_{\nu}
\nonumber\\
N_{12,7} = u_{\mu} d\cdot u \tilde{u}^{\mu} d\cdot\tilde{u}^{\rho} \qquad
N_{12,8} = u_{\mu} d_{\alpha}u_{\beta} \tilde{u}^{\mu} d^{\alpha}\tilde{u}^{\beta}
\nonumber\\
N_{12,9} = u_{\mu} d_{\alpha}u_{\beta} \tilde{u}^{\mu} d^{\beta}\tilde{u}^{\alpha} \qquad
N_{12,10} = u_{\mu} d\cdot u \tilde{u}^{\nu} d^{\mu}\tilde{u}_{\nu}
\nonumber\\
N_{12,11} = u_{\mu} d_{\nu}u_{\rho} \tilde{u}^{\rho} d^{\mu}\tilde{u}^{\nu} \qquad
N_{12,12} = u_{\mu} d_{\rho}u_{\nu} \tilde{u}^{\rho} d^{\rho}\tilde{u}_{\nu}
\nonumber\\
N_{12,13} = u_{\mu} d\cdot u \tilde{u}_{\nu} d^{\nu}\tilde{u}^{\mu} \qquad
N_{12,14} = u_{\mu} d_{\nu}u_{\rho} \tilde{u}^{\rho} d^{\nu}\tilde{u}^{\mu}
\nonumber\\
N_{12,15} = u_{\mu} d_{\rho} u_{\nu} \tilde{u}^{\mu} d^{\nu}\tilde{u}^{\mu}
\nonumber\\
\nonumber\\
N_{13,1} = u_{\mu} d^{\mu}d^{\alpha}u^{\beta} \tilde{u}^{\alpha} \tilde{u}_{\beta} \qquad
N_{13,2} = u_{\mu} d^{\nu}d\cdot u \tilde{u}^{\mu} \tilde{u}^{\nu}. 
\label{N-10-13}
\eea

The next step is to consider the gauge variations
$
d_{Q}N_{j}
$
for all
$
43 + 83 + 23 = 149
$
terms from the three sectors described above. Some of these terms can be eliminated with coboundaries of the type
$
d_{Q}B;
$
there are $26$ such coboundaries: we can make 
\bea
c_{2,2} = 0
\nonumber\\
c_{3,k} = 0,\quad k = 5,8
\nonumber\\
c_{4,k} = 0,\quad k = 3,10,11,12
\nonumber\\
c_{5,k} = 0,\quad k = 14,\dots,19
\nonumber\\
c_{6,k} = 0,\quad k = 4,5,13
\nonumber\\
c_{7,k} = 0,\quad k = 1,3,4,6,8,11,12,16,17
\eea
so we are left with $123$ independent terms. We impose the condition that the sum 
$
\sum c_{j,k}d_{Q}N_{j,k}
$
is a total derivative. In principle we should make an ansatz for the total derivative also, but we can simplify
the analysis is we observe that one can write
\be
d_{Q}N = i~u_{\mu}~X^{\mu} + i~h_{\alpha\beta}~X^{\alpha\beta} + {\rm total~derivative};
\label{dQN}
\ee
indeed for the terms in the sector (a) we have to ``move" all derivatives appearing on the $u$ factor on the $h$
factors, up to a total derivative. In the (c) sector we ``move" all derivatives appearing on the 
$
h_{\alpha\beta}
$ 
factor on the 
$
u, \tilde{u}
$
factors, up to a total derivative. Sector (b) is a mixted one i.e. we have to perform both tricks on various terms.

Let us make a general ansatz 
\be
N^{\mu} = u_{\nu}~R^{\mu\nu} + d_{\nu}u_{\rho}~R^{\mu\nu\rho} + h_{\alpha\beta}~S^{\mu,\alpha\beta}
+ d_{\nu}h_{\alpha\beta}~S^{\mu\nu,\alpha\beta}
\label{N-R}
\ee
with 
$
R^{\mu\nu}, R^{\mu\nu\rho} \sim h h h
$
and
$
S^{\mu,\alpha\beta}, S^{\mu\nu,\alpha\beta} \sim u u \tilde{u}.
$
It is convenient to split 
\be
R^{\mu\nu\rho} = R_{+}^{\mu\nu\rho} + R_{-}^{\mu\nu\rho}
\ee
where 
$
R_{\pm}^{\mu\nu\rho}
$
are symmetric (resp. anti-symmetric) in
$
\mu \leftrightarrow \nu.
$
Then we write
\bea
d_{\nu}u_{\rho}~R_{-}^{\mu\nu\rho} = d_{\nu}(u_{\rho} R_{-}^{\mu\nu\rho}) - u_{\rho} d_{\nu}R_{-}^{\mu\nu\rho}
\nonumber
\eea
and notice that the first term can be neglected because it gives a null contribution in
$
d_{\mu}N^{\mu}
$
and the second term can be absorbed in the first term of (\ref{N-R}). So, we can assume that 
$
R^{\mu\nu\rho}
$
is symmetric in
$
\mu \leftrightarrow \nu.
$
In a similar way we can argue that
$
S^{\mu\nu,\alpha\beta}
$
can be chosen symmetric in
$
\mu \leftrightarrow \nu.
$
It means that we can write
\bea
R^{\mu\nu\rho} = \eta^{\mu\nu}~R^{\rho} + \tilde{R}^{\mu\nu\rho} 
\nonumber\\
S^{\mu\nu,\alpha\beta} = \eta^{\mu\nu}~S^{\alpha\beta} + \tilde{S}^{\mu\nu,\alpha\beta}
\eea
where 
$
\tilde{R}^{\mu\nu\rho} 
$
and
$
\tilde{S}^{\mu\nu,\alpha\beta}
$
do {\bf not} contain the factor 
$
\eta^{\mu\nu}.
$

Then the equation (\ref{sN-a}) becomes equivalent to the following system:
\bea
X^{\mu} = d_{\nu}R^{\nu\mu}
\nonumber\\
R^{\mu\nu} + d_{\rho}R^{\rho\mu\nu} = 0
\nonumber\\
\tilde{R}^{\mu\nu\rho} = 0
\nonumber\\
X^{\alpha\beta} = d_{\mu}S^{\mu,\alpha\beta}
\nonumber\\
S^{\mu,\alpha\beta} + d_{\nu}S^{\nu\mu,\alpha\beta} = 0
\nonumber\\
\tilde{S}^{\mu\nu,\alpha\beta} = 0.
\eea
It easy to derive from this system that
\be
X^{\mu} = - \Box R^{\mu}, \qquad
X^{\alpha\beta} = - \Box S^{\alpha\beta}.
\label{X}
\ee

So, we must consider first, the most general forms for 
$
R^{\mu}
$
and 
$
S^{\alpha\beta}.
$
We have 
\be
R^{\mu} = \sum \lambda_{j} R^{\mu}_{j}
\ee
where
\bea
R^{\mu}_{1} = d^{\mu}h h^{2}, \quad 
R^{\mu}_{2} = d_{\nu}h h h^{\mu\nu},
\nonumber\\
R^{\mu}_{3} = d_{\mu}h^{\mu\nu} h^{2}, \quad 
R^{\mu}_{4} = d^{\mu}h h_{\alpha\beta} h^{\alpha\beta},
\nonumber\\
R^{\mu}_{5} = d_{\mu}h^{\alpha\beta} h_{\alpha\beta} h, \quad 
R^{\mu}_{6} = d^{\alpha}h^{\mu\beta} h_{\alpha\beta} h,
\nonumber\\
R^{\mu}_{7} = d^{\rho}h_{\nu\rho} h^{\mu\nu} h, \quad 
R^{\mu}_{8} = d^{\nu}h h^{\mu\rho} h_{\nu\rho},
\nonumber\\
R^{\mu}_{9} = d^{\mu}h^{\alpha\beta} h_{\alpha\lambda} {h_{\beta}}^{\lambda}, \quad 
R^{\mu}_{10} = d_{\nu}h^{\mu\nu} h_{\alpha\beta} h^{\alpha\beta},
\nonumber\\
R^{\mu}_{11} = d^{\alpha}h^{\mu\beta} h_{\alpha\lambda} {h_{\beta}}^{\lambda}, \quad 
R^{\mu}_{12} = d_{\nu}h^{\alpha\beta} h^{\mu\nu} h_{\alpha\beta},
\nonumber\\
R^{\mu}_{13} = d_{\alpha}h_{\nu\beta} h^{\mu\nu} h^{\alpha\beta}, \quad 
R^{\mu}_{14} = d_{\sigma}h^{\rho\sigma} h^{\mu\nu} h_{\nu\rho}.
\eea

Also we have 
\be
S^{\alpha\beta} = \sum \omega_{j} S^{\alpha\beta}_{j}
\ee
where
\bea
S^{\alpha\beta}_{1} = {\cal S}_{\alpha\beta} (u^{\alpha} d^{\beta}u^{\mu} \tilde{u}_{\mu}), \quad 
S^{\alpha\beta}_{2} = {\cal S}_{\alpha\beta} (u^{\alpha} d^{\mu}u^{\beta} \tilde{u}_{\mu}),
\nonumber\\
S^{\alpha\beta}_{3} = {\cal S}_{\alpha\beta} (u^{\alpha} d \cdot u \tilde{u}^{\beta}), \quad 
S^{\alpha\beta}_{4} = {\cal S}_{\alpha\beta} (u^{\mu} d^{\alpha}u^{\beta} \tilde{u}_{\mu}),
\nonumber\\
S^{\alpha\beta}_{5} = {\cal S}_{\alpha\beta} (u^{\mu} d^{\alpha}u_{\mu} \tilde{u}^{\beta}), \quad 
S^{\alpha\beta}_{6} = {\cal S}_{\alpha\beta} (u_{\mu} d^{\mu}u^{\alpha} \tilde{u}^{\beta}),
\nonumber\\
S^{\alpha\beta}_{7} = \eta^{\alpha\beta} u^{\mu} d_{\mu}u_{\nu} \tilde{u}^{\nu}, \quad 
S^{\alpha\beta}_{8} = \eta^{\alpha\beta} u^{\mu} d_{\nu}u_{\mu} \tilde{u}^{\nu},
\nonumber\\
S^{\alpha\beta}_{9} = \eta^{\alpha\beta} u^{\mu} d\cdot u \tilde{u}_{\mu}, \quad 
S^{\alpha\beta}_{10} = {\cal S}_{\alpha\beta} (u^{\alpha} u_{\mu} d^{\beta}\tilde{u}_{\mu}),
\nonumber\\
S^{\alpha\beta}_{11} = {\cal S}_{\alpha\beta} (u^{\alpha} u^{\mu} d_{\mu}\tilde{u}^{\beta}, \quad 
S^{\alpha\beta}_{12} = \eta^{\alpha\beta} u^{\mu} u^{\nu} d_{\mu}\tilde{u}_{\nu}.
\eea

Next, we are left with a routine operation: to consider all $123$ variables
$
c_{j,k}
$
from the expresion of $N$ and the $14 + 12$ variables
$
\lambda_{j}, \omega_{k},
$
exhibit
$
d_{Q}N
$
in the form (\ref{dQN}) and solve the system (\ref{X}). We are left with a highly over-determined 
system of $256$ equations for the $149$ variables 
$
c_{j,k}, \lambda_{j}, \omega_{k}.
$
After tedious computations one is left with $14$ solutions
\bea
N = a_{1}~N_{1} + a_{2}~( N_{3,1} + 2 N_{3,3} ) + a_{3}~( 2 N_{3,3} + N_{3,6} )
\nonumber\\
+ a_{4}~N_{4,5} + a_{5}~N_{4,8} + a_{6}~(2 N_{5,2} + N_{5,8})
\nonumber\\
+ a_{7}~(2 N_{5,9} + N_{5,12}) + a_{8}~(N_{6,1} + 2 N_{8,9}) + a_{9}~(N_{6,2} + 2 N_{8,10}) 
\nonumber\\
+ a_{10}~( N_{6,16} + N_{8,17} + N_{8,27}) + a_{11}~( N_{6,18} + N_{8,33} + N_{8,35})
\nonumber\\
+ a_{12}~( N_{10,4} + N_{12,8} - N_{12,14})
+ a_{13}~(- N_{10,1} + N_{11,2} + N_{12,12} - N_{12,15}) 
\nonumber\\
+ a_{14}~( N_{12,7} + N_{12,11} - N_{12,12} - N_{12,13} - N_{13,1} + N_{13,2})
\label{N-1-13}
\eea
and one can prove that they are total derivatives:
\be
N = d_{\mu}N^{\mu}
\ee
where
\bea
N^{\mu}_{1} = {1\over 3}~h^{3} d^{\mu}h, \quad 
N^{\mu}_{2} = h d^{\mu}h h_{\alpha\beta} h^{\alpha\beta},
\nonumber\\
N^{\mu}_{3} = h^{2} h_{\alpha\beta}d^{\mu}h^{\alpha\beta}, \quad 
N^{\mu}_{4} = {1\over 3}~d^{\mu}h h_{\lambda\alpha} {h^{\lambda}}_{\beta} h^{\alpha\beta},
\nonumber\\
N^{\mu}_{5} = {1\over 2}~\Bigl( h h_{\lambda\alpha} {h^{\lambda}}_{\beta} d^{\mu}h^{\alpha\beta}
- {1\over 3}~d^{\mu}h h_{\lambda\alpha} {h^{\lambda}}_{\beta} h^{\alpha\beta}\Bigl), \quad 
N^{\mu}_{6} = d^{\mu}h^{\alpha\beta} h_{\alpha\rho} h_{\beta\sigma} h^{\rho\sigma},
\nonumber\\
N^{\mu}_{7} = h^{\rho\sigma} h_{\rho\sigma} h_{\alpha\beta} d^{\mu}h^{\alpha\beta}, \quad
N^{\mu}_{8} = {1\over 2}~\Bigl( u_{\nu} d^{\mu}\tilde{u}^{\nu} h^{2}
+ 2 u_{\nu} \tilde{u}^{\nu} h d^{\mu}h  - d^{\mu}u^{\nu} \tilde{u}_{\nu} h^{2}\Bigl)
\nonumber\\
N^{\mu}_{9} = {1\over 2}~\Bigl( u_{\nu} d^{\mu}\tilde{u}^{\nu} h^{\alpha\beta} h_{\alpha\beta}
+ 2 u_{\nu} \tilde{u}^{\nu} h_{\alpha\beta} d^{\mu}h^{\alpha\beta}  
- d^{\mu}u^{\nu} \tilde{u}_{\nu} h^{\alpha\beta} h_{\alpha\beta} \Bigl)
\nonumber\\
N^{\mu}_{10} = {1\over 2}~\Bigl( u^{\nu} d^{\mu}\tilde{u}^{\rho} h_{\nu\rho} h
+ u_{\nu} \tilde{u}_{\rho} d^{\mu}h^{\nu\rho} h +  u_{\nu} \tilde{u}_{\rho} h^{\nu\rho} d^{\mu}h 
- d^{\mu}u^{\nu} \tilde{u}^{\rho} h_{\nu\rho} h \Bigl)
\nonumber\\
N^{\mu}_{11} = {1\over 2}~\Bigl( u_{\nu} d^{\mu}\tilde{u}^{\rho} h^{\nu\sigma} h_{\rho\sigma}
+ u_{\nu} \tilde{u}^{\rho} d^{\mu}h^{\nu\sigma} h_{\rho\sigma} 
+  u_{\nu} \tilde{u}^{\rho} h^{\nu\sigma} d^{\mu}h_{\rho\sigma} 
- d^{\mu}u^{\nu} \tilde{u}_{\rho} h_{\nu\sigma} h^{\rho\sigma} \Bigl)
\nonumber\\
N^{\mu}_{12} = {1\over 2}~\Bigl( u_{\rho} d^{\mu}u_{\nu} \tilde{u}^{\rho} \tilde{u}^{\nu}
+ u_{\rho} u_{\nu} d^{\mu}\tilde{u}^{\rho} \tilde{u}^{\nu} 
+ u_{\rho} u_{\nu} \tilde{u}^{\rho} d^{\mu}\tilde{u}^{\nu}
- d^{\mu}u_{\rho} u_{\nu} \tilde{u}^{\rho} \tilde{u}^{\nu} \Bigl)  
\nonumber\\
N^{\mu}_{13} = u_{\alpha} u_{\beta} \tilde{u}^{\mu} d^{\alpha}\tilde{u}^{\beta}, \quad 
N^{\mu}_{14} = u_{\nu} d\cdot u \tilde{u}^{\mu} \tilde{u}^{\nu} 
- u^{\mu} d_{\alpha}u_{\beta} \tilde{u}^{\alpha} \tilde{u}^{\beta}.
\label{N-mu-1-13}
\eea

The expresions 
$
N_{1} - N_{12}
$
from (\ref{N-1-13}) are grouping in a clever way all terms of the form 
$$
a_{1} a_{2} d_{\mu}a_{3} d^{\mu}a_{4}
$$
from the lists (\ref{N-1-5}), (\ref{N-6-9}) and (\ref{N-10-13}) with the exception of
$
N_{12,5}.
$
In fact the general structure is
\bea
a_{1} ( d^{\mu}a_{2} d_{\mu}a_{3} a_{4} + a_{2} d_{\mu}a_{3} d^{\mu}a_{4} + d^{\mu}a_{2} a_{3} d_{\mu}a_{4})
\nonumber\\
\simeq {1\over 2}~a_{1}~\Box ( a_{2} a_{3} a_{4}) = {\rm total~divergence} + {1\over 2}~\Box a_{1}~a_{2} a_{3} a_{4}
\simeq {\rm total~divergence}
\eea
where, as before, $\simeq$ means modulo equations of motion (Klein-Gordon).
\newpage
So we have the important result:
\begin{thm}
Let $N$ be a quadri-linear polynomial in the variables verifying
\be
\omega(N) = 6, \qquad gh(N) = 0
\ee
and the gauge invariance condition
\be
d_{Q}N = i~d_{\mu}N^{\mu}
\label{sN-b}
\ee
for some 
$
N^{\mu}.
$
Then $N$ is a coboundary i.e.
\be
N \simeq d_{Q}B + d_{\mu}B^{\mu}.
\ee
\end{thm}

This theorem proves that, after we eliminate the anomaly (\ref{st2-2-1aa}) with the finite renormalization (\ref{second-order}),
the arbitrariness of the chronological product
$
T(T(x_{1}),T(x_{2}))
$
is trivial in the quadri-linear sector i.e. a coboundary. Presumably this is true also for all
$
N^{I}.
$
So, the only arbitrariness left is in the tri-linear sector where we obtain the interaction Lagrangian (\ref{T}) $+$ (\ref{T-int}).
This arbitrariness amounts to a redefinition of the coupling constant. The details are in 
\begin{thm}
Let $N$ be a tri-linear polynomial in the variables verifying
\be
\omega(N) = 6, \qquad gh(N) = 0
\ee
and the gauge invariance condition (\ref{sN-b})
Then $N$ is a coboundary i.e.
\be
N = d_{Q}B + d_{\mu}B^{\mu}.
\ee
\end{thm}
{\bf Proof:} (i) First, we note that for tri-linear finite renormalizations we can have in (\ref{R-IJ}) a supplementary term 
\be
R(T^{I}(x_{1}),T^{J}(x_{2})) = \cdots + 
\partial_{\mu}\partial_{\nu}\partial_{\rho}(x_{1} - x_{2})~N(T^{I},T^{J})^{\mu\nu\rho}(x_{2})
\label{R-IJ-d}
\ee
where 
\be
\omega(N(T^{I},T^{J})^{\mu\nu\rho}) = 3, \qquad gh(N(T^{I},T^{J})^{\mu\nu\rho}) = |I| + |J|.
\ee
First we consider the expression
$
N(T,T)^{\mu\nu\rho};
$
because is of ghost number $0$ it must have the form:
\be
N(T,T)^{\mu\nu\rho} \sim h h h + h u \tilde{u}.
\ee
The existence of such an expression is prevented by Lorentz covariance, so 
\be
N(T,T)^{\mu\nu\rho} = 0.
\ee
Similarly we must have 
\be
N(T^{\alpha},T)^{\mu\nu\rho} \sim u h h + u u \tilde{u}
\ee
and, again, the existence of such an expression is prevented by Lorentz covariance, so 
\be
N(T^{\alpha},T)^{\mu\nu\rho} = 0.
\ee
Next we have
\be
N(T^{\alpha},T^{\beta})^{\mu\nu\rho},\quad N(T^{\alpha\beta},T)^{\mu\nu\rho} \sim u u h
\ee
and, as before, the existence of such an expression is prevented by Lorentz covariance, so 
\be
N(T^{\alpha},T^{\beta})^{\mu\nu\rho} = 0,\qquad N(T^{\alpha\beta},T)^{\mu\nu\rho} = 0.
\ee
Finally, 
\be
N(T^{\alpha\beta},T^{\gamma})^{\mu\nu\rho},\quad  N(T^{\alpha\beta\gamma},T)^{\mu\nu\rho} \sim u u u
\ee
and the existence of such an expression is prevented by Lorentz covariance, so 
\be
N(T^{\alpha\beta},T^{\gamma})^{\mu\nu\rho} = 0,\qquad N(T^{\alpha\beta\gamma},T)^{\mu\nu\rho} = 0.
\ee
So we have:
\be
N(T^{I},T^{J})^{\mu\nu\rho} = 0.
\ee

(ii) Next, we observe that the assertion of theorem \ref{NIJ2} stays true for the case when the expression
$
N(T^{I},T^{J})^{\mu\nu}
$
is tri-linear in the fields variables. Indeed because 
$
gh(N(T^{I},T^{J}))^{\mu\nu} = |I| + |J|
$
we must have
\be
N(T^{I},T^{J})^{\mu\nu} = 0, \qquad |I| + |J| \geq 4
\ee
because an expression tri-linear in the fields can have the ghost number at most $3$. It follows that we have only the cases
(i) - (vi) of theorem \ref{NIJ2}; but in these cases we did not used the explicit form of 
$
N(T^{I},T^{J})^{\mu\nu};
$
we have proved that we can fix
\be
N(T^{I},T^{J})^{\mu\nu} = 0
\ee
only through general cohomological arguments. The same observation applies to the expressions
$
N(T^{I},T^{J})^{\mu};
$
we need only the steps (i) - (vi) of theorem \ref{NIJ3} so we also have
\be
N(T^{I},T^{J})^{\mu} = 0
\ee
in this case. It follows that we are reduced to the study of equation (\ref{sN-a}) for a tri-linear expression $N$. Such an equation
leads to a solution cohomological to (\ref{T}) $+$ (\ref{T-int}) - see \cite{Sc2}, \cite{massive}.
$\qed$

So, at least in the second order of perturbation theory, we have proved that quantum gravity is renormalizable as in the 
pure Yang-Mills case!

These is a slight discrepancy with the pure Yang-Mills case. In \cite{wick+hopf} we have proved that the finite renormalizations
used to eliminate the anomalies in the second order of the perturbation theory can be obtained by a clever redefinition of the 
chronological products
$
T(\xi_{a,\mu}(x_{1}),\xi_{b,\nu}(x_{2})). 
$

In the quantum gravity formalism used above we have the follow possible redefinitions:
\bea
N_{1}(h_{\mu_{1}\nu_{1},\lambda_{1}}(x_{1}),h_{\mu_{2}\nu_{2},\lambda_{2}}(x_{2}))
= {1\over 2}~(\eta_{\mu_{1}\mu_{2}}~\eta_{\nu_{1}\nu_{2}} + \eta_{\mu_{1}\nu_{2}}~\eta_{\nu_{1}\mu_{2}})~
\eta_{\lambda_{1}\lambda_{2}}~\delta(x_{1} - x_{2})
\nonumber\\
N_{2}(h_{\mu_{1}\nu_{1},\lambda_{1}}(x_{1}),h_{\mu_{2}\nu_{2},\lambda_{2}}(x_{2}))
= \eta_{\mu_{1}\nu_{1}}~\eta_{\mu_{2}\nu_{2}}~\eta_{\lambda_{1}\lambda_{2}}~\delta(x_{1} - x_{2})
\nonumber\\
N_{3}(h_{\mu_{1}\nu_{1},\lambda_{1}}(x_{1}),h_{\mu_{2}\nu_{2},\lambda_{2}}(x_{2}))
= {1\over 4}~(\eta_{\mu_{1}\lambda_{1}}~\eta_{\mu_{2}\lambda_{2}}~\eta_{\nu_{1}\nu_{2}}
+ \eta_{\nu_{1}\lambda_{1}}~\eta_{\mu_{2}\lambda_{2}}~\eta_{\mu_{1}\nu_{2}}
\nonumber\\
+ \eta_{\mu_{1}\lambda_{1}}~\eta_{\nu_{2}\lambda_{2}}~\eta_{\nu_{1}\mu_{2}}
+ \eta_{\nu_{1}\lambda_{1}}~\eta_{\nu_{2}\lambda_{2}}~\eta_{\mu_{1}\mu_{2}})~\delta(x_{1} - x_{2})
\nonumber\\
N_{4}(h_{\mu_{1}\nu_{1},\lambda_{1}}(x_{1}),h_{\mu_{2}\nu_{2},\lambda_{2}}(x_{2}))
= {1\over 4}~(\eta_{\mu_{1}\lambda_{2}}~\eta_{\mu_{2}\lambda_{1}}~\eta_{\nu_{1}\nu_{2}}
+ \eta_{\nu_{1}\lambda_{2}}~\eta_{\mu_{2}\lambda_{1}}~\eta_{\mu_{1}\nu_{2}}
\nonumber\\
+ \eta_{\mu_{1}\lambda_{2}}~\eta_{\nu_{2}\lambda_{1}}~\eta_{\nu_{1}\mu_{2}}
+ \eta_{\nu_{1}\lambda_{2}}~\eta_{\nu_{2}\lambda_{1}}~\eta_{\mu_{1}\mu_{2}})~\delta(x_{1} - x_{2})
\nonumber\\
N_{5}(h_{\mu_{1}\nu_{1},\lambda_{1}}(x_{1}),h_{\mu_{2}\nu_{2},\lambda_{2}}(x_{2}))
= {1\over 4}~(\eta_{\mu_{1}\nu_{1}}~\eta_{\mu_{2}\lambda_{2}}~\eta_{\nu_{2}\lambda_{1}}
+ \eta_{\mu_{1}\nu_{1}}~\eta_{\nu_{2}\lambda_{2}}~\eta_{\mu_{2}\lambda_{1}}
\nonumber\\
+ \eta_{\mu_{2}\nu_{2}}~\eta_{\mu_{1}\lambda_{1}}~\eta_{\nu_{1}\lambda_{2}}
+ \eta_{\mu_{2}\nu_{2}}~\eta_{\nu_{1}\lambda_{1}}~\eta_{\mu_{1}\lambda_{2}})~\delta(x_{1} - x_{2})
\eea
\bea
N_{6}(u_{\mu_{1},\nu_{1}}(x_{1}),\tilde{u}_{\mu_{2},\nu_{2}}(x_{2}))
= \eta_{\mu_{1}\mu_{2}}~\eta_{\nu_{1}\nu_{2}}~\delta(x_{1} - x_{2})
\nonumber\\
N_{7}(u_{\mu_{1},\nu_{1}}(x_{1}),\tilde{u}_{\mu_{2},\nu_{2}}(x_{2}))
= \eta_{\mu_{1}\nu_{2}}~\eta_{\nu_{1}\mu_{2}}~\delta(x_{1} - x_{2})
\nonumber\\
N_{8}(u_{\mu_{1},\nu_{1}}(x_{1}),\tilde{u}_{\mu_{2},\nu_{2}}(x_{2}))
= \eta_{\mu_{1}\nu_{1}}~\eta_{\mu_{2}\mu_{2}}~\delta(x_{1} - x_{2})
\eea
\bea
N_{9}(u_{\mu,\alpha\beta}(x_{1}),\tilde{u}_{\nu}(x_{2}))
= \eta_{\mu\nu}~\eta_{\alpha\beta}~\delta(x_{1} - x_{2})
\nonumber\\
N_{10}(u_{\mu,\alpha\beta}(x_{1}),\tilde{u}_{\nu}(x_{2}))
= {1\over 2}~(\eta_{\mu\alpha}~\eta_{\nu\beta} + \eta_{\mu\beta}~\eta_{\nu\alpha})~\delta(x_{1} - x_{2})
\eea
\bea
N_{11}(u_{\mu,\alpha\beta}(x_{1}),\tilde{u}_{\nu,\rho}(x_{2}))
= {1\over 2}~\eta_{\mu\nu}~
(\eta_{\alpha\rho}~\partial_{\beta} + \eta_{\beta\rho}~\partial_{\alpha})~\delta(x_{1} - x_{2})
\nonumber\\
N_{12}(u_{\mu,\alpha\beta}(x_{1}),\tilde{u}_{\nu,\rho}(x_{2}))
= {1\over 2}~\eta_{\mu\rho}~
(\eta_{\alpha\nu}~\partial_{\beta} + \eta_{\beta\nu}~\partial_{\alpha})~\delta(x_{1} - x_{2})
\nonumber\\
N_{13}(u_{\mu,\alpha\beta}(x_{1}),\tilde{u}_{\nu,\rho}(x_{2}))
= {1\over 2}~\eta_{\nu\rho}~
(\eta_{\alpha\mu}~\partial_{\beta} + \eta_{\beta\mu}~\partial_{\alpha})~\delta(x_{1} - x_{2})
\nonumber\\
N_{14}(u_{\mu,\alpha\beta}(x_{1}),\tilde{u}_{\nu,\rho}(x_{2}))
= \eta_{\mu\nu}~\eta_{\alpha\beta}~\partial_{\rho}~\delta(x_{1} - x_{2})
\nonumber\\
N_{15}(u_{\mu,\alpha\beta}(x_{1}),\tilde{u}_{\nu,\rho}(x_{2}))
= \eta_{\mu\rho}~\eta_{\alpha\beta}~\partial_{\nu}~\delta(x_{1} - x_{2})
\nonumber\\
N_{16}(u_{\mu,\alpha\beta}(x_{1}),\tilde{u}_{\nu,\rho}(x_{2}))
= \eta_{\nu\rho}~\eta_{\alpha\beta}~\partial_{\mu}~\delta(x_{1} - x_{2})
\nonumber\\
N_{17}(u_{\mu,\alpha\beta}(x_{1}),\tilde{u}_{\nu,\rho}(x_{2}))
= {1\over 2}~(\eta_{\mu\alpha}~\eta_{\nu\beta} + \eta_{\mu\beta}~\eta_{\nu\alpha})~
\partial_{\rho}~\delta(x_{1} - x_{2})
\nonumber\\
N_{18}(u_{\mu,\alpha\beta}(x_{1}),\tilde{u}_{\nu,\rho}(x_{2}))
= {1\over 2}~(\eta_{\mu\alpha}~\eta_{\rho\beta} + \eta_{\mu\beta}~\eta_{\rho\alpha})~
\partial_{\nu}~\delta(x_{1} - x_{2})
\nonumber\\
N_{19}(u_{\mu,\alpha\beta}(x_{1}),\tilde{u}_{\nu,\rho}(x_{2}))
= {1\over 2}~(\eta_{\nu\alpha}~\eta_{\rho\beta} + \eta_{\nu\beta}~\eta_{\rho\alpha})~
\partial_{\mu}~\delta(x_{1} - x_{2})
\eea

However, we did not succeed to exhibit the finite renormalization (\ref{second-order})
using these redefinitions, as in the pure YM case.

\newpage
\section{Conclusions}

The main result of this paper is that perturbative quantum gravity is in fact a renormalizable theory, at least in the second order 
of the perturbation theory. Mathematically it means that a certain (relative) cohomology problem is trivial. The proof is brute force, based 
on the elaboration of a long list of possible terms appearing in the generic form of a cocyle and reducing the cocyle equation to 
a long system of linear equations which must be solved. It is clear that this method becomes unmanageable in higher orders of the 
perturbation theory, so the next goal would be to find more sophisticated ways to solve the cohomology problem appearing in 
higher orders of perturbation theory.


\begin{thebibliography}{99}

\bibitem{BS}
N. N. Bogoliubov, D. Shirkov,
``{\it Introduction to the Theory of Quantized Fields}",
John Wiley and Sons, 1976 (3rd edition)

\bibitem{BLOT}
N. N. Bogolubov, A. A. Logunov, A.I. Oksak, I. Todorov, 
``{\it General Principles of Quantum Field Theory}", Kluwer 1989

\bibitem{D}
M. D\"utsch, ``{\it From Classical Field Theory to Perturbative Quantum Field Theory}", 
Progress in Mathematical Physics {\bf 74}, Springer 2019

\bibitem{DF}
M. D\"utsch, K. Fredenhagen,
``{\it Algebraic Quantum Field Theory, Perturbation Theory,
and the Loop Expansion}",
arXiv: hep-th/0001129, Commun. Math. Phys. {\bf 219} (2001) 5 - 30

\bibitem{EG}
H. Epstein, V. Glaser,
``{\it The R\^ole of Locality in Perturbation Theory}",
Ann. Inst. H. Poincar\'e {\bf 19 A} (1973) 211-295

\bibitem{Gl}
V. Glaser,
``{\it Electrodynamique Quantique}",
L'enseignement du 3e cycle de la physique en Suisse Romande (CICP), Semestre
d'hiver 1972/73

\bibitem{gravity}
D. R. Grigore,
``{\it On the Quantization of the Linearized Gravitational Field}", \\
hep-th/9905190, Class. Quant. Grav. {\bf 17} (2000) 319-344

\bibitem{algebra}
D. R. Grigore,
``{\it A Generalization of Gauge Invariance}", 
arxiv: hep-th/1612.04998, Journal of Mathematical Physics {\bf 58} (2017) 082303

\bibitem{ano-free}
D. R. Grigore, ``{\it Anomaly-Free Gauge Models: A Causal Approach}", 
hep-th/1804.08276, Romanian Journ. Phys. {\bf 64} (2019) 102

\bibitem{sr-gr} 
D. R. Grigore,
``{\it On the Super-Renormalizablity of Quantum Gravity in the Linear Approximation}",
arXiv:1905.05410v1 [hep-th], Romanian Journ. Phys. {\bf 65} (2020) 101

\bibitem{wick+hopf}
D. R. Grigore,
 ``{\it Wick Theorem and Hopf Algebra Structure in Causal Perturbative Quantum Field Theory}",
arXiv:2202.08056v2 [hep-th]

\bibitem{massive}
D. R. Grigore, G. Scharf,
``{\it Massive Gravity as a Quantum Gauge Theory}", \\
arXiv: hep-th/0404157, General Relativity and Gravitation {\bf 37} (2005) 1075-1096

\bibitem{H}
K. Hepp, ``{\it Renormalization Theory}", in ``{\it Statistical Mechanics and Quantum Field Theory}" pp. 429 - 500,
(Les Houches 1970), C. DeWitt-Morette, Raymond Stora (eds.), Gordon and Breach 1971

\bibitem{P}
J. Polchinski, ``{\it Renormalization and Effective Lagrangians}",
Nucl. Phys. {\bf B 231} (1984) 269 - 295

\bibitem{PS}
G. Popineau, R. Stora, 
``{\it A Pedagogical Remark on the Main Theorem of Perturbative Renormalization Theory}", 
Nuclear Physics {\bf B 912} (2016) 70 - 78 

\bibitem{S}
M. Salmhofer, ``{\it Renormalization: An Introduction}", (Theoretical and Mathematical Physics) Springer 1999

\bibitem{Sc1}
G. Scharf,
``{\it Finite Quantum Electrodynamics: The Causal Approach}",
(second edition) Springer, 1995; (third edition) Dover, 2014

\bibitem{Sc2}
G. Scharf,
``{\it Quantum Gauge Theories. A True Ghost Story}",
John Wiley, 2001,
``{\it Quantum Gauge Theories - Spin One and Two}",
Google books, 2010
and
``{\it Gauge Field Theories: Spin One and Spin Two, 100 Years After General Relativity}", Dover 2016

\bibitem{Sto1}
R. Stora,
``{\it Lagrangian Field Theory}",
Les Houches lectures, Gordon and Breach, N.Y., 1971, 
C. De Witt, C. Itzykson eds.

\bibitem{St1}
O. Steinmann,
``{\it Perturbation Expansions in Axiomatic Field Theory}",
Lect. Notes in Phys. {\bf 11}, Springer, 1971

\bibitem{WG}
A. S. Wightman, L. G\aa rding,
``{\it Fields as Operator-Valued Distributions in Relativistic Quantum Field
Theory}", Arkiv Fysik {\bf 28} (1965) 129-184

\end{thebibliography}
\end{document}